\PassOptionsToPackage{colorlinks,citecolor=blue,urlcolor=blue}{hyperref}
\documentclass[11pt]{article}

\usepackage[margin=1in]{geometry}
\usepackage{lipsum}
\usepackage{amsthm,amsmath,amsfonts,amssymb}
\usepackage[authoryear]{natbib}
\usepackage{xcolor}
\usepackage{float}
\usepackage{booktabs}
\usepackage{tikz}
\usepackage{orcidlink}
\usetikzlibrary{bayesnet}
\usetikzlibrary{positioning}
\usepackage[colorlinks,citecolor=blue,urlcolor=blue]{hyperref}
\usepackage{graphicx}
\usepackage{url}

\usepackage[T1]{fontenc}
\usepackage[utf8]{inputenc} % safe even if modern LaTeX ignores it
\usepackage{float}
\usepackage{algorithm}
\usepackage{algpseudocode}
\usepackage{empheq}
\usepackage{empheq}
\usepackage{booktabs}   % for \toprule, |rule, \bottomrule
\usepackage{multirow}   % for \multirow
\usepackage{etoolbox}
\usepackage{bbm}
\usepackage{tikz}
\usepackage{bm}
\usetikzlibrary{positioning,decorations.pathreplacing}

\definecolor{darkblue}{RGB}{0,51,102}
\definecolor{dodgerblue}{RGB}{30,144,255}
\definecolor{darkgreen}{RGB}{0,100,0}
\definecolor{ygreen}{RGB}{34,139,34}
\definecolor{darkorange}{RGB}{255,140,0}

\def\oper#1{\csdef{#1}{\operatorname{#1}}}
\forcsvlist\oper{GL,SL,deg,Hom,End,Vect,Mod,Rep,Ker,Im,Id}

\def\defbb#1{\csdef{b#1}{\mathbb{#1}}}
\forcsvlist\defbb{A,B,C,D,E,G,H,I,J,K,L,M,N,O,Q,R,S,T,U,V,W,X,Y,Z}

\def\defcal#1{\csdef{c#1}{\mathcal{#1}}}
\forcsvlist\defcal{A,B,C,D,E,G,H,I,J,K,L,M,N,O,Q,R,S,T,U,V,W,X,Y,Z}

\theoremstyle{plain}

\theoremstyle{definition}

\title{Sequential Bayesian analysis of time-varying transmission in stratified epidemic renewal models}

\author{
  Dhorasso Temfack\,\orcidlink{0009-0007-0630-7504}\thanks{School of Computer Science and Statistics, Trinity College Dublin, Ireland. Email: \href{mailto:temfackd@tcd.ie}{temfackd@tcd.ie}} \and
 Jason Wyse\,\orcidlink{0000-0003-1391-7371}\footnotemark[1]
}

\date{} % or specify a date

\begin{document}

\maketitle

%Abstract

\begin{abstract} 
Understanding how transmission varies across age groups is important for epidemic surveillance and targeted intervention, but estimating age-specific transmission in near real time is challenging when infections are only partially observed, observations are delayed, and transmission varies across age groups. We develop a Bayesian age-stratified renewal model that jointly estimates latent infections, time-varying age-specific susceptibility, and age-specific contributions to transmission within an empirically specified contact structure. To enable sequential inference under these processes, we use a block particle filtering approach embedded within a sequential Monte Carlo squared framework, with fixed-lag smoothing to account for temporal dependence and delayed observations. From the inferred transmission process, we derive a next-generation matrix, effective reproduction number, and reproduction contributions. Applied to age-stratified COVID-19 surveillance data from Ireland during the Omicron dominant period, the framework reveals heterogeneity in susceptibility and transmission, with older adults exhibiting higher susceptibility, while younger and working-age groups make larger contributions to transmission because of their higher contact rates. These results demonstrate how statistical inference can provide age-specific transmission estimates to support targeted epidemic monitoring and intervention.
\end{abstract}

% % Keywords
% \noindent\textbf{Keywords:} age-stratified epidemics, renewal process models, sequential Monte Carlo

\section{Introduction}

Real-time estimation of age-stratified transmission from surveillance data is central to epidemic monitoring. Age structure shapes infectious disease dynamics through heterogeneous contacts, differential reporting, and age-specific responses to interventions. Observed incidence therefore reflects interacting transmission processes across subpopulations, and ignoring this structure can obscure epidemic drivers, bias intervention assessment, and delay detection of resurgence. For public-health decision-making, understanding how transmission varies across age groups is particularly important because transmission occurs across settings such as households, schools, and workplaces. Such information can support targeted interventions and surveillance while reducing reliance on broad population-level measures. The COVID-19 pandemic provided a clear example of this relevance. In the United States, \cite{monod_age_2021} found that adults aged 20--49 accounted for the majority of onward SARS-CoV-2 transmission during resurgence, suggesting that targeting these groups could help control epidemics while facilitating the safe reopening of schools. Estimating age-specific transmission in near real time is challenging because infection processes are only partially observed and are further complicated by reporting delays and heterogeneity in surveillance across age groups.

Two broad modelling frameworks can be used to represent age-specific heterogeneity. Age-structured compartmental models capture it explicitly through stratified compartments and contact matrices \citep{bouman_bayesian_2024, bouranis_bayesian_2025}, but typically require high-dimensional latent representations and correspondingly complex inference procedures. Semi-mechanistic renewal models offer a more parsimonious alternative \citep{bhatt_semimechanistic_2023}. By expressing incidence as a convolution of past infections with a generation-time distribution, they provide an interpretable representation of transmission that can, under suitable assumptions, be derived as a reduced form of compartmental dynamics \citep{champredon_equivalence_2018}. They are also widely used for estimating reproduction numbers and short-term forecasting \citep{cori_new_2013, abbott_estimating_2020}. However, standard renewal models assume population-level homogeneity, whereas extending them to age-stratified settings introduces substantial inferential challenges. Recent work has extended renewal models to structured populations through multi-type renewal processes and estimated effective reproduction numbers in heterogeneous populations \citep{green_inferring_2022, bouros_misspecification_2026, huang_evaluating_2026}.  However, these developments do not address the additional challenge of sequentially estimating age-specific transmission as it evolves over time within a Bayesian framework.

From an inferential perspective, age-stratified renewal models involve high-dimensional latent infection trajectories coupled through generation-time convolution and age-specific transmission. Their posterior therefore combines latent incidence states with time-varying transmission parameters, often resulting in strong temporal dependence and cross-age coupling. Markov chain Monte Carlo (MCMC), a widely used approach for Bayesian inference \citep{abbott_estimating_2020, huang_evaluating_2026}, can become increasingly computationally demanding as epidemic time series and latent dimensions grow, particularly under strong posterior correlations and non-Gaussianity. Moreover, MCMC generally requires refitting the model as new surveillance data arrive, limiting its suitability for real-time monitoring. Approximate Bayesian computation avoids explicit likelihood evaluation but relies on carefully designed summary statistics and repeated simulation, which can be difficult for structured epidemic models and does not naturally support sequential updating \citep{pooley_estimation_2022}. Sequential Monte Carlo (SMC) methods, or particle filters, instead represent the posterior through weighted particles that are updated recursively as data accrue, making them well suited to real-time inference \citep{watson_jointly_2024, steyn_primer_2025}. However, standard particle filters applied jointly across age groups can suffer severe particle degeneracy in high-dimensional and strongly coupled state spaces. A small number of poorly matched age-specific components can cause the importance weights of the entire particle system to collapse. This motivates a more targeted block-structured approach to sequential inference for age-stratified epidemic models.

In this paper, we develop a Bayesian age-stratified renewal model in which age-specific susceptibility and contact structure jointly determine transmission dynamics through an empirically specified contact matrix, together with a sequential inference scheme tailored to this structure. To mitigate particle degeneracy while retaining sequential updating, we use block particle filtering, resampling within age-specific blocks rather than across the full state, thereby improving particle efficiency while preserving cross-age dependence through joint propagation of the coupled transmission process \citep{rebeschini_can_2015, ionides_iterated_2024}. Fixed-lag smoothing further mitigates path degeneracy and accounts for delayed observations arising from the reporting process and renewal structure. Because static transmission parameters must also be learned as data accrue, the block particle filter is embedded within a sequential Monte Carlo squared (SMC$^2$) framework \citep{chopin_smc2_2013}, with particle Metropolis-within-Gibbs moves used to rejuvenate parameter particles when required. The observation model accounts for reporting delays and day-of-week effects in age-stratified case notifications. The resulting transmission process defines a time-varying next-generation matrix, from which we derive the effective reproduction number and age-specific reproduction contributions to quantify overall transmission potential and the relative contribution of different age groups over time. We illustrate the approach using age-stratified COVID-19 surveillance data from Ireland during the Omicron dominant period, demonstrating its ability to infer structured transmission dynamics from partially observed epidemic data.

% The remainder of the paper is organised as follows. Section~\ref{sec:data} describes the epidemiological setting, data, and contact structure. Section~\ref{sec:model} introduces the age-stratified renewal state-space model, and Section~\ref{sec:inf_method} presents the block SMC$^2$ framework. Section~\ref{sec:result} presents the Irish COVID-19 application and short-term forecasts, and Section~\ref{sec:discussion} discusses the findings and limitations. Additional diagnostics, sensitivity analyses, and simulation results are provided in the supplementary material.

\section{Data sources} \label{sec:data}

\subsection{Daily counts}

We analyse the COVID-19 epidemic in Ireland from March to July 2022, a period dominated by Omicron subvariants following the relaxation of non-pharmaceutical interventions. Daily counts of confirmed SARS-CoV-2 infections stratified by age group were obtained from the GeoHive platform maintained by the Health Protection Surveillance Centre\footnote{\url{https://www.geohive.ie/datasets/d8eb52d56273413b84b0187a4e9117be_0/about}}, which constitutes the official repository of COVID-19 surveillance statistics in Ireland and the primary data source for model calibration. Reported cases are subject to age-dependent ascertainment and delays between infection and laboratory confirmation; these features are explicitly accounted for in the observation model in Section~\ref{sec:obs_model}. Age-specific population denominators were obtained from the Population and Migration Estimates (April 2022) published by the Central Statistics Office\footnote{\url{https://www.cso.ie/en/releasesandpublications/ep/p-pme/populationandmigrationestimatesapril2022/data/}}.

\subsection{Age structuring and social mixing}
\label{subsec:age}

\begin{figure}[H]
    \centering
    \includegraphics[width=0.74\textwidth]{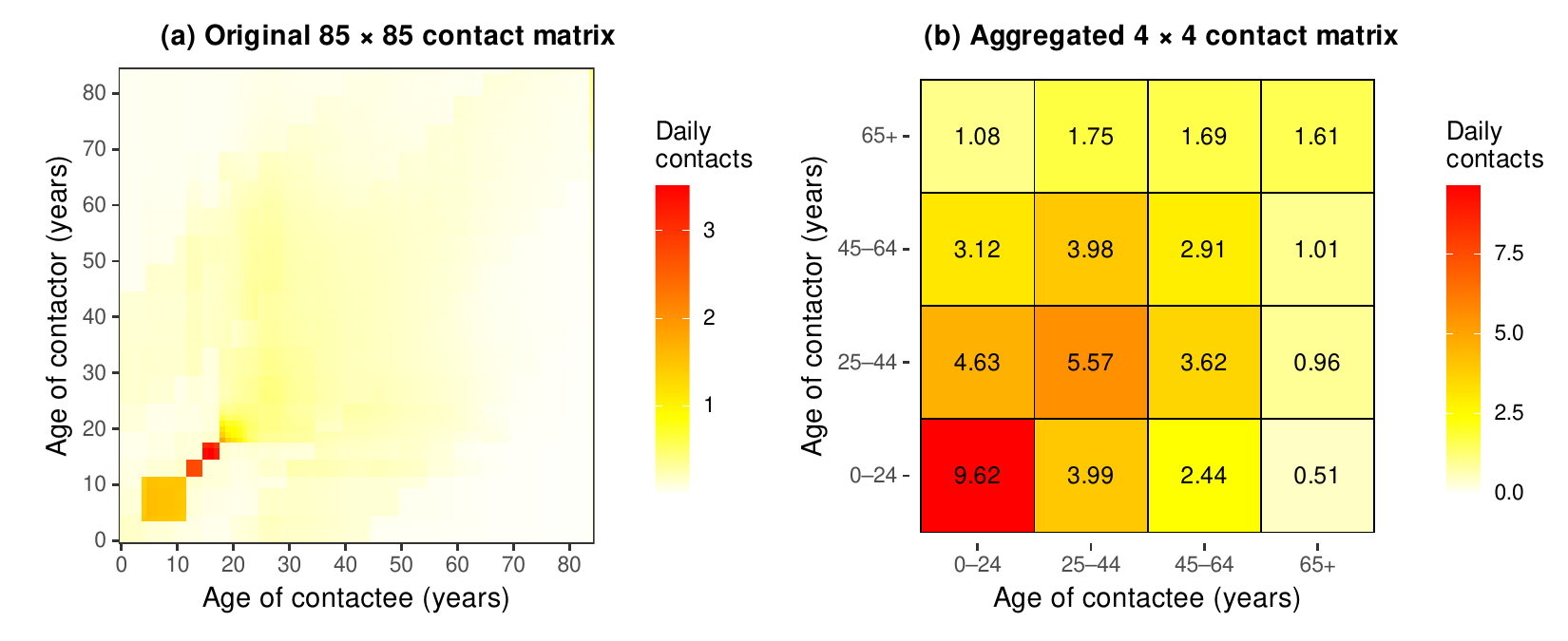}
    \caption{\footnotesize Age-structured contact matrices for Ireland. Panel~(a): original $85 \times 85$ matrix at single-year age resolution. Panel~(b): aggregated $4 \times 4$ matrix after population-weighted aggregation and reciprocity correction.}
    \label{fig:contact}
\end{figure}

To represent age-dependent heterogeneity in transmission, social mixing, and morbidity, we partition the population into four age groups: 0--24, 25--44, 45--64, and 65+ years. This stratification reflects broad differences in contact behaviour and COVID-19 vulnerability across the life course \citep{lai_epidemiological_2020, ofori_agestratified_2023}. The 0--24 group reflects greater mixing in educational and social settings, while the 25--44 and 45--64 groups distinguish working-age populations with increasing age-related risk. The 65+ group represents a population with distinct mixing patterns and greater vulnerability to severe COVID-19. Age-specific mixing is represented by a $4 \times 4$ contact matrix $\mathbf{M}$, where $M_{ab}$ denotes the average daily number of contacts from group $a$ to group $b$. We construct $\mathbf{M}$ from the synthetic age-structured contact matrices of \cite{mistry_inferring_2021}, which provide country-specific estimates of contacts across home, school, workplace, and community settings. The original $85 \times 85$ single-year matrix is first adjusted to satisfy reciprocity and then aggregated using population-weighted averaging, giving $N_a M_{ab}=N_b M_{ba}$ at the group level. Figure~\ref{fig:contact} shows moderate assortative mixing, with stronger within-group contacts among younger groups and reduced interaction involving the 65+ group. Although these four age groups are used in the analysis, the framework can accommodate alternative age partitions to reflect the epidemiological question and available data, without altering the underlying renewal model described in Section~\ref{sec:renewal_model}.

\section{Model} \label{sec:model}
\subsection{Hierarchical state-space model}

We consider a population partitioned into $A$ age groups and observed over times $t=1,\ldots,T$. Let $y_t=(y_{1,t},\dots,y_{A,t})^\top$ denote the vector of observed data at time $t$, where $y_{a,t}=C_{a,t}$ represents the number of reported cases in age group $a$ at time $t$. Let $I_t=(I_{1,t},\dots,I_{A,t})^\top$ denote the corresponding latent incidence process. We define $\beta_t=(\beta_{1,t},\dots,\beta_{A,t})^\top$ as age-specific, time-varying susceptibility multipliers that modulate the infection risk associated with contact. Thus, $\beta_{a,t}$ captures temporal variation in the effective susceptibility of age group $a$ within the transmission process, conditional on the age-specific contact structure.

Throughout, for integers $s\leq t$, we use the shorthand $s{:}t=\{s,s+1,\ldots,t\}$ and $x_{s:t}=(x_s,x_{s+1},\ldots,x_t)$. The model is specified as the following hierarchical state-space system:
\begin{subequations}
\begin{align}
\beta_{t} 
  &\sim P_\beta(.|\beta_{1:t-1}, \theta_\beta),
    & \rhd & \text{ susceptibility dynamics} \label{eq:state_beta} \\
I_{t}
  &\sim P_{_I}(.|\beta_{t}, I_{1:t-1}),
    & \rhd & \text{ infection process} \label{eq:state_I} \\
y_{t} 
  &\sim P_y(.|I_{1:t}, \theta_y),
    & \rhd & \text{ observation model} \label{eq:obs}
\end{align}
\end{subequations}
where $\theta=(\theta_\beta, \theta_y)$ denotes the collection of static model parameters. Since susceptibility and infectivity cannot be separately identified from incidence data alone without additional information or structural assumptions, all age-specific heterogeneity in transmission is represented through the susceptibility multipliers $\beta_{a,t}$. The transmission process is therefore determined by the interaction between the contact matrix and the age-specific susceptibility multipliers \citep{bouranis_bayesian_2025}.

\begin{figure}[H]
    \centering
    \begin{tikzpicture}[
        scale=0.46,
        transform shape,
        font=\normalsize,
        >=latex,
        state/.style={
            draw=blue!70!black,
            line width=1.4pt,
            rounded corners,
            minimum width=4.7cm,
            minimum height=2.1cm
        },
        betastate/.style={
            draw=orange!90!black,
            line width=1.6pt,
            rounded corners,
            minimum width=4.7cm,
            minimum height=2.0cm
        },
        obs/.style={
            draw=green!30!black,
            line width=1.4pt,
            rounded corners,
            minimum width=5cm,
            minimum height=2.0cm
        },
        icircle/.style={
            draw=dodgerblue,
            line width=1.5pt,
            dashed,
            minimum width=1.6cm,
            minimum height=1.6cm,
            text width=1.4cm,
            align=center,
            circle,
            inner sep=0pt,
            outer sep=0pt
        },
        betacircle/.style={
            draw=orange!85!black,
            line width=1.5pt,
            dashed,
            minimum width=1.6cm,
            minimum height=1.6cm,
            text width=1.4cm,
            align=center,
            circle,
            inner sep=0pt,
            outer sep=0pt
        },
        ycircle/.style={
            draw=green!60!black,
            line width=1.3pt,
            minimum width=1.6cm,
            minimum height=1.6cm,
            text width=1.4cm,
            align=center,
            circle,
            inner sep=0pt,
            outer sep=0pt
        },
        arrow/.style={->, line width=1pt, draw=gray!80},
        dashedarrow/.style={->, line width=1pt, dashed, draw=gray!80}
    ]
    
    % ================= I states =================
    \node[state] (Itm1) {};
    \node[state, right=2cm of Itm1] (It) {};
    \node[state, right=2cm of It] (Itp1) {};
    
    % ================= Beta layer =================
    \node[betastate, above=2.5cm of Itm1] (Betatm1) {};
    \node[betastate, above=2.5cm of It]   (Betat) {};
    \node[betastate, above=2.5cm of Itp1] (Betatp1) {};
    
    % --- Beta circles ---
    \node[betacircle] at ([xshift=-1.3cm]Betatm1.center) {\Large$\bm{\beta_{1,t-1}}$};
    \node[betacircle] at ([xshift= 1.3cm]Betatm1.center) {\Large$\bm{\beta_{A,t-1}}$};
    \node at (Betatm1.center) {\Large$\bm{\cdots}$};
    
    \node[betacircle] at ([xshift=-1.3cm]Betat.center) {\Large$\bm{\beta_{1,t}}$};
    \node[betacircle] at ([xshift= 1.3cm]Betat.center) {\Large$\bm{\beta_{A,t}}$};
    \node at (Betat.center) {\Large$\bm{\cdots}$};
    
    \node[betacircle] at ([xshift=-1.3cm]Betatp1.center) {\Large$\bm{\beta_{1,t+1}}$};
    \node[betacircle] at ([xshift= 1.3cm]Betatp1.center) {\Large$\bm{\beta_{A,t+1}}$};
    \node at (Betatp1.center) {\Large$\bm{\cdots}$};
    
    % --- Beta overbraces (above beta boxes) ---
    \node[above=0mm of Betatm1]
    {\huge$\overbrace{\phantom{\rule{3.4cm}{0pt}}}^{\Huge\textcolor{orange!90!black}{\bm{\beta_{t-1}}}}$};
    
    \node[above=0mm of Betat]
    {\huge$\overbrace{\phantom{\rule{3.4cm}{0pt}}}^{\Huge\textcolor{orange!90!black}{\bm{\beta_{t}}}}$};
    
    \node[above=0mm of Betatp1]
    {\huge$\overbrace{\phantom{\rule{3.4cm}{0pt}}}^{\Huge\textcolor{orange!90!black}{\bm{\beta_{t+1}}}}$};
    
    % --- Vertical arrows beta -> I (edge-to-edge, no brace conflict) ---
    \draw[arrow] (Betatm1.south) -> (Itm1.north);
    \draw[arrow] (Betat.south)   -> (It.north);
    \draw[arrow] (Betatp1.south) -> (Itp1.north);
    
    % ================= I circles =================
    \node[icircle] at ([xshift=-1.3cm]Itm1.center) {\Large$\bm{I_{1,t-1}}$};
    \node[icircle] at ([xshift= 1.3cm]Itm1.center) {\Large$\bm{I_{A,t-1}}$};
    \node at (Itm1.center) {\Large$\bm{\cdots}$};
    
    \node[icircle] at ([xshift=-1.3cm]It.center) {\Large$\bm{I_{1,t}}$};
    \node[icircle] at ([xshift= 1.3cm]It.center) {\Large$\bm{I_{A,t}}$};
    \node at (It.center) {\Large$\bm{\cdots}$};
    
    \node[icircle] at ([xshift=-1.3cm]Itp1.center) {\Large$\bm{I_{1,t+1}}$};
    \node[icircle] at ([xshift= 1.3cm]Itp1.center) {\Large$\bm{I_{A,t+1}}$};
    \node at (Itp1.center) {\Large$\bm{\cdots}$};
    
    % --- I overbraces (tight to box, below beta arrows) ---
    \node[above=0mm of Itm1]
    {\huge$\overbrace{\phantom{\rule{3.4cm}{0pt}}}^{\Huge\textcolor{blue!60!black}{\bm{I_{t-1}}}}$};
    
    \node[above=0mm of It]
    {\huge$\overbrace{\phantom{\rule{3.4cm}{0pt}}}^{\Huge\textcolor{blue!60!black}{\bm{I_{t}}}}$};
    
    \node[above=0mm of Itp1]
    {\huge$\overbrace{\phantom{\rule{3.4cm}{0pt}}}^{\Huge\textcolor{blue!60!black}{\bm{I_{t+1}}}}$};
    
    % ================= y states =================
    \node[obs, below=3cm of Itm1] (ytm1) {};
    \node[obs, below=3cm of It]   (yt) {};
    \node[obs, below=3cm of Itp1] (ytp1) {};
    
    % --- y circles ---
    \node[ycircle] at ([xshift=-1.3cm]ytm1.center) {\Large$\bm{y_{1,t-1}}$};
    \node[ycircle] at ([xshift= 1.3cm]ytm1.center) {\Large$\bm{y_{A,t-1}}$};
    \node at (ytm1.center) {\Large$\bm{\cdots}$};
    
    \node[ycircle] at ([xshift=-1.3cm]yt.center) {\Large$\bm{y_{1,t}}$};
    \node[ycircle] at ([xshift= 1.3cm]yt.center) {\Large$\bm{y_{A,t}}$};
    \node at (yt.center) {\Large$\bm{\cdots}$};
    
    \node[ycircle] at ([xshift=-1.3cm]ytp1.center) {\Large$\bm{y_{1,t+1}}$};
    \node[ycircle] at ([xshift= 1.3cm]ytp1.center) {\Large$\bm{y_{A,t+1}}$};
    \node at (ytp1.center) {\Large$\bm{\cdots}$};
    
    % --- Underbraces y ---
    \node[below=0mm of ytm1]
    {\huge$\underbrace{\phantom{\rule{3.4cm}{0pt}}}_{\textcolor{green!40!black}{\Huge\bm{y_{t-1}}}}$};
    
    \node[below=0mm of yt]
    {\huge$\underbrace{\phantom{\rule{3.4cm}{0pt}}}_{\textcolor{green!40!black}{\Huge\bm{y_{t}}}}$};
    
    \node[below=0mm of ytp1]
    {\huge$\underbrace{\phantom{\rule{3.4cm}{0pt}}}_{\textcolor{green!40!black}{\Huge\bm{y_{t+1}}}}$};
    
    % ================= Temporal + emission arrows =================
    \draw[arrow] (Itm1) -> (It);
    \draw[arrow] (It)   -> (Itp1);
    
    \draw[arrow] (Betatm1) -> (Betat);
    \draw[arrow] (Betat)   -> (Betatp1);
    
    \draw[arrow] (Itm1) -> (ytm1);
    \draw[arrow] (It)   -> (yt);
    \draw[arrow] (Itp1) -> (ytp1);
    
    \draw[dashedarrow] (Itm1) -> (yt);
    \draw[dashedarrow] (Itm1) -> (ytp1);
    \draw[dashedarrow] (It)   -> (ytp1);
    
    % ================= Invisible Ai / Af =================
    \node[left=4cm of Betatm1] (Bi) {};
    \node[right=2cm of Betatp1] (Bf) {};
    
    \node[left=4cm of Itm1] (Ai) {};
    \node[right=2cm of Itp1] (Af) {};
    
    \draw[arrow] (Ai) -> (Itm1);
    \draw[arrow] (Itp1) -> (Af);
    
    \draw[arrow] (Bi) -> (Betatm1);
    \draw[arrow] (Betatp1) -> (Bf);
    
    \draw[dashedarrow] (Ai) -> (ytm1);
    \draw[dashedarrow] (Ai) -> (yt);
    \draw[dashedarrow] (Ai) -> (ytp1);
    
    \end{tikzpicture}

\caption{\footnotesize
Graphical representation of the hierarchical state-space model. The age-specific susceptibility process $\beta_t$ influences the latent infection process $I_t$, which evolves dynamically over time. Hidden states (dashed circles) generate the observed data (solid circles) through the observation model. Dashed arrows indicate additional dependencies arising from reporting delays or cross-temporal structure. For clarity, arrows representing dependence on previous states (see Eqs.~\eqref{eq:state_beta}--\eqref{eq:state_I}) are omitted. Hyperparameters $\theta$ (not shown) govern all conditional distributions.
}
    \label{fig:ssm_age}
\end{figure}
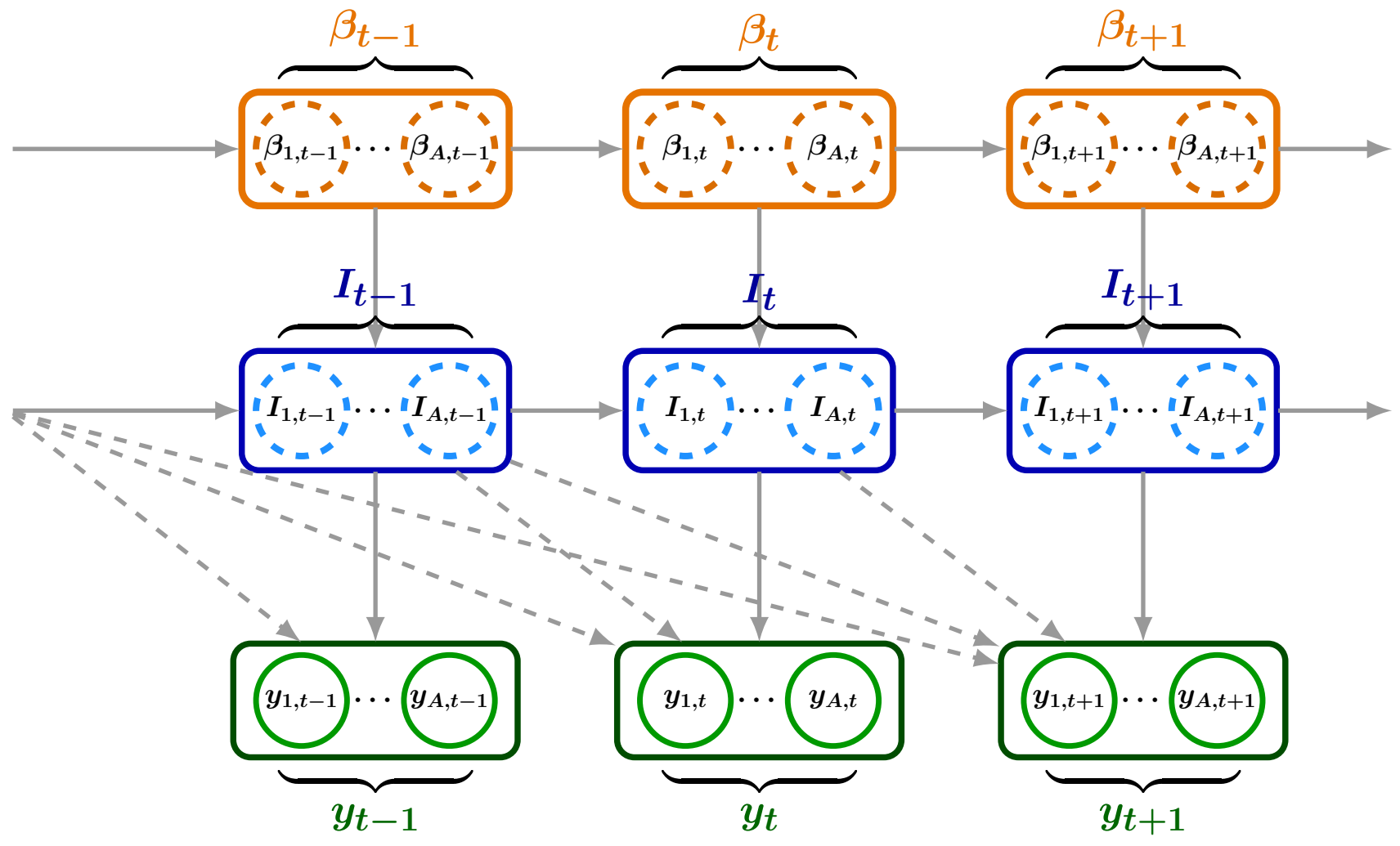

Figure~\ref{fig:ssm_age} provides a graphical representation of the state-space hierarchy. The transition distribution $P_\beta$ governs the evolution of age-specific susceptibility, $P_{_I}$ describes the age-structured renewal process, and $P_y$ links latent infections to reported cases. Although the renewal process is not Markovian, SMC methods remain applicable by retaining the required latent state histories over a fixed lag of $L$ time steps \citep{steyn_primer_2025}, as described in Section~\ref{sec: fixed_lagBPF}. This formulation enables sequential Bayesian inference for the joint posterior distribution $P(\beta_{1:t},I_{1:t},\theta |y_{1:t})$, $t\leq T$.

\subsection{Transmission process} \label{sec:renewal_model}

Let the total population be $N=\sum_{a=1}^A N_a$, where $N_a$ denotes the number of individuals in age group $a$. Interactions between age groups are described by the contact matrix $\mathbf{M}$, where $M_{a,b}$ represents the average daily number of contacts from individuals in source age group $a$ to individuals in recipient age group $b$. Transmission dynamics are determined by the contact structure and the age-specific susceptibility multipliers $\beta_{a,t}$, which scale the risk of infection following contact. We model temporal variation in the susceptibility multipliers using a second-order Gaussian random walk (RW2) prior on the log scale:
\begin{subequations}\label{eq:rw2_beta}
\begin{align}
\beta_{a,1} &\sim \pi(\beta_{a,1}), \\
\beta_{a,2} | \beta_{a,1} &\sim \mathrm{exp}(\log(\beta_{a,1}) + \varepsilon_{a,2}), \\
\beta_{a,t} | \beta_{a,t-1}, \beta_{a,t-2} 
&\sim \mathrm{exp}(2 \log(\beta_{a,t-1}) - \log(\beta_{a,t-2}) + \varepsilon_{a,t}), \quad t \geq 3,
\end{align}
\end{subequations}
where $\varepsilon_{a,t}\sim \mathcal{N}(0,\sigma_a^2)$ and $\sigma_a$ controls the degree of temporal smoothing within age group $a$. The RW2 prior allows gradual changes over time while avoiding the excessive variability associated with a first-order random walk. The initial distribution $\pi(\beta_{a,1})$ specifies prior information on the baseline susceptibility level. 
% Other smoothness priors, such as integrated Brownian motion, could also be considered but generally introduce additional latent variables and increase computational complexity \citep{sebastian_gaussian_2025}.

The number of new infections in each age group follows a multitype renewal process: 
\begin{align}\label{eq:renewal_infection}
I_{a,t}| \beta_{a,t}, I_{1:t-1}
\sim \mathrm{Poisson}\!\left( \beta_{a,t}
\sum_{b=1}^A M_{b,a}
\sum_{\tau=1}^{m_g} I_{b,t-\tau}\, g_{\tau}
\right),
\end{align}
 The generation interval $\{g_\tau\}_{\tau=1}^{m_g}$ is a discrete probability mass function truncated at a maximum lag $m_g$, with $g_{\tau}$ representing the probability that transmission occurs $\tau$ days after infection and satisfying $\sum_{\tau=1}^{m_g} g_\tau = 1$. A common generation interval is assumed across age groups. Allowing for age-specific generation-interval distributions is a natural extension but is beyond the scope of this study.

The effective reproduction number, $R_t$, is defined as the expected number of secondary infections generated by a typical newly infected individual at time $t$. It is given by the dominant eigenvalue (spectral radius) of the time-varying next-generation matrix \citep{green_inferring_2022, bouros_misspecification_2026},
\begin{align}
\mathbf{K}(t)=\operatorname{diag}(\beta_t)\mathbf{M}^{\!\top},
\end{align}
where $K_{a,b}(t)=\beta_{a,t}M_{b,a}$ denotes the expected number of infections generated in recipient group $a$ by a single infectious individual in source group $b$; rows index recipients and columns index sources. The effective reproduction number is therefore
\begin{align}
R_t=\rho\!\left(\mathbf{K}(t)\right),
\end{align}
where $\rho(\cdot)$ denotes the spectral radius. To characterise heterogeneity in transmission across age groups by source (infectious) group, we define the age-specific reproduction contribution as the column sum of the next-generation matrix \citep{bouros_misspecification_2026}:
\begin{align}
R_{a,t}
=
\sum_{b=1}^{A}K_{b,a}(t)
=
\sum_{b=1}^{A}\beta_{b,t}M_{a,b}.
\end{align}
This quantity decomposes transmission potential by the age group of the infectious (source) individuals. Specifically, $R_{a,t}$ is the expected number of secondary infections generated across all recipient groups by a single infectious individual in age group $a$, i.e., the total contribution of column $a$ of $\mathbf{K}(t)$. It identifies which age groups contribute most to onward transmission through their role as sources of infection.

\subsection{Observation process}\label{sec:obs_model}

The observation model links latent infections $I_{a,t}$ to reported cases $C_{a,t}$ through infection-to-reporting delays and overdispersed observation noise.

Given latent incidence, the expected number of reported cases in age group $a$ at time $t$ is given by the discrete convolution
\begin{align}
\mu_{a,t} = \rho_a \sum_{\tau=1}^{m_r} I_{a,t-\tau}\, f_{\tau},
\label{eq:cases_mean_conv}
\end{align}
where $f_{\tau}$ denotes the infection-to-reporting delay distribution, obtained by convolving the incubation period and reporting delay distributions. The distribution is defined on $\tau=1,\ldots,m_r$, where $m_r$ is the maximum infection-to-reporting delay, and satisfies $\sum_{\tau=1}^{m_r}f_{\tau}=1$. The parameter $\rho_a\in(0,1]$ denotes the age-specific ascertainment ratio, that is, the proportion of infections in age group $a$ captured by the surveillance system and subsequently reported.

Conditional on the latent incidence, reported case counts are assumed to be independent across age groups and days and to follow a Negative Binomial distribution, which accommodates overdispersion arising from reporting, testing variability, and stochasticity in surveillance data:
\begin{align}
C_{a,t}| I_{a,1:t}, \rho_a, \kappa \sim \mathrm{NegBin}\!\left(\omega_{a,d(t)}\,\mu_{a,t},\, \kappa\right),
\label{eq:cases_obs}
\end{align}
with mean $\omega_{a,d(t)}\mu_{a,t}$ and variance $\omega_{a,d(t)}\mu_{a,t}+\bigl(\omega_{a,d(t)}\mu_{a,t}\bigr)^2/\kappa$, where $\kappa>0$ is the dispersion parameter. Smaller values of $\kappa$ correspond to greater overdispersion, and the Poisson limit is recovered as $\kappa\to\infty$. The factor $\omega_{a,d(t)}$ is a multiplicative day-of-the-week effect on reporting intensity, where $d(t)\in\{1,\ldots,7\}$ indexes Monday through Sunday. If $d_1$ denotes the weekday index of the first observation, then $d(t)=((d_1+t-2)\bmod 7)+1$.

Following \cite{plank_multipathogen_2026}, day-of-the-week effects are estimated empirically from the observed case counts and treated as fixed adjustments. For each age group $a$ and day $t$, we define
\begin{align}
\omega_{a,d(t)}
= \frac{
7 \sum_{s \in \mathcal{T}:d(s)=d(t)} C_{a,s}
}{
\sum_{s \in \mathcal{T}} C_{a,s}
},
\label{eq:dow_weights}
\end{align}
where $\mathcal{T}$ is the set of days in the 16 complete weeks ending at the last observed time point $T$. Because each weekday appears equally often in $\mathcal{T}$, these weights have mean one across the seven weekdays and hence preserve the average weekly reporting intensity. This formulation assumes that weekday reporting patterns are approximately constant over the calibration window, which may not hold if testing practices or reporting systems change. We do not estimate the weekday effects jointly with the epidemic parameters, as is done in the homogeneous model (Section~\ref{app:simple_model} of the supplementary material), since this would increase model complexity and risk confounding short-term reporting variation with genuine changes in transmission. The weights are estimated once from data up to time $T$ and held fixed. Earlier filtering steps therefore use weekday information from later time points, and the earliest steps precede the calibration window altogether. In a prospective application, the weights would be updated at each time point using only the data then available.

\section{Sequential inference} \label{sec:inf_method}
We perform sequential Bayesian inference for the latent age-stratified infection dynamics and model parameters under the state-space formulation in \eqref{eq:state_beta}--\eqref{eq:obs}. Let the latent state at time $t$ be $X_t=(\beta_t,I_t)$. To obtain an approximate sequential posterior in this high-dimensional setting, we use a block-localised particle approximation. Inference proceeds in two stages: (i) conditional on $\theta$, a block particle filter approximates the filtering distribution $\tilde{P}(X_t|y_{1:t},\theta)$ and the block-localised predictive likelihood $\tilde{P}(y_t|y_{1:t-1},\theta)$; (ii) this procedure is embedded within an SMC$^2$-inspired algorithm to sequentially update the parameter distribution while accounting for latent-state uncertainty.

\subsection{State inference via block particle filtering}\label{sec: fixed_lagBPF}

The latent state is high-dimensional, comprising age-specific transmissibility and incidence processes for each of the 
$A$ age groups, which are coupled across groups through the contact matrix governing the epidemic dynamics. A standard particle filter propagates and resamples the full joint state at each time point \citep{gordon_novel_1993}. As the dimension of the state increases, the importance weights can become increasingly concentrated, leading to particle degeneracy and loss of particle diversity \citep{rebeschini_can_2015}. In the present setting, observations are age-specific, so global resampling can discard useful state components for some age groups when the corresponding joint particle receives a low overall weight. The Block Particle Filter (BPF) addresses this by localising the resampling step to subsets of the state while retaining the coupled epidemic dynamics during propagation \citep{rebeschini_can_2015, ionides_iterated_2024}.

We partition the latent state into age-specific blocks, where $X_{a,t}=(\beta_{a,t},I_{a,t})$ denotes the state associated with age group $a$. The blocks are used for computational localisation rather than treated as independent components, since transmission between age groups remains coupled through the contact matrix. At time $t$, each particle $j$ retained after the previous resampling step is propagated forward by sampling its current state from the transition model. Denoting the resulting propagated state by $\widetilde{X}_{t}^{(j)}$, the transition density for the joint state can be expressed as
\begin{align}\label{coupled_dyn}
\tilde{X}_t^{(j)}
\sim P(X_t|X_{1:t-1}^{(j)},\theta)
=
\prod_{a=1}^{A}
P(\beta_{a,t}|\beta_{a,t-1}^{(j)},\beta_{a,t-2}^{(j)},\sigma_a)
P(I_{a,t}|\beta_{a,t}^{(j)},I_{1:t-1}^{(j)}),
\end{align}
Upon receiving the observation $y_t$, the propagated particles are weighted using the observation density. Given the conditional independence of observations across age groups, the joint observation density factorises as
\begin{align}
P(y_t|X_{1:t},\theta)
=\prod_{a=1}^{A}
P(y_{a,t}|X_{a,1:t},\theta).
\end{align}
The factorisation of the observation model allows age-specific likelihood contributions to be used for block-wise weighting. For each particle and age group, the local weights are
\begin{align}
w_{a,t}^{(j)}
=P(y_{a,t}|\tilde{X}_{a,1:t}^{(j)},\theta)
=P(C_{a,t}|\tilde{I}_{a,1:t}^{(j)},\rho_a,\kappa).
\end{align}
The updated particles $X_{a,t}^{1:N_x}$ for each age block are obtained by resampling $\tilde{X}_{a,t}^{1:N_x}$ according to the normalised weights $\bar{w}_{a,t}^{1:N_x}$. The resulting particles 
$X_t^{1:N_x}=\{X_{1,t}^{1:N_x},\dots,X_{A,t}^{1:N_x}\}$
provide a Monte Carlo approximation to the block-localised filtering distribution 
$\tilde{P}(X_t|y_{1:t},\theta)$. The corresponding filtering recursion and the derivation of the block-localised likelihood approximation are provided in Section~\ref{app:bpf_likelihood} of the supplementary material. Although particles are propagated jointly through the coupled transmission model in Eqs.~\eqref{eq:rw2_beta}--\eqref{eq:renewal_infection}, separate resampling within age blocks introduces a controlled approximation to the dependence structure between blocks.

To account for the temporal delay between infections and their subsequent observations, we incorporate a fixed-lag smoothing approach \citep{watson_jointly_2024, steyn_primer_2025}. Since infections at previous time points contribute to future observations through the convolution structure in Eq.~\eqref{eq:cases_mean_conv}, particle trajectories are retained over a window of $L$ time steps. At time $t$, the algorithm maintains an approximation to the joint distribution $P(X_{t-L+1:t}|y_{1:t},\theta)$ by propagating and resampling the retained particle trajectories within this lag window. Combined with block localisation, this yields the approximation $\tilde{P}(X_{t-L+1:t}|y_{1:t},\theta)$. The lag length $L$ is chosen to exceed the effective support of both the generation interval $g_\tau$ and the reporting delay distribution $f_\tau$, ensuring that observations outside the retained window have negligible influence on inference for the current state \citep{steyn_primer_2025}.

\begin{algorithm}[H]
\caption{Fixed-Lag Block Particle Filter (step $t$)} \label{BPF}
Operations involving index $j$ must be performed for $j = 1,\dots, N_x$ and $a = 1,\dots, A$.

\textbf{Inputs:} Observations $y_{t}$, number of particles $N_x$, fixed-lag $L$ and state particles $X_{1:t-1}^{1:N_x}$.

\textbf{Outputs:} State particles $X_{1:t}^{1:N_x}$, block-localised predictive likelihood approximation
$\tilde{P}(y_t|y_{1:t-1},\theta)$ and block-localised marginal likelihood approximation $\tilde{P}(y_{1:t}|\theta)$.

\hrulefill 
\begin{algorithmic}[1]
        \State \textbf{Prediction:} Sample $\tilde{X}_{t}^{(j)} \sim P(.| X^{(j)}_{t-L:t-1}, \theta)$
        
        \State \textbf{Block-wise weighting and resampling:} 
        \Statex \hspace{\algorithmicindent} Compute $w_{a,t}^{(j)} = P(y_{a,t} | \tilde{X}_{a,t-L+1:t}^{(j)},\theta)$ and $\bar{w}^{(j)}_{a,t} = w_{a,t}^{(j)}/\sum_{k=1}^{N_x} w_{a,t}^{k}$
        \Statex \hspace{\algorithmicindent} Resample indices
        $X_{a,t-L+1:t}^{1:N_x} \sim \mathrm{Resampling} \left(\tilde{X}_{a,t-L+1:t}^{1:N_x}, ~\bar{w}_{a,t}^{1:N_x}\right)$
        \Statex \hspace{\algorithmicindent} Recombine blocks $X_{t-L+1:t}^{(j)} = \{X_{1,t-L+1:t}^{(j)}, \dots, X_{A,t-L+1:t}^{(j)}\}$
        
        \State \textbf{Predictive likelihood estimate:} $\tilde{P}(y_t|y_{1:t-1},\theta) =\frac{1}{N_x} \sum_{j=1}^{N_x}\left(\prod_{a=1}^{A}  w_{a,t}^{(j)}\right)$
        \State \textbf{Marginal likelihood estimate:} $\tilde{P}(y_{1:t}|\theta)=\tilde{P}(y_t|y_{1:t-1},\theta) \cdot \tilde{P}(y_{1:t-1}|\theta)$
\end{algorithmic}
\end{algorithm}

Step $t$ of the fixed-lag BPF is summarised in Algorithm~\ref{BPF}. At initialisation ($t=0$), particles $X_0^{(j)}$ are sampled independently from the initial distribution $P(X_0)$ for $j=1,\ldots,N_x$, and the marginal likelihood is initialised as $\tilde{P}(y_0 | \theta)=1$. The block-localised likelihood approximation does not generally provide an unbiased estimator of the exact marginal likelihood because the block resampling step modifies the dependence structure between age-specific components. Consequently, the subsequent parameter learning procedure targets the posterior induced by the block-localised fixed-lag likelihood approximation $\tilde{P}(\theta|y_{1:t})\propto \tilde{P}(y_{1:t}|\theta)P(\theta)$, rather than the exact posterior based on the full likelihood. Section~\ref{app:pmcmc} of the supplementary material further investigates the effect of block-wise resampling on particle degeneracy and compares its performance with that of a standard bootstrap particle filter, including an empirical evaluation of the likelihood estimator.

\subsection{Parameter inference via SMC\texorpdfstring{$^2$}{2}}\label{sec:smc2}
To sequentially learn $\theta$ while simultaneously estimating latent states with fixed-lag resampling, we adopt SMC$^2$ \citep{chopin_smc2_2013}. Because the inner likelihood estimator is based on the block-localised particle filter, the resulting algorithm should be interpreted as an approximate SMC$^2$ procedure. Each parameter particle is associated with a fixed-lag BPF of lag $L$. 

Let $\{\theta^{m}, u_{t-1}^{m}\}_{m=1}^{N_{\theta}}$ denote the parameter particles and their associated weights at time $t-1$. Each parameter particle $\theta^{m}$ has $N_x$ associated BPF particles that approximate the fixed-lag distribution $P(X_{t-L:t-1} | y_{1:t-1}, \theta^{m})$. Upon receiving $y_t$, the BPF is advanced for each $\theta^{m}$, producing an estimated predictive likelihood $\tilde{P}(y_t | y_{1:t-1}, \theta^{m})$. The parameter weights are then updated according to $u_t^{m} \propto u_{t-1}^{m} \, \tilde{P}(y_t | y_{1:t-1}, \theta^{m})$. When the effective sample size (ESS) falls below a threshold, particle Metropolis within Gibbs rejuvenation \citep[PMwG,][]{gunawan_efficient_2017} is performed using data up to $t$ to restore particle diversity. We group the model parameters into $A+1$ blocks, with $\theta_b=(\sigma_b,\rho_b)$ for $b=1,\ldots,A$, corresponding to the age-specific parameters and $\theta_{A+1}=\kappa$, corresponding to the observation-dispersion parameter. 

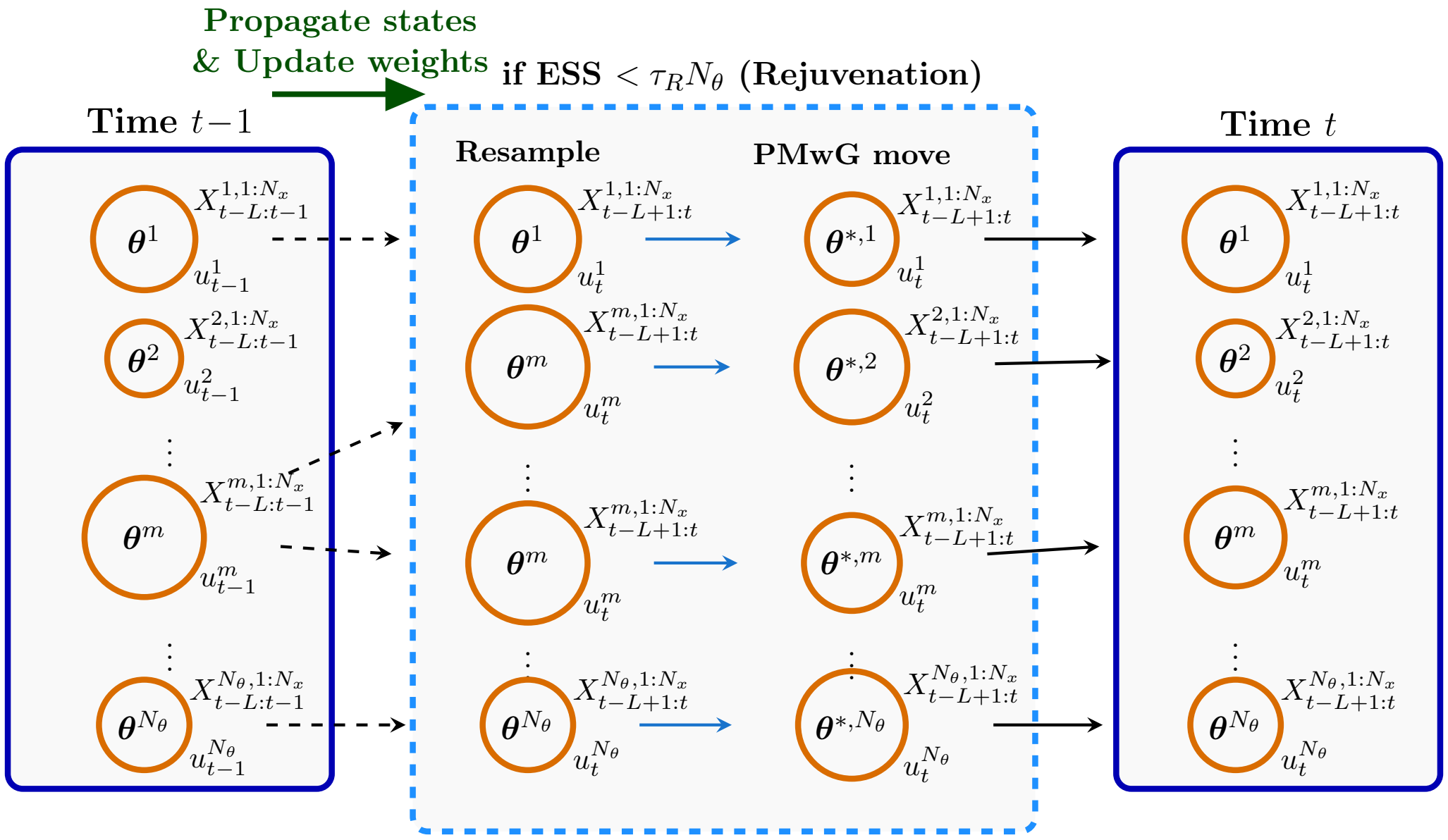
\begin{figure}[H]
\centering
\begin{tikzpicture}[scale=0.8, transform shape,
    stage/.style={rectangle, rounded corners, draw=blue!70!black, minimum width=3.8cm, minimum height=7.5cm, line width=1.6pt, fill=gray!5},
    rejuv/.style={rectangle, rounded corners, draw=dodgerblue, dashed, minimum width=7.3cm, minimum height=8.5cm, line width=1.6pt, fill=gray!5},
    particle/.style={circle, draw=orange!85!black, line width=1.6pt, font=\normalsize},
    arrow/.style={->, >=stealth, thick},
    label/.style={font=\small, text=black}
]

% Stage 1: Time t-1
\node[stage] (s1) at (0, -0.5) {};
\node[above, font=\large\bfseries] at (s1.north) {Time $t\!-\!1$};

% Particles with different sizes based on weights
\node[particle, minimum size=1.2cm] (th1) at (-0.3, 2.2) {$\bm{\theta}^{1}$};
\node[label, anchor=west] at (th1.north east) {$X_{t-L:t-1}^{1,1:N_x}$};
\node[label, anchor=west] at (th1.south east) {$u_{t-1}^{1}$};

\node[particle, minimum size=0.8cm] (th2) at (-0.3, 0.8) {$\bm{\theta}^{2}$};
\node[label, anchor=west] at (th2.north east) {$X_{t-L:t-1}^{2,1:N_x}$};
\node[label, anchor=west] at (th2.south east) {$u_{t-1}^{2}$};

\node at (0, -0.2) {$\vdots$};

\node[particle, minimum size=1.4cm] (thm) at (-0.3, -1.3) {$\bm{\theta}^{m}$};
\node[label, anchor=west] at (thm.north east) {$X_{t-L:t-1}^{m,1:N_x}$};
\node[label, anchor=west] at (thm.south east) {$u_{t-1}^{m}$};

\node at (0, -2.6) {$\vdots$};

\node[particle, minimum size=0.7cm] (thN) at (-0.3, -3.5) {$\bm{\theta}^{N_{\theta}}$};
\node[label, anchor=west] at (thN.north east) {$X_{t-L:t-1}^{N_{\theta},1:N_x}$};
\node[label, anchor=west] at (thN.south east) {$u_{t-1}^{N_{\theta}}$};

% Label between Stage 1 and Rejuvenation block
\node[font=\normalsize\bfseries, darkgreen!80!black, align=center] at (2, 4.5) {Propagate states\\ \& Update weights};
% Horizontal arrow with weight update formula
\draw[->, line width=1.5pt, darkgreen!80!black] (1.2,3.9) --  (3, 3.9);

% Stage 2: Rejuvenation
\node[rejuv] (s2) at (6.5, -0.5) {};
\node[above, font=\bfseries] at (s2.north) {\quad if ESS $< \tau_R N_{\theta}$ (Rejuvenation)};

% Title for Resample (inside box)
\node[label, align=center, font=\small\bfseries] at (4.2, 3.2) {Resample};
% Left part: After resampling
\node[particle, minimum size=1.2cm] (r1) at (4.2, 2.2) {$\bm{\theta}^{1}$};
\node[label, anchor=west] at (r1.north east) {$X_{t-L+1:t}^{1,1:N_x}$};
\node[label, anchor=west] at (r1.south east) {$u_t^1$};

\node[particle, minimum size=1.4cm] (r2) at (4.2, 0.7) {$\bm{\theta}^{m}$};
\node[label, anchor=west] at (r2.north east) {$X_{t-L+1:t}^{m,1:N_x}$};
\node[label, anchor=west] at (r2.south east) {$u_t^m$};

\node at (4.2, -0.5) {$\vdots$};

\node[particle, minimum size=1.4cm] (rm) at (4.2, -1.6) {$\bm{\theta}^{m}$};
\node[label, anchor=west] at (rm.north east) {$X_{t-L+1:t}^{m,1:N_x}$};
\node[label, anchor=west] at (rm.south east) {$u_t^m$};

\node at (4.2, -2.7) {$\vdots$};

\node[particle, minimum size=0.7cm] (rN) at (4.2, -3.5) {$\bm{\theta}^{N_{\theta}}$};
\node[label, anchor=west] at (rN.north east) {$X_{t-L+1:t}^{N_{\theta},1:N_x}$};
\node[label, anchor=west] at (rN.south east) {$u_t^{N_{\theta}}$};

% Title for PMwG (inside box)
\node[label, align=center, font=\small\bfseries] at (8, 3.2) {PMwG move};

% Right part: After PMwG
\node[particle, minimum size=0.95cm] (p1) at (8, 2.2) {$\bm{\theta}^{*,1}$};
\node[label, anchor=west] at (p1.north east) {$X_{t-L+1:t}^{1,1:N_x}$};
\node[label, anchor=west] at (p1.south east) {$u_t^1$};

\node[particle, minimum size=1.3cm] (p2) at (8, 0.7) {$\bm{\theta}^{*,2}$};
\node[label, anchor=west] at (p2.north east) {$X_{t-L+1:t}^{2,1:N_x}$};
\node[label, anchor=west] at (p2.south east) {$u_t^2$};

\node at (8, -0.5) {$\vdots$};

\node[particle, minimum size=1.1cm] (pm) at (8, -1.6) {$\bm{\theta}^{*,m}$};
\node[label, anchor=west] at (pm.north east) {$X_{t-L+1:t}^{m,1:N_x}$};
\node[label, anchor=west] at (pm.south east) {$u_t^{m}$};

\node at (8, -2.7) {$\vdots$};

\node[particle, minimum size=1.25cm] (pN) at (8, -3.5) {$\bm{\theta}^{*,N_{\theta}}$};
\node[label, anchor=west] at (pN.north east) {$X_{t-L+1:t}^{N_{\theta},1:N_x}$};
\node[label, anchor=west] at (pN.south east) {$u_t^{N_{\theta}}$};

% Stage 3: Time t
\node[stage] (s3) at (13, -0.5) {};
\node[above, font=\large\bfseries] at (s3.north) {Time $t$};

\node[particle, minimum size=1.2cm] (f1) at (12.5, 2.2) {$\bm{\theta}^{1}$};
\node[label, anchor=west] at (f1.north east) {$X_{t-L+1:t}^{1,1:N_x}$};
\node[label, anchor=west] at (f1.south east) {$u_t^{1}$};

\node[particle, minimum size=0.85cm] (f2) at (12.5, 0.8) {$\bm{\theta}^{2}$};
\node[label, anchor=west] at (f2.north east) {$X_{t-L+1:t}^{2,1:N_x}$};
\node[label, anchor=west] at (f2.south east) {$u_t^{2}$};

\node at (12.5, -0.2) {$\vdots$};

\node[particle, minimum size=1.15cm] (fm) at (12.5, -1.3) {$\bm{\theta}^{m}$};
\node[label, anchor=west] at (fm.north east) {$X_{t-L+1:t}^{m,1:N_x}$};
\node[label, anchor=west] at (fm.south east) {$u_t^{m}$};

\node at (12.5, -2.6) {$\vdots$};

\node[particle, minimum size=0.95cm] (fN) at (12.5, -3.5) {$\bm{\theta}^{N_{\theta}}$};
\node[label, anchor=west] at (fN.north east) {$X_{t-L+1:t}^{N_{\theta},1:N_x}$};
\node[label, anchor=west] at (fN.south east) {$u_t^{N_{\theta}}$};

% VERY SHORT Dashed arrows from stage 1 to resampled particles
\draw[arrow, dashed, shorten >=0.7cm, shorten <=0.7cm] (th1) -- (r1);
\draw[arrow, dashed, shorten >=0.7cm, shorten <=0.9cm] (thm) -- (r2);
\draw[arrow, dashed, shorten >=0.7cm, shorten <=0.7cm] (thm) -- (rm);
\draw[arrow, dashed, shorten >=0.7cm, shorten <=0.7cm] (thN) -- (rN);

% VERY SHORT Arrows from resampled to PMwG
\draw[arrow, dodgerblue!80!black, shorten >=0.6cm, shorten <=0.6cm] (r1) -- (p1);
\draw[arrow, dodgerblue!80!black, shorten >=0.6cm, shorten <=0.6cm] (r2) -- (p2);
\draw[arrow, dodgerblue!80!black, shorten >=0.6cm, shorten <=0.6cm] (rm) -- (pm);
\draw[arrow, dodgerblue!80!black, shorten >=0.6cm, shorten <=0.6cm] (rN) -- (pN);

% VERY SHORT Arrows from PMwG to final stage
\draw[arrow, shorten >=0.8cm, shorten <=0.8cm] (p1) -- (f1);
\draw[arrow, shorten >=0.8cm, shorten <=0.8cm] (p2) -- (f2);
\draw[arrow, shorten >=0.8cm, shorten <=0.8cm] (pm) -- (fm);
\draw[arrow, shorten >=0.8cm, shorten <=0.8cm] (pN) -- (fN);

\end{tikzpicture}
\caption{\footnotesize SMC$^2$ workflow for parameter learning. At time $t-1$, each parameter particle is associated with $N_x$ state particles. Upon receiving $y_t$, BPF advances forward in time for each $\theta^m$, producing likelihood estimates that update weights. When ESS drops below the threshold, parameter particles are resampled and rejuvenated via the PMwG kernel to maintain diversity. Here, $X_{s:t}^{m,1:N_x}=\big(X_{s:t}^{m,1},\ldots,X_{s:t}^{m,N_x}\big)$ denotes the collection of $ N_x$ state particle trajectories associated with the $m$-th parameter particle over the time points $s,\ldots,t$.}
\label{fig:smc2_workflow}
\end{figure}

For any block $b\in\{1,\ldots,A+1\}$, let $\theta_{-b}$ denote the collection of all remaining parameter blocks. Given the current value of $\theta_b$, a proposed value $\theta_b^*$ is drawn from a proposal density $q(\theta_b^* | \theta_b)$. In our setting, $q(\cdot | \theta_b):= \mathcal{N}\left(\theta_b,~(2.38)^2/d_b\cdot \hat{\Sigma}_{\theta,t}^{(b)}\right)$, where $\hat{\Sigma}_{\theta,t}^{(b)}$ is the empirical covariance matrix of the current particles for block $\theta_b$, with $d_b$ the dimension of the block. The BPF is rerun to produce the likelihood estimate $\tilde{P}(y_{1:t} | \theta_b^*, \theta_{-b})$, and the proposal is accepted with probability
\begin{align}\label{accept_ratio}
\alpha(\theta_b^*| \theta_b) = \min\left\{1, \;\frac{\tilde{P}(y_{1:t} | \theta_b^*, \theta_{-b}) \, P(\theta_b^*)}{\tilde{P}(y_{1:t} | \theta_b, \theta_{-b}) \, P(\theta_b)} \times \frac{q(\theta_b | \theta_b^*)}{q(\theta_b^* | \theta_b)}\right\}.
\end{align}
PMwG exploits the natural block structure induced by age-specific parameters, improving mixing efficiency in this high-dimensional setting. The resulting parameter particles provide approximate samples from the posterior distribution induced by the block-localised likelihood approximation. A schematic overview is provided in Figure~\ref{fig:smc2_workflow}.

As described above, SMC$^2$ provides samples from the posterior distribution of the model parameters, which can subsequently be used to infer the latent process. Conditional on a given parameter value, the associated block particle filter provides samples from the posterior distribution of the latent states given the observations. Marginalizing this conditional distribution over the posterior uncertainty in the parameters gives
\begin{align}
\tilde{P}(X_{t-L+1:t}| y_{1:t})
=
\mathrm{E}_{\theta | y_{1:t}}
\big[
\tilde{P}(X_{t-L+1:t}| y_{1:t}, \theta)
\big].
\end{align}
This accounts for uncertainty in both the parameters and latent states, allowing reconstruction of the unobserved epidemic trajectory and a posterior description of the transmission dynamics over time.

% \textcolor{red}{At time $t$, each BPF associated with a parameter particle $\theta^m$ provides samples from the block-localised conditional fixed-lag smoothing distribution $\tilde{P}(X_{t-L+1:t} | y_{1:t}, \theta^m)$. In principle, the equally weighted parameter particles after resampling could be used directly to approximate the marginal block-localised fixed-lag smoothing distribution
% \begin{align*}
%     \tilde{P}(X_{t-L+1:t}| y_{1:t})
%     =
%     \mathrm{E}_{\tilde{P}(\theta | y_{1:t})}
%     \big[
%     \tilde{P}(X_{t-L+1:t}| y_{1:t}, \theta)
%     \big].
% \end{align*}
% However, early in the time series the data are often insufficient to identify $\theta$ accurately, making the online approximation of $\tilde{P}(\theta|y_{1:t})$ subject to increased Monte Carlo variability. Instead, after processing the full data set $y_{1:T}$, we use the final parameter approximation $\tilde{P}(\theta|y_{1:T})$ and select $N_s \ll N_{\theta}$ parameter particles according to their final weights. For each selected parameter particle, the fixed-lag BPF is rerun over the observations, and the resulting trajectories provide a Monte Carlo approximation to the marginal block-localised fixed-lag smoothing distribution. In our experiments, $N_s=100$ provided stable posterior summaries.}

\begin{algorithm}[H]
\caption{SMC$^2$ with block update parameters (step $t$)} \label{SMC2}
Operations involving indexes $m$ must be performed for $m = 1, \dots, N_{\theta}$.

\textbf{Inputs:} Observations $y_{1:t}$, weighted particles $\{\theta^{m}, u_{t-1}^{m}\}$, associated BPF particles $X_{t-L:t-1}^{m,1:N_x}$.

\textbf{Outputs:} Weighted particles $\{\theta^{m}, u_t^{m}\}$, updated BPF ensemble members $X_{t-L+1:t}^{m,1:N_x}$.

\hrulefill
\begin{algorithmic}[1]
    \State \textbf{State propagation and weight updating:}
    \Statex \hspace{\algorithmicindent} Perform iteration $t$ of BPF with inputs $\{\theta^{m}, X_{t-L:t-1}^{m,1:N_x}, y_t\}$
  and get $X_{t-L+1:t}^{m,1:N_x}, ~\tilde{P}(y_t|y_{1:t-1},\theta^{m})$
    \Statex \hspace{\algorithmicindent} Compute weights $u_t^{m} = \bar{u}_{t-1}^{m} \, \tilde{P}(y_t | y_{1:t-1}, \theta^{m})$ and $\bar{u}^{m}_{t} = u_{t}^{m}/\sum_{k=1}^{N_{\theta}}u_{t}^{k}$
    \Statex \hspace{\algorithmicindent} Compute $\mathrm{ESS} = 1/\sum_{k=1}^{N_{\theta}}(\bar{u}_{t}^{k})^2$
     \Statex \textbf{if} $\mathrm{ESS} < \tau_R N_{\theta}$ \textbf{then}
        \State \hspace{\algorithmicindent} \textbf{Resample:} 
        \Statex \hspace{\algorithmicindent}\hspace{\algorithmicindent} 
        $\{ \theta^{1:N_{\theta}}, X_{t-L+1:t}^{1:N_{\theta},1:N_x}, \tilde{P}(y_t | \theta^{m}) \} 
        \sim \mathrm{Resampling}(\{ \theta^{1:N_{\theta}}, X_{t-L+1:t}^{1:N_{\theta},1:N_x}, \tilde{P}(y_t | \theta^{m}) \}, \bar{u}_t^{1:N_{\theta}})$
        \Statex \hspace{\algorithmicindent}\hspace{\algorithmicindent} Reset weights:  $u_t^{m} = 1/N_{\theta}$ 
    
        \State \hspace{\algorithmicindent} \textbf{PMwG move:} \Comment{$N_{\text{move}}$ repetitions}
         \Statex \hspace{\algorithmicindent}\hspace{\algorithmicindent} \textbf{for} each block $b  \in \{1, \ldots, A+1\}$ \textbf{do}
            \Statex \hspace{\algorithmicindent}\hspace{\algorithmicindent} \qquad Propose new value $\theta_b^* \sim q(\cdot | \theta_b)$
            \Statex \hspace{\algorithmicindent}\hspace{\algorithmicindent} \qquad Perform iterations $1, \ldots, t$ of Algorithm \ref{BPF} to obtain likelihood $\tilde{P}(y_{1:t} | \theta_b^*, \theta_{-b})$
            \Statex \hspace{\algorithmicindent}\hspace{\algorithmicindent} \qquad Accept/reject according to the Metropolis-Hastings ratio in Eq. \eqref{accept_ratio}
        \Statex \hspace{\algorithmicindent}\hspace{\algorithmicindent} \textbf{end for}

    \Statex\textbf{end if}
\end{algorithmic}
\end{algorithm}

\subsection{Practical considerations}
For missing observations, the corresponding likelihood contribution is set to one, so that unobserved $C_{a,t}$ values do not contribute to the likelihood. After the last observed time point, probabilistic forecasts are generated by forward simulation from the posterior particle approximation at the forecast origin, with latent infection counts propagated recursively via the renewal model and observations simulated from the observation model. During the forecast horizon, the age-specific susceptibility parameters $\beta_{a,t}$ are held at their posterior values at the forecast origin. This assumes that susceptibility remains at its estimated level over the forecast horizon, as future changes in factors affecting infection risk cannot be inferred from the observed data. Such an assumption is consistent with short-term epidemic forecasting approaches that project recent estimates forward \citep{abbott_estimating_2020}. The resulting forecasts are posterior predictive projections conditional on the estimated parameters at the forecast origin, capturing stochastic uncertainty in the epidemic and observation processes but not structural changes in transmission dynamics. They should therefore be interpreted as conditional forecasts rather than counterfactual predictions under changing epidemiological regimes.

To maintain stable likelihood estimation as data accumulate, the particle count $N_x$ must grow as $O(t)$ to keep the variance of the log-likelihood estimator bounded \citep{andrieu_particle_2010}. We implement this adaptively: $N_x$ is initialised at 300 with a cap of $N_x^{\max}=1\,000$. The dominant computational cost arises from running an inner particle filter for each parameter particle, yielding a per-time-step complexity of $\mathcal{O}(N_{\theta} N_xL)$; these inner filters are fully parallelisable across parameter particles, with only resampling and weight normalisation requiring global synchronisation. After each rejuvenation step, if the log-likelihood variance $\hat{\sigma}_{N_x}^2$ evaluated at the mean of the current parameter particles exceeds $1.5$, the particle set for each $\theta^{m}$ is replaced by a larger ensemble via generalised importance sampling with an artificial backward kernel yielding unit incremental weights \citep{botha_automatically_2023}. The updated count is $N_x^{\text{new}} = \min\{\lceil \hat{\sigma}_{N_x}^2 N_x\rceil, N^{\text{max}}_x \}$.  The number of parameter particles is fixed at $N_{\theta}=800$ throughout, and rejuvenation in Algorithm~\ref{SMC2} is triggered whenever the effective sample size falls below the threshold $\tau_R=0.4$. To assess the accuracy of the sequential inference, we compare the marginal filtered parameter estimate from SMC$^{2}$ at the final time point $T$ with that from an independent particle MCMC (PMCMC) run targeting the approximate posterior distribution $\tilde{P}(\theta| y_{1:T})$ \citep[see supplementary material Section~\ref{app:pmcmc}]{andrieu_particle_2010}.

The generation interval, incubation period, and onset-to-report delay are modelled as discretised Gamma distributions, truncated at fixed upper bounds and converted to probability mass functions using a one-day shift. Distributional parameters and prior specifications are summarised in Table~\ref{tab:model_params}. The mean generation interval is specified using estimates from \citep{xu_assessing_2023, watson_jointly_2024}, while the mean incubation period is based on the estimate reported by \citet{zeng_serial_2023}. Age-specific case ascertainment ratios $\rho_a$ are assigned Beta priors; the rationale for these choices and model initialization are provided in supplementary material Section~\ref{app:prior_initialisation}. Prior predictive simulations were used during model specification to ensure that the chosen priors produced plausible epidemic trajectories. Sensitivity analyses of alternative delay specifications and priors for $\beta_{a,1}$ are reported in supplementary material Section~\ref{app:sensitivity}. Posterior inference is robust to reporting-delay and prior assumptions, although estimates of $R_t$ show moderate sensitivity to the assumed generation interval. Sequential inference uses fixed-lag resampling with $L=31$ days, restricting resampling at each update to the most recent 31 days of the latent history.

\begin{table}[H]
\centering
\small
\caption{ Prior specifications, fixed parameters, and epidemiological delay distributions. Delay distributions are reported as mean $\pm$ standard deviation (in days), together with the truncation maximum. Prior distributions are expressed in standard probabilistic notation.}
\label{tab:model_params}
\begin{tabular}{p{8.5cm} r}
\hline
Description & Value / Prior \\
\hline
Initial susceptibility $\beta_{a,1}$, $a=1,\ldots,4$ & $\mathrm{Beta}(1,\,10)$ \\
RW2 innovation standard deviation $\sigma_a$, $a=1,\ldots,4$ & $\mathrm{LogNormal}(-3.9,\,0.5)$ \\
Case ascertainment ratio $\rho_1$ (0--24 years) & $\mathrm{Beta}(42,\,193)$ \\
Case ascertainment ratio $\rho_2$ (25--44 years) & $\mathrm{Beta}(60,\,213)$ \\
Case ascertainment ratio $\rho_3$ (45--64 years) & $\mathrm{Beta}(77,\,164)$ \\
Case ascertainment ratio $\rho_4$ (65+ years) & $\mathrm{Beta}(83,\,115)$ \\
Negative Binomial dispersion $\kappa$ & $\mathrm{Exp}(0.1)$ \\[6pt]
Generation interval & $3.3 \pm 2.1$ days (max 14 days) \\
Incubation period & $3.0 \pm 2.3$ days (max 14 days) \\
Symptom onset to report delay & $2 \pm 1.0$ days (max 10 days) \\
% Fixed resampling lag $L$ & 31 days \\
\hline
\end{tabular}
\end{table}

\section{Results}
\label{sec:result}
We first evaluated the proposed Bayesian age-stratified renewal model using simulated data generated under known parameter values and latent transmission trajectories. The results show that the proposed framework can recover both age-specific and population-level transmission dynamics. Detailed results from the simulation study are reported in Section~\ref{app:supp_sim} of the supplementary material.

We next applied the proposed framework to Irish COVID-19 surveillance data from 1 March to 17 July 2022, covering the peak and decline of the Omicron BA.2 wave and the emergence of the BA.4/BA.5 wave. During this period, national testing guidance shifted from PCR testing of all symptomatic individuals to testing focused on higher-risk groups, affecting the age distribution of notified cases\footnote{\url{https://www.hpsc.ie/a-z/respiratory/coronavirus/novelcoronavirus/surveillance/epidemiologyofcovid-19inirelandweeklyreports/2022/}}. Forecasts were then generated for the subsequent 14 days beyond the final observation date. A single run of the inference procedure over the full study period required approximately 11 hours on a Linux server using 16 CPU cores. The implementation and code used to reproduce the analyses are openly available at \href{https://github.com/Dhorasso/Age-Renewal-SMC.git}{https://github.com/Dhorasso/Age-Renewal-SMC.git}.

\subsection{Age-specific susceptibility and reproduction contributions}
\label{sec:beta}

\begin{figure}[H]
    \centering

    \includegraphics[width=\linewidth]{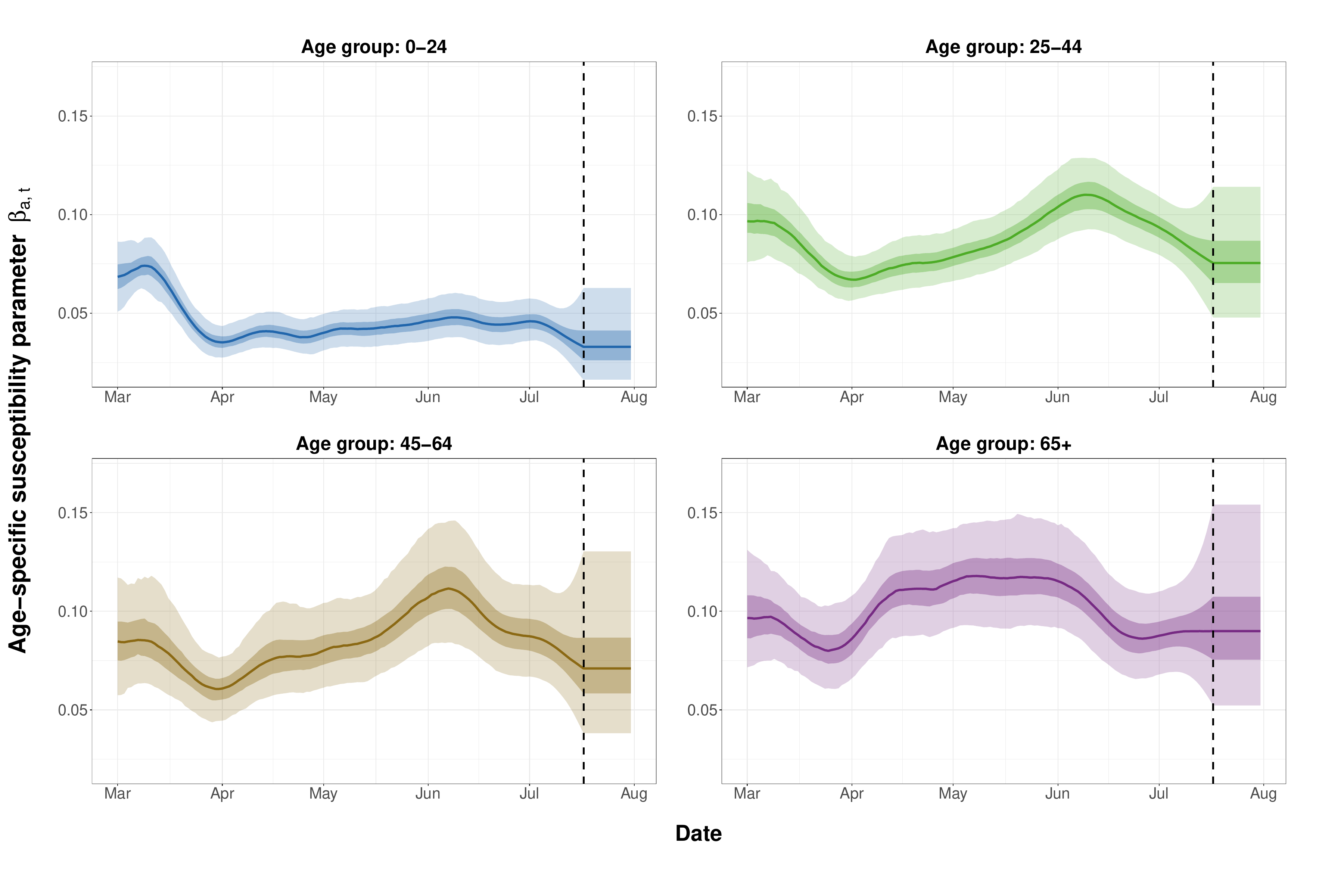}
    \caption{\footnotesize Marginal filtered estimates of the age-specific susceptibility $\beta_{a,t}$, $a=1,\ldots,4$. Solid lines denote posterior medians and shaded regions indicate 50\% and 95\% credible intervals. The vertical dashed line marks the forecast origin; beyond this point, $\beta_{a,t}$ is held fixed and uncertainty reflects stochastic propagation over the 14-day horizon.}
    \label{fig:beta_age}
\end{figure}

\begin{figure}[H]
  \centering
  \includegraphics[width=\linewidth]{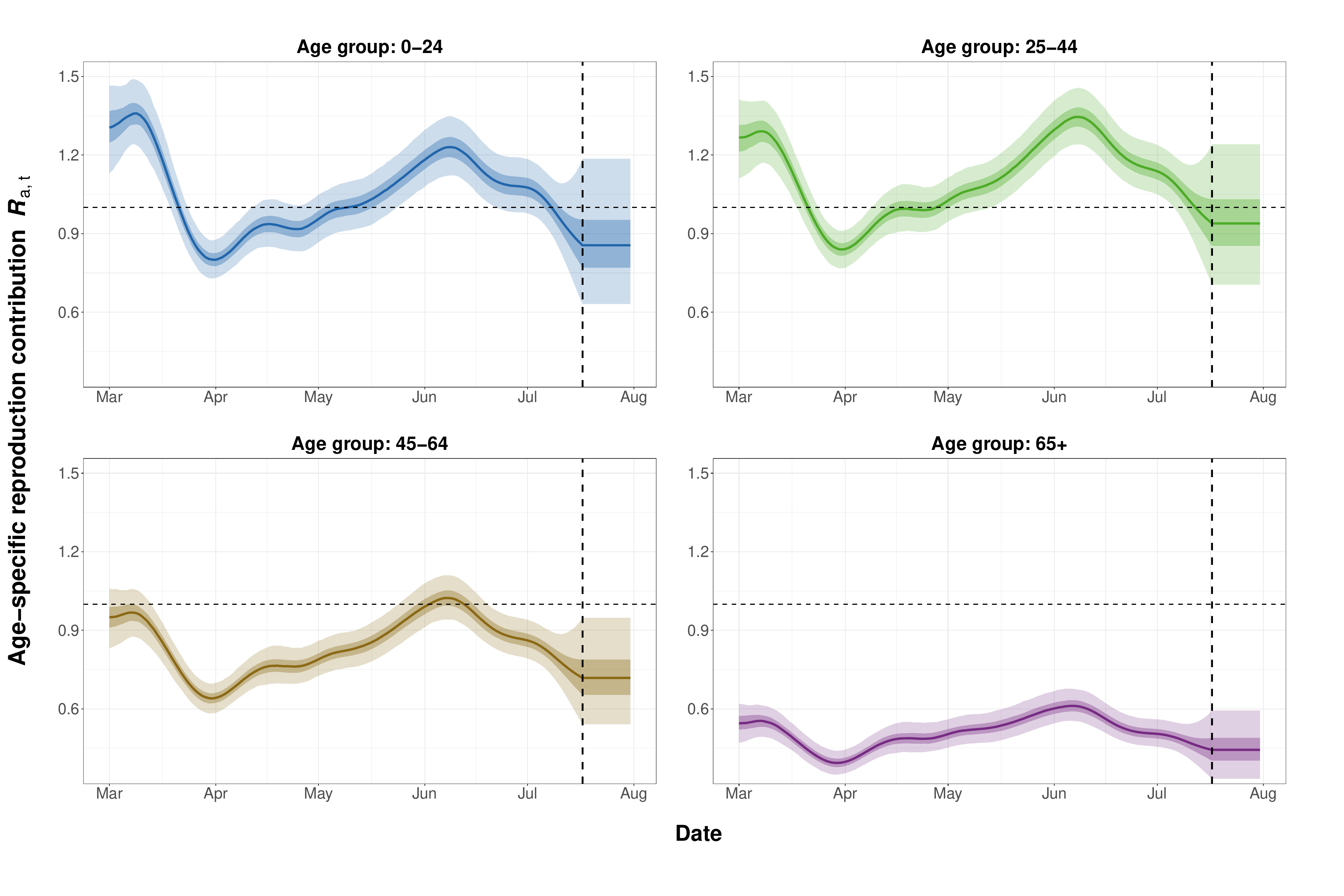}
  \caption{\footnotesize Marginal filtered estimates of the age-specific reproduction number contribution $R_{a,t}$, $a=1,\ldots,4$. Solid lines denote posterior medians and shaded regions indicate 50\% and 95\% credible intervals. The dashed horizontal line marks $R_{a,t}=1$; the vertical dashed line indicates the forecast origin.}
  \label{fig:Rt_age}
\end{figure}

Figures~\ref{fig:beta_age} and~\ref{fig:Rt_age} show marginal filtered estimates of the age-specific susceptibility parameter $\beta_{a,t}$ and the reproduction contribution $R_{a,t}$. These quantities capture different aspects of transmission: $\beta_{a,t}$ reflects the per-contact infection risk for age group $a$, whereas $R_{a,t}$ incorporates the contact structure through $\mathbf{M}$ to quantify the contribution of each group, as a source of infection, to onward transmission.

The susceptibility trajectories (Figure~\ref{fig:beta_age}) show a broadly stable ordering of the posterior medians, with the highest values in the 65+ group for most of the study period, intermediate and overlapping values in the 25--44 and 45--64 groups, and the lowest values in the 0--24 group throughout. The 25--44 and 45--64 groups decline to a trough in early April, rise to a peak in June, and decrease thereafter, whereas the 65+ group shows a shallower and earlier dip before plateauing at elevated values from April to June and declining gradually from July onwards. In contrast, after an initial decline in March, the 0--24 group remains approximately constant from April onwards, with only a small increase in June. These trajectories suggest an age gradient in the model-implied per-contact infection risk, although they may also reflect age-specific ascertainment effects, such as differences in symptom-driven testing. The ordering is consistent with previous findings \citep{bouranis_bayesian_2025}.

Reproduction number contributions (Figure~\ref{fig:Rt_age}) show a different pattern after accounting for age-specific mixing. All groups exhibit a broadly similar shape, with a relative peak in early March near the height of the BA.2 wave, a decline to a trough in late March/early April, and a resurgence from May onwards associated with the emergence of BA.4/BA.5, though only the 0--24 and 25--44 groups cross above one during the March peak. The largest increases are observed for the 0--24 and 25--44 groups, which return above one from May and remain elevated into July. The 0--24 group reaches the highest initial peak but declines more sharply thereafter, whereas the 25--44 group attains a markedly higher summer peak and maintains elevated values for longer. The 45--64 group briefly exceeds one around late May to mid-June, while the 65+ group remains well below one throughout, peaking at about 0.61 in June.

These differences between $\beta_{a,t}$ and $R_{a,t}$ arise from the interaction between susceptibility and the mixing structure in $\mathbf{M}$ (Figure~\ref{fig:contact}). The 0--24 group has the highest within-group contact rate and contact volume, contributing to its pronounced March peak in $R_{0-24,t}$; however, this contribution declines as cross-group mixing becomes insufficient to sustain transmission despite the low $\beta_{0-24,t}$. The 25--44 group combines substantial within and between group contact intensity with moderate susceptibility, resulting in the largest sustained contribution to transmission. Conversely, although the 65+ group has the highest susceptibility, its lower contact volume and weaker within-group mixing constrain its reproduction contribution, which is largely associated with interactions with the 25--44 and 45--64 groups.

These results highlight that age-specific transmission contributions are shaped by both susceptibility and contact patterns. Consequently, interventions targeting highly connected groups may have a larger impact on transmission potential, whereas reducing infection risk remains particularly relevant for groups with higher susceptibility. A sensitivity analysis to the contact matrix (Section~\ref{app:sensitivity_contact} of the supplementary material) shows that $\pm 30\%$ one-at-a-time perturbations of the school, work and community weights shift $\beta_{a,t}$ by up to about 20\%, reflecting its structural non-identifiability with $\mathbf{M}$. Nonetheless, the age ordering of susceptibility is largely preserved and $R_{a,t}$ changes by at most about 9\%, indicating limited sensitivity to the assumed contacts.

\subsection{Latent infections and model fit}
\label{sec:fit}

\begin{figure}[H]
  \centering
  \includegraphics[width=\linewidth]{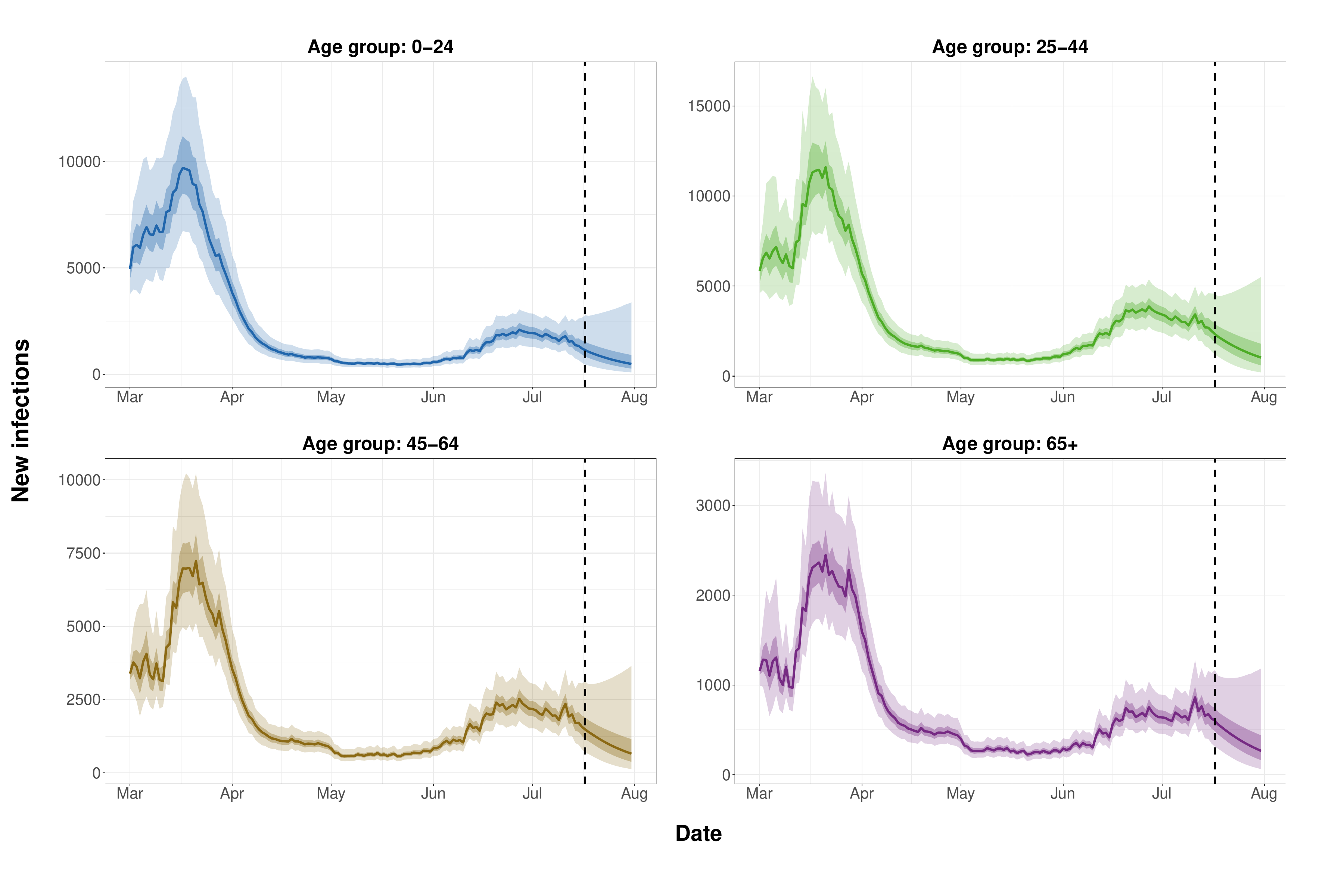}
  \caption{\footnotesize Marginal filtered estimates of latent daily infections $I_{a,t}$, $a=1,\ldots,4$. Solid lines denote posterior medians and shaded regions indicate 50\% and 95\% posterior credible intervals. The vertical dashed line marks the forecast origin.}
  \label{fig:infections}
\end{figure}

Posterior distributions for latent infections, $I_{a,t}$, are shown in Figure~\ref{fig:infections}. The inferred infection trajectories recover the principal features of the epidemic over the study period. Infection incidence peaks in late March, during the late growth and peak phase of the BA.2 wave. It then declines to a trough during May and early June before increasing again with the emergence of BA.4/BA.5. The largest infection burdens are estimated in the 25--44 year group, followed by the 0--24 and 45--64 year groups, whereas the 65+ group exhibits a substantially attenuated trajectory throughout.

Static parameters reported in Table~\ref{tab:model_params} are estimated sequentially within the filtering framework, and their evolution is illustrated in Figure~\ref{fig:supp_param_covid} in the supplementary material. As additional data are assimilated, the corresponding posterior distributions progressively concentrate, reflecting a gradual reduction in parameter uncertainty over time. The final filtered estimates are in close agreement with the corresponding posterior summaries obtained from the PMMH algorithm, providing empirical support for the consistency between sequential filtering and full batch inference. Further diagnostic evidence from the SMC$^2$ implementation is provided in Figure~\ref{fig:diag_smc2} in the supplementary material, confirming stable performance of the sequential inference procedure.

\begin{figure}[H]
  \centering
  \includegraphics[width=\linewidth]{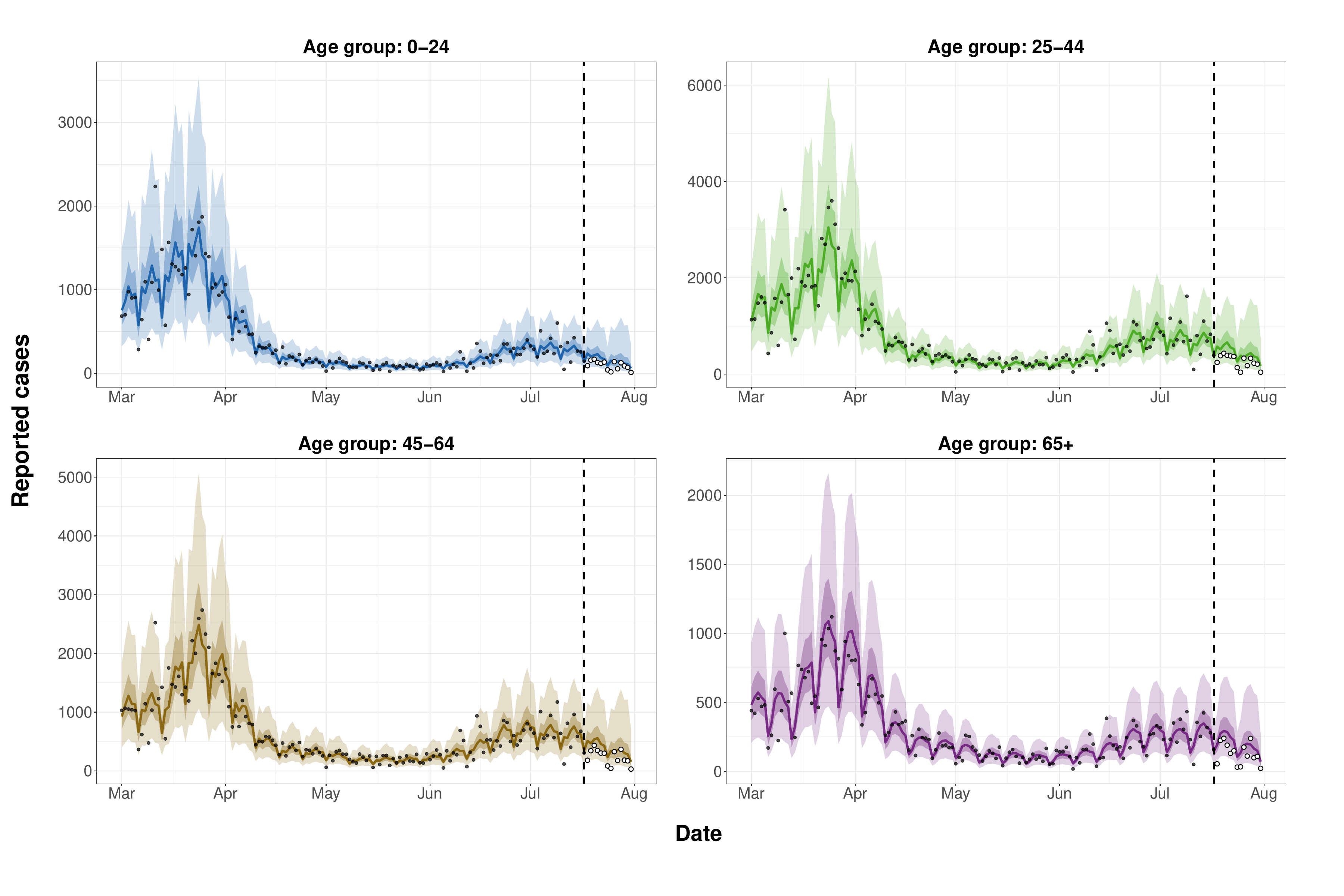}
\caption{\footnotesize Marginal filtered distributions for reported daily cases $C_{a,t}$, $a=1,\ldots,4$. Solid lines denote posterior medians and shaded regions indicate 50\% and 95\% posterior credible intervals. Black dots represent observations used in filtering. The vertical dashed line marks the forecast origin. White circles denote out-of-sample observations used for forecast evaluation.}
  \label{fig:cases}
\end{figure}

Using the inferred parameters, including time-varying susceptibility, we generate probabilistic estimates of daily reported cases and compare these with the observed data (Figure~\ref{fig:cases}). Parameter uncertainty is propagated through the model to obtain posterior predictive distributions. Observed counts are generally well captured across age groups, with the posterior median reproducing both the decline following the BA.2 wave and the subsequent increase associated with BA.4/BA.5. Credible intervals widen during periods of increasing incidence, reflecting greater uncertainty when case numbers change rapidly. The model also captures short-term variation in reported cases associated with differences in testing and reporting intensity through the day-of-week adjustment.

\begin{figure}[H]
  \centering
  \includegraphics[width=0.82\linewidth]{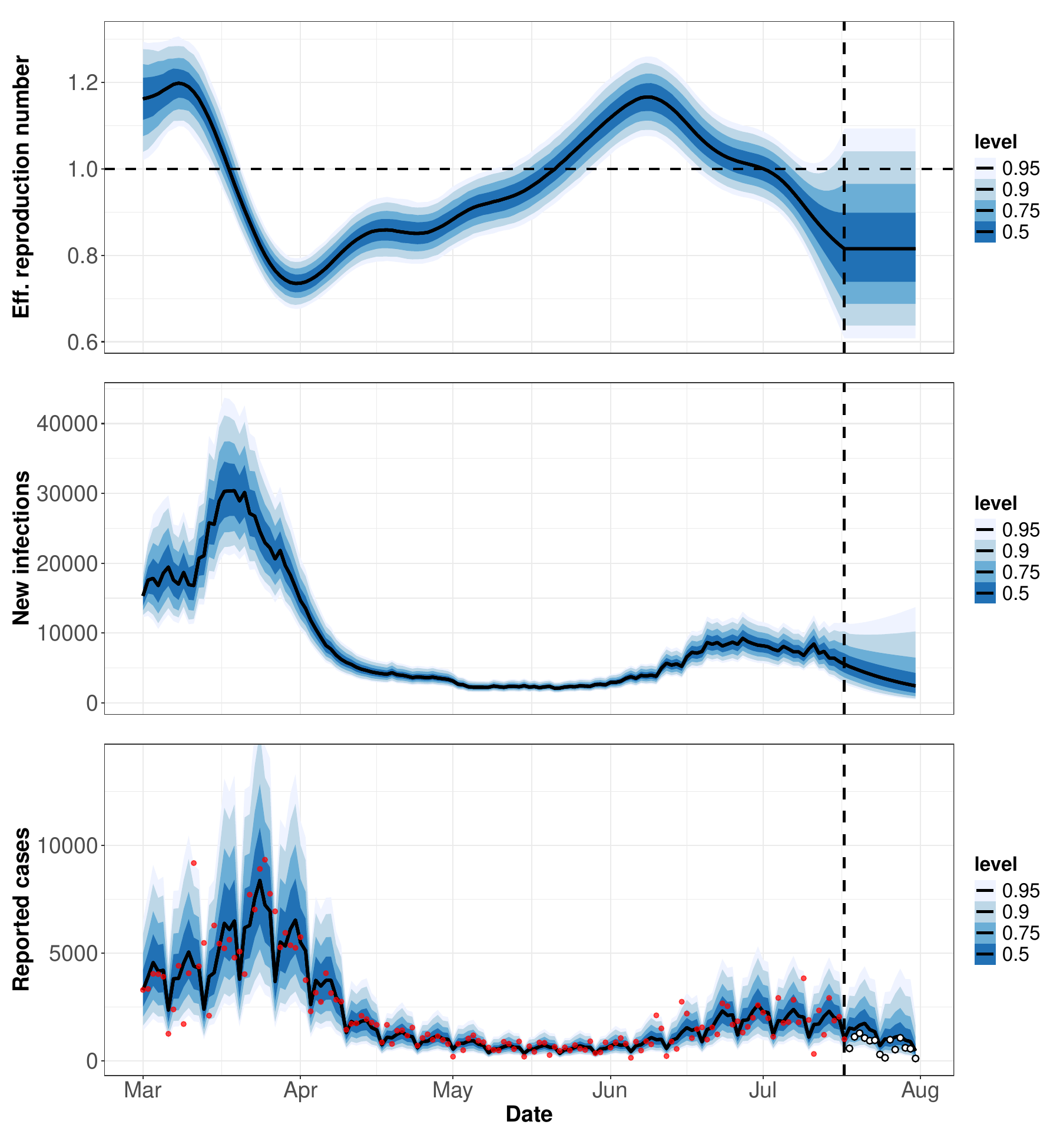}
  \caption{\footnotesize Aggregate posterior distributions for the effective reproduction number $R_t$ (top), total latent infections $I_t$ (middle), and total reported cases $C_t$ (bottom). Solid lines denote posterior medians and shaded regions indicate posterior credible intervals. The vertical dashed line marks the forecast origin. White circles denote observations not assimilated during filtering.}
  \label{fig:aggregate}
\end{figure}

The corresponding population-level summaries are presented in Figure~\ref{fig:aggregate}. The estimated effective reproduction number, $R_t$, closely tracks the inferred epidemic dynamics. $R_t$ exceeds one at the start of the study period, declines below the epidemic threshold in late March to a minimum of about 0.75 in early April, and then rises gradually, remaining below one until mid-to-late May. It crosses the threshold again in late May and peaks at around 1.15 in early June, ahead of the observed increase in reported cases. It then declines, falling back below one in early July.

Forecasts for the 14-day period following 17 July 2022 are shown after the vertical dashed line in Figures~\ref{fig:cases}--\ref{fig:aggregate}. The model exhibits good short-term predictive performance, with most observations lying within the 95\% predictive intervals, indicating reasonable calibration, with some under-coverage in the older age groups. A mild tendency toward overestimation is observed, consistent with a slowdown in BA.4/BA.5 transmission relative to the dynamics inferred at the forecast origin; that is, the model extrapolates recent trends that slightly exceed the subsequently realised incidence. As expected in a renewal-based state-space framework, predictive uncertainty increases with forecast horizon across all age groups, reflecting accumulated uncertainty in latent incidence and transmission intensity. We further assess probabilistic forecast accuracy using the continuous ranked probability score (CRPS), computed against observed data using 2,000 posterior predictive particles and log-transformed case counts to reduce the influence of extreme observations and improve comparability across time and age strata.

\begin{figure}[H]
\centering
\includegraphics[width=0.63\linewidth]{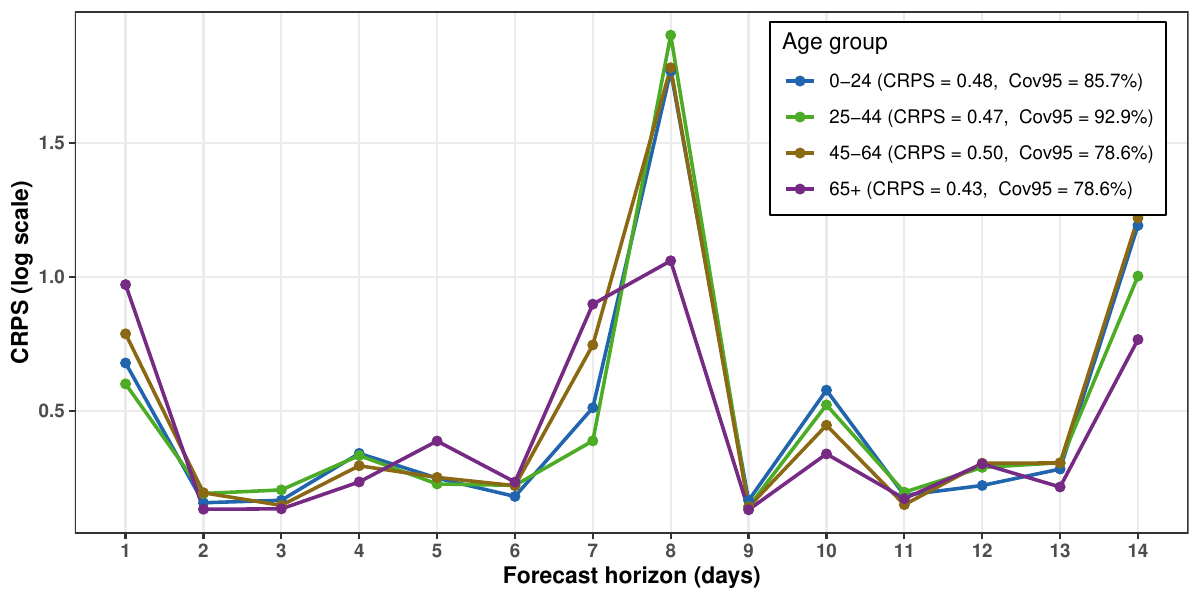}
\caption{\footnotesize Continuous ranked probability score (CRPS) by forecast horizon and age group, computed on log-transformed data from the 14-day forecast issued on 17 July 2022. Cov95 denotes the proportion of the 14 held-out observations within the 95\% posterior predictive interval.}
\label{fig:crps}
\end{figure}

Figure~\ref{fig:crps} reports the CRPS across the 14 forecast horizons for each age group, based on the forecasts issued at the final observation date (17 July 2022). Overall, predictive accuracy is broadly comparable across age groups, with the mean CRPS ranging from 0.43 for the 65+ group to 0.50 for the 45--64 group. In contrast, empirical coverage of the 95\% predictive intervals (Cov95) is more heterogeneous. Coverage ranges from 78.6\% for both the 45--64 and 65+ groups to 92.9\% for the 25--44 group, while the 0--24 group has intermediate coverage of 85.7\%. Thus, the empirical coverage is below the nominal 95\% level for all four age groups, with the greatest degree of under-coverage observed for the 45--64 and 65+ groups. However, these coverage estimates should be interpreted with caution, as they are based on only 14 forecast horizons per age group. Consequently, a difference of just one or two observations falling outside the predictive interval can result in a relatively large change in the estimated coverage. The under-coverage may partly reflect the assumption that $\beta_{a,t}$ is held fixed at the forecast origin. Since $\beta_{a,t}$ was still declining from its June peak, this assumption may have led to overestimated transmission, consistent with the mild overestimation noted above.

CRPS does not increase monotonically with forecast horizon but shows sharp peaks at horizons 8 and 14, corresponding to the weekly reporting cycle evident in Figure~\ref{fig:cases}, where low-reporting days inflate the score independently of forecast calibration. Outside these horizons, CRPS remains stable and comparable across age groups, indicating consistent short-term predictive performance.

\subsection{Comparison with homogeneous models}
\label{sec:Rt}

\begin{figure}[H]
  \centering
  \includegraphics[width=0.8\linewidth]{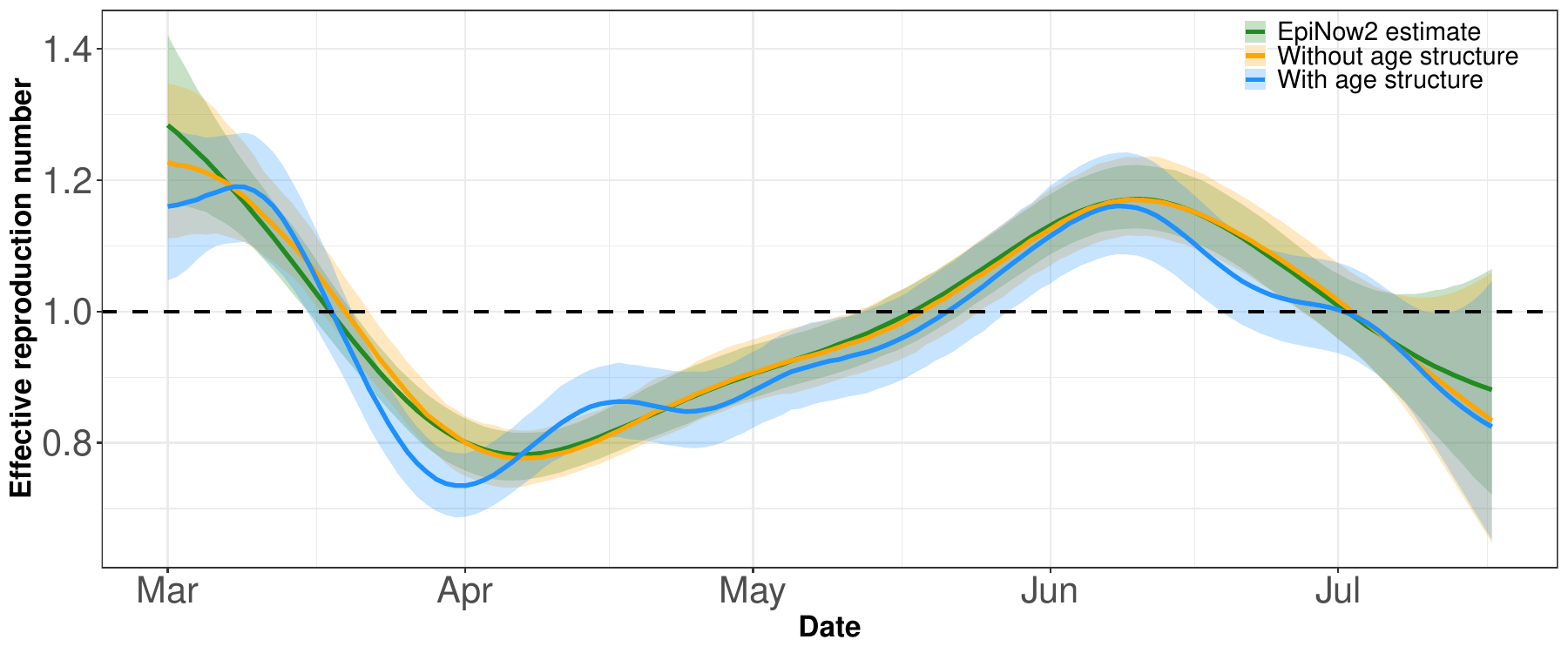}
  \caption{\footnotesize Comparison of filtered estimates of the effective reproduction number $R_t$ from the age-stratified model (blue), EpiNow2 (green), and our homogeneous renewal model (orange). Solid lines denote posterior medians and shaded regions indicate 90\% posterior credible intervals.}
  \label{fig:Rt_comp}
\end{figure}

Figure~\ref{fig:Rt_comp} compares estimates of the effective reproduction number obtained from the proposed age-stratified model, a homogeneous version of our model (see Section~\ref{app:simple_model} of the supplementary material), and EpiNow2 \citep{abbott_estimating_2020}, a widely used renewal-equation-based tool for real-time estimation of the time-varying reproduction number from case data, which models $R_t$ non-parametrically using a Gaussian process with a Mat\'ern~(3/2) kernel. Like our homogeneous model, EpiNow2 assumes a well-mixed population without demographic stratification, providing a natural benchmark for assessing the effect of age structure. The homogeneous model closely tracks the EpiNow2 estimates throughout the study period, with both capturing the main temporal features of transmission: the decline below one following the peak of the BA.2 wave, the sub-threshold period until mid-May, the subsequent rise to a peak in early June associated with the emergence of BA.4/BA.5, and the decline back towards one in July. This agreement provides reassurance that the proposed renewal framework produces estimates consistent with an established method in the absence of demographic structure.
 The age-stratified model recovers the same overall temporal pattern but differs slightly in the magnitude of $R_t$ and in the associated uncertainty. The largest departures from the homogeneous and EpiNow2 estimates occur in March and April, around the peak and decline of the BA.2 wave. The age-stratified estimate starts lower in early March, reaches a lower trough in early April, and rises earlier towards the sub-threshold plateau in late April. Smaller differences appear after the early June peak, where the age-stratified estimate declines slightly earlier. The 90\% credible intervals are generally wider under the age-stratified model, particularly around the April trough, reflecting the additional uncertainty associated with the age-specific susceptibility parameters and the structured transmission dynamics. During periods of more stable transmission (late May to early June and July), the three models produce similar estimates. Overall, the comparison suggests that accounting for age structure matters most near epidemic turning points, when differences in contact patterns and transmission intensity across age groups (Figure~\ref{fig:Rt_age}) most strongly influence population-level dynamics.

\section{Discussion}\label{sec:discussion}

We developed a Bayesian age-stratified renewal model for sequential inference of epidemic transmission dynamics from routinely collected surveillance data. By combining age-structured transmission, contact patterns, and sequential Monte Carlo methods within a hierarchical state-space framework, the model jointly estimates latent infections and transmission indicators while coherently propagating uncertainty. Simulation results support the ability of the model to recover both age-specific and population-level transmission dynamics under known conditions.

Application to Irish COVID-19 surveillance data illustrates the value of incorporating age structure into epidemic renewal models. Although estimates of the effective reproduction number are broadly consistent with those from a homogeneous formulation, the age-structured model provides additional insight into the sources of transmission heterogeneity. The analysis identifies distinct age-specific contributions to transmission: individuals aged 25--44 act as persistent drivers of spread, those aged 0--24 contribute most strongly at the start of the study period (the late growth phase of the BA.2 wave) and during the BA.4/BA.5 resurgence before declining more sharply than the 25--44 group, the 45--64 group remains closer to the epidemic threshold, and the 65+ group remains furthest from it despite its higher susceptibility. These patterns differ from susceptibility estimates, which are highest among older adults and lowest among younger groups. Instead, transmission contributions are shaped by the interaction between susceptibility and contact behaviour, with younger and working-age groups combining moderate susceptibility with greater contact rates. This separation highlights how aggregate transmission measures may obscure important demographic heterogeneity and is consistent with evidence that contact patterns are key determinants of epidemic spread \citep{monod_age_2021}. The age stratification used here reflects differences in morbidity, social mixing, and transmission risk across the population, while retaining sufficient resolution to capture the main demographic patterns in the epidemic. The same modelling framework can accommodate alternative age partitions, depending on the epidemiological question and available data.

The proposed framework is sequential by design, carrying particle states and parameter information forward as new observations become available. When rejuvenation is triggered, the parameter update uses all observations available up to the current time point; thus, although inference proceeds sequentially, the standard SMC$^2$ implementation is not fully \textit{online} in the strict sense, as rejuvenation becomes more costly with accumulated data. This cost can be limited by restricting rejuvenation to a fixed window of recent observations \citep{temfack_sequential_2025}. The sequential formulation nevertheless provides coherent posterior updating of latent states and model parameters as observations accumulate, supporting timely assessment of epidemic dynamics with uncertainty quantification. As in other renewal models \citep{abbott_estimating_2020, steyn_primer_2025}, inference is affected by generation-time and infection-to-observation delays, which are incorporated alongside observation variation to support inference from partially observed surveillance data.

Several limitations motivate future extensions. Reported cases are influenced not only by transmission dynamics but also by testing behaviour and reporting practices. In particular, changes in testing guidance during 2022 may have affected the observed age-specific patterns. Although observation noise, reporting delays, and age-specific ascertainment are modelled explicitly, transmission and ascertainment remain only partially identifiable from case data alone \citep{bouman_bayesian_2024}. Furthermore, immunity from prior infection or vaccination is not explicitly represented, and incorporating additional data sources such as serological, vaccination, or mobility information could improve identifiability \citep{bhatt_semimechanistic_2023}. Other assumptions could also be relaxed, including the fixed contact matrix and constant generation-time and reporting-delay distributions. Sensitivity analyses (Section~\ref{app:sensitivity} of the supplementary material) indicate limited sensitivity to the assumed priors, delay distributions, and composition of contacts. They also suggest that additional flexibility may improve performance under more substantial deviations. From a computational perspective, the block particle filter improves scalability in high-dimensional latent systems, although the PMwG rejuvenation step remains computationally demanding. Future improvements could include more efficient particle filtering strategies, adaptive proposal mechanisms, and delayed-acceptance kernels that reduce the computational cost of evaluating proposed parameter updates.

The Irish application focuses on a limited period spanning the transition between Omicron waves. While this analysis illustrates the inferential capability of the framework under realistic surveillance conditions, evaluating sustained performance over longer observation periods and across multiple epidemic waves will be important for further assessing its performance in operational surveillance settings. Although developed for age structure, the framework extends naturally to other forms of heterogeneity, including geographic, behavioural, and socioeconomic variation. For example, incorporating mobility or commuting data could support spatial renewal models with joint inference across regions. More broadly, this work contributes to structured renewal modelling by combining demographic heterogeneity with sequential Bayesian inference, providing a flexible framework for investigating transmission dynamics and integrating richer data sources in future epidemic analyses.

\section*{Acknowledgments}
This publication has emanated from research conducted with the financial support of Taighde Éireann – Research Ireland under Grant number 21/FFP-P/10123. The authors would also like to thank Eishita Yadav for helpful discussions regarding the social contact matrix.

%Bibliography here
% \bibliographystyle{agsm}
\bibliographystyle{apalike-doi} 
\bibliography{reference}

\appendix
\renewcommand{\thefigure}{\thesection\arabic{figure}}
\renewcommand{\thetable}{\thesection\arabic{table}}
\renewcommand{\thealgorithm}{\thesection\arabic{algorithm}}
\counterwithin{figure}{section}
\counterwithin{table}{section}

% ============================================================
% supplementary material — Simulation Experiment
% Journal of the Royal Statistical Society: Series A
% ============================================================
\newpage

\section{Filtering recursion and likelihood approximation }
\label{app:bpf_likelihood}
For a fixed parameter vector $\theta$, let
$\pi(X_{1:t})=P(X_{1:t}| y_{1:t},\theta)$
denote the filtering distribution of the latent particle trajectories. The filtering recursion relies on the conditional independence assumptions that $y_t$ is conditionally independent of $y_{1:t-1}$ given the latent state history $X_{1:t}$, and that $X_t$ is conditionally independent of $y_{1:t-1}$ given the previous latent history $X_{1:t-1}$. Under these assumptions, the filtering distribution can be recursively obtained from Bayes' rule as
\begin{align}
\pi(X_{1:t})
&=
P(X_{1:t}| y_{1:t},\theta) \nonumber\\
&\propto
P(y_t| X_{1:t},\theta)
P(X_{1:t}| y_{1:t-1},\theta) \nonumber\\
&=
P(y_t| X_{1:t},\theta)
P(X_t| X_{1:t-1},\theta)
P(X_{1:t-1}| y_{1:t-1},\theta) \nonumber\\
&=
P(y_t| X_{1:t},\theta)
P(X_t| X_{1:t-1},\theta)
\pi(X_{1:t-1}).
\end{align}
Let $q(X_{1:t})=q(X_{1:t}|y_{1:t},\theta)$ denote the proposal distribution used to generate the particle trajectories. We define it recursively as
\begin{align}
q(X_{1:t})=q(X_t| X_{1:t-1},y_{t},\theta)q(X_{1:t-1}).
\end{align}
The corresponding importance weights are given by the ratio of the target filtering distribution to the proposal distribution,
\begin{align}
w_t^{(j)} = \frac{\pi(X^{(j)}_{1:t})}{q_t(X_{1:t}^{(j)})} 
\propto \frac{ P(y_t| X_{1:t}^{(j)},\theta) P(X_t^{(j)}| X_{1:t-1}^{(j)},\theta) }
{q(X_t^{(j)}| X_{1:t-1}^{(j)},y_{t},\theta)}w_{t-1}^{(j)} . 
\end{align}
Choosing the proposal distribution to be the state transition density, $q(X_t|X_{1:t-1},y_{t},\theta)=P(X_t|X_{1:t-1},\theta)$, yields the bootstrap particle filter, for which the transition terms cancel. Moreover, following the resampling step, all particles are assigned equal weights, and the recursive weights simplify to
\begin{align}
w_t^{(j)}
=
P(y_t|X_{1:t}^{(j)},\theta)
=
\prod_{a=1}^{A}
P(y_{a,t}|X_{a,1:t}^{(j)},\theta),
\end{align}
where the factorisation follows from conditional independence of observations across age groups. The corresponding block-local likelihood contributions used for age-specific resampling are therefore
\begin{align}
w_{a,t}^{(j)}
=P(y_{a,t}|X_{a,1:t}^{(j)},\theta).
\end{align}
With the fixed-lag implementation, resampling is applied to the most recent $L$ states within each particle trajectory. The retained particles therefore approximate the lag-window smoothing distribution
$P(X_{t-L+1:t}|y_{1:t},\theta)$,
rather than the full trajectory distribution $P(X_{1:t}|y_{1:t},\theta)$. States outside the retained window are not directly updated by future observations and affect inference only through their contribution to the propagated trajectories. Under standard filtering stability conditions, the approximation error decreases as $L$ increases \citep{doucet_tutorial_}. Combined with block-wise resampling, this yields the block-localised fixed-lag particle approximation rather than the exact joint filtering distribution.

The predictive likelihood required for sequential parameter learning is obtained by integrating the observation density with respect to the predictive distribution of the retained latent trajectories,
\begin{align}
P(y_t|y_{1:t-1},\theta)
&=
\int
P(y_t|X_{t-L+1:t},\theta)
P(X_{t-L+1:t}|y_{1:t-1},\theta)
\mathrm{d}X_{t-L+1:t}.
\end{align}
The propagated particles $\{\tilde{X}_{t-L+1:t}^{(j)}\}_{j=1}^{N_x}$ provide an empirical approximation of the predictive distribution,
\begin{align}
P(X_{t-L+1:t}|y_{1:t-1},\theta)
\approx
\frac{1}{N_x}
\sum_{j=1}^{N_x}
\delta_{\tilde{X}_{t-L+1:t}^{(j)}}(X_{t-L+1:t}),
\end{align}
which leads to the Monte Carlo approximation
\begin{align}\label{inc_lik}
\tilde{P}(y_t|y_{1:t-1},\theta)
&=
\frac{1}{N_x}
\sum_{j=1}^{N_x}
P(y_t|\tilde{X}_{t-L+1:t}^{(j)},\theta)\nonumber\\
&=
\frac{1}{N_x}
\sum_{j=1}^{N_x}
\prod_{a=1}^{A}
P(y_{a,t}|\tilde{X}_{a,t-L+1:t}^{(j)},\theta)\nonumber\\
&=
\frac{1}{N_x}
\sum_{j=1}^{N_x}
\prod_{a=1}^{A}w_{a,t}^{(j)} .
\end{align}
% The corresponding block-localised approximation to the marginal likelihood up to time $t$ is obtained through the sequential decomposition,
% \begin{align}
% \tilde{P}(y_{1:t}|\theta)
% =
% \prod_{t=1}^{t}
% \tilde{P}(y_s|y_{1:s-1},\theta).
% \end{align}
% This approximation introduces two sources of error. First, the block particle filter replaces the full particle approximation by age-specific resampling based on local likelihood contributions. Although propagation follows the joint transition distribution, block-wise resampling modifies the dependence structure between age-specific components and therefore introduces an approximation due to the coupling between age groups through the contact matrix. Second, the fixed-lag approximation truncates the dependence between current observations and infections occurring more than $L$ time steps in the past. Under standard filtering stability conditions, this truncation error decreases as the lag length increases. Consequently, $\tilde{P}(y_{1:t}|\theta)$ should be interpreted as an approximation to the exact marginal likelihood, with accuracy controlled by the number of particles $N_x$, the block decomposition, and the lag length $L$. For further discussion of block particle approximations and their application in high-dimensional state-space models, we refer the reader to \citet{asfaw_tutorial_2024}.
The corresponding block-localised approximation to the marginal likelihood up to time $t$ is obtained through the sequential decomposition,
\begin{align}\label{eq:app_bloc_mglik}
\tilde{P}(y_{1:t}|\theta)
=
\prod_{s=1}^{t}
\tilde{P}(y_s|y_{1:s-1},\theta).
\end{align}
This approximation introduces two sources of error. First, block-wise resampling modifies the dependence structure between age-specific components induced by the contact matrix, replacing the full joint particle approximation with age-specific resampling based on local likelihood contributions. Second, the fixed-lag approximation truncates the dependence between current observations and infections occurring more than $L$ time steps in the past. Consequently, $\tilde P(y_{1:t}|\theta)$ is an approximation to the marginal likelihood, with accuracy controlled by $N_x$, the lag $L$, and the block decomposition.

% For further discussion of block particle approximations in high-dimensional state-space models, we refer the reader to \citet{asfaw_tutorial_2024}.

This construction extends naturally to joint sequential learning of $\theta$ and the latent states, following the SMC$^2$ framework of \citet{chopin_smc2_2013}. Each of $N_{\theta}$ parameter particles $\theta^{m}$, $m=1,\dots,N_{\theta}$, carries its own instance of the block-localised fixed-lag filter above, and the resulting estimate $\tilde{P}(y_{1:t}|\theta^{m})$ is used to weight and resample the parameter particles at each time step.

Let $\psi_{t,\theta}(\tilde{X}_{1:t}^{1:N_x},X_{1:t}^{1:N_x})$ denote the joint density of all random variables generated by the state particle filter
up to time $t$. This density includes the propagated trajectories $\tilde{X}_{1:t}^{1:N_x}$ and the resampled trajectories $X_{1:t}^{1:N_x}$, with the corresponding ancestry variables generated during the block-wise resampling steps implicitly included in the particle-filter construction. The SMC$^2$ targets the extended distribution associated with the block-localised likelihood approximation. Denoting this distribution by $\pi_t(\theta,\tilde{X}_{1:t}^{1:N_x},X_{1:t}^{1:N_x}|y_{1:t})$, it is defined as
\begin{align}
\pi_t\big(\theta,\tilde{X}_{1:t}^{1:N_x},X_{1:t}^{1:N_x}\big|y_{1:t}\big)
&\propto
P(\theta)\,\tilde{P}(y_{1:t}|\theta,\tilde{X}_{1:t}^{1:N_x})\,
\psi_{t,\theta}\big(\tilde{X}_{1:t}^{1:N_x},X_{1:t}^{1:N_x}\big) \nonumber\\
&\propto
\pi_{t-1}\big(\theta,\tilde{X}_{1:t-1}^{1:N_x},X_{1:t-1}^{1:N_x}\big|y_{1:t-1}\big)\,
\tilde{P}(y_t|y_{1:t-1},\theta,\tilde{X}_{1:t}^{1:N_x}) \nonumber\\
&\qquad\times
\psi_{t,\theta}\big(\tilde{X}_t^{1:N_x},X_t^{1:N_x}\big|X_{1:t-1}^{1:N_x}\big),
\end{align}
where $P(\theta)$ is the prior density on $\theta$, and $\tilde{P}(y_t|y_{1:t-1},\theta,\tilde{X}_{1:t}^{1:N_x})$ is the Monte Carlo predictive likelihood increment defined above, computed from the block-local weights $w_{a,t}^{(j)}$ in Eq. \eqref{inc_lik}.

This construction is not a proper SMC$^2$ scheme in the sense of \citet{chopin_smc2_2013}: the inner filter targets an approximation to the filtering distribution rather than the exact one, since block-wise resampling breaks the joint dependence between age groups and the fixed-lag truncation discards the exact dependence of $y_t$ on states more than $L$ steps in the past. Accordingly, $\pi_t(\theta, X_{1:t})$ is best understood as an SMC$^2$-\emph{like} target: the outer sampler still weights $\theta$-particles consistently by the block-localised approximate likelihood $\tilde{P}(y_{1:t}|\theta)$, but the resulting joint distribution over $(\theta,X_{1:t})$ inherits the same two sources of approximation error as the inner filter, controlled by the block decomposition.

\section{Prior specification and model initialisation}

\subsection*{Age-specific case ascertainment priors}\label{app:prior_initialisation}

Reported case counts under-represent true infections, and this deficit is not uniform across age: from January 2022, Irish national testing guidance restricted state-funded PCR testing for individuals aged 4--39 without an underlying risk condition, directing this group toward self-managed antigen testing, while PCR access was retained for children under 4, adults aged 40 and over, and clinically vulnerable individuals\footnote{\url{https://www.hpsc.ie/a-z/respiratory/coronavirus/novelcoronavirus/surveillance/epidemiologyofcovid-19inirelandweeklyreports/2022/}}. As no stratified serosurvey covering the study period is available for Ireland, the ascertainment ratios $\rho_a$, $a=1,\ldots,4$, are not identifiable from case counts alone. In the absence of an auxiliary data stream that is not connected with testing behaviour, $\rho_a$ can trade off with the negative binomial dispersion parameter $\kappa$ in the observation process. We therefore assign each $\rho_a$ an informative $\mathrm{Beta}(\alpha_a,b_a)$ prior, with target mean and standard deviation set from (i) the age-tiered testing policy above, and (ii) a national seroprevalence survey of individuals aged 12--69 (June--July 2020) estimating true infections at approximately three times confirmed PCR notifications \citep{heavey_study_2021}, taken as an upper bound given subsequent reductions in testing eligibility. This motivates lower ascertainment for the 0--24 and 25--44 groups (self-managed antigen pathway) and higher ascertainment for the 45--64 and 65+ groups (retained PCR pathway, higher clinical risk). For each group, a target mean $m$ and standard deviation $s$ (Table~\ref{tab:rho_moments}) were converted to Beta shape parameters by moment-matching,
\begin{align}
    \ell = \frac{m(1-m)}{s^2} - 1, \qquad
    a = \ell\, m, \qquad
    b = \ell\, (1-m),
\end{align}
giving $\rho \sim \mathrm{Beta}(a, b)$ as reported in Table~\ref{tab:model_params}. Standard deviations were kept tight ($s \approx 0.025$--$0.035$, i.e.\ $a+b \approx 200$--$270$) rather than diffuse, so that $\rho$ remains anchored to the evidence above throughout inference rather than drifting under a weakly informative prior. A prior predictive check (Section~\ref{app:sensitivity}) confirms these priors yield case trajectories consistent with the observed HPSC series.

\begin{table}[H]
\centering
\caption{Target mean, standard deviation, and Beta shape parameters for $\rho_a$.}
\label{tab:rho_moments}
\begin{tabular}{lccc}
\hline
Age group & $m$ & $s$ & $(a,b)$ \\
\hline
$\rho_1$ (0--24)  & 0.18 & 0.025 & (42, 193) \\
$\rho_2$ (25--44) & 0.22 & 0.025 & (60, 213) \\
$\rho_3$ (45--64) & 0.32 & 0.030 & (77, 164) \\
$\rho_4$ (65+)    & 0.42 & 0.035 & (83, 115) \\
\hline
\end{tabular}
\end{table}

\subsection*{Particle initialisation}

Each parameter particle $m = 1,\ldots,N_\theta$ initialises its static parameters $\theta^{m}$, including $\rho_a^{m}$, by sampling from the priors in Table~\ref{tab:model_params}; these are subsequently updated within the SMC$^2$ algorithm as observations accumulate. The initial susceptibility $\beta_{a,1}^{(m, j)}$ is drawn from its $\mathrm{Beta}(1,10)$ prior independently for each age group. For each parameter particle $m$, the initial latent infection count of state particle $j = 1,\ldots,N_x$ is obtained by rescaling the mean reported cases over the first five days of the study period, $\bar Y_a$, by that parameter particle's own draw of the ascertainment ratio,
\begin{align}
    I_{a,1}^{(m,j)} \sim \mathrm{Poisson}\!\left(\max\!\left(\frac{\bar Y_{a}}{\rho_a^{m}},\,1\right)\right), \qquad a = 1,\ldots,4,
\end{align}
so that the latent state is initialised on the scale of true infections rather than reported cases, consistent with the observation model in Section~\ref{sec:obs_model}.

\section{Estimation of model parameters}\label{app:pmcmc}

To validate the posterior estimates obtained from the SMC$^2$ algorithm in Section~\ref{sec:smc2}, we compare them with those obtained using a Particle Marginal Metropolis--Hastings (PMMH) sampler \citep{andrieu_particle_2010}. Since the likelihood is not available in closed form due to the latent infection process, PMMH is implemented using the block-localised fixed-lag likelihood approximation,
\begin{align}
\tilde{P}(y_{1:T}|\theta)
=
\prod_{t=1}^{T}
\tilde{P}(y_t|y_{1:t-1},\theta)
=
\prod_{t=1}^{T}
\left(
\frac{1}{N_x}
\sum_{j=1}^{N_x}
\prod_{a=1}^{A}w_{a,t}^{(j)}
\right).
\end{align}
Algorithm~\ref{PMMH} embeds this estimator within the Metropolis--Hastings framework to generate samples from the posterior induced by the block-localised likelihood approximation,
\begin{align}
\tilde{P}(\theta|y_{1:T})
\propto
\tilde{P}(y_{1:T}|\theta)P(\theta).
\end{align}
These samples provide a reference for assessing the accuracy of the sequential parameter estimates obtained with SMC$^2$.
\begin{algorithm}[H]
\caption{Particle Marginal Metropolis-Hastings (PMMH)}\label{PMMH}
\textbf{Inputs:} Observations $y_{1:T}$, prior $P(\theta)$, proposal $q(\theta'|\theta)$, iterations $M$, particles $N_x$.\\
\textbf{Outputs:} Posterior samples $\{\theta^{(i)}\}_{i=1}^{M}$.

\hrulefill
\begin{algorithmic}[1]
\State \textbf{Initialisation:} Sample $\theta^{(1)} \sim P(\theta)$; run Algorithm~\ref{BPF} with $(\theta^{(1)}, y_{1:T})$ to obtain $\tilde{P}(y_{1:T}|\theta^{(1)})$.
\For{$i = 2, \dots, M$}
    \State \textbf{Propose:} $\theta^{*} \sim q(\cdot|\theta^{(i-1)})$.
    \State \textbf{Estimate likelihood:} Run Algorithm~\ref{BPF} with $(\theta^{*}, y_{1:T})$ to obtain $\tilde{P}(y_{1:T}|\theta^{*})$.
    \State \textbf{Compute acceptance ratio:}
        \[
        \alpha = \min\!\left(1,\;
        \frac{\tilde{P}(y_{1:T}|\theta^{*})\,P(\theta^{*})}{\tilde{P}(y_{1:T}|\theta^{(i-1)})\,P(\theta^{(i-1)})}
        \times
        \frac{q(\theta^{(i-1)}|\theta^{*})}{q(\theta^{*}|\theta^{(i-1)})}
        \right)
        \]

    \State \textbf{If} {$u \sim \mathcal{U}(0,1) < \alpha$} \textbf{then}
    \Statex \hspace{\algorithmicindent} \qquad $\theta^{(i)} \leftarrow \theta^{*}$ and  $\widehat{P}(y_{1:T}|\theta^{(i)}) \leftarrow \widehat{P}(y_{1:T}|\theta^{*})$
    \Statex \hspace{\algorithmicindent} \textbf{else}
        \Statex \hspace{\algorithmicindent} \qquad  $\theta^{(i)} \leftarrow \theta^{(i-1)}$ and  $\widehat{P}(y_{1:T}|\theta^{(i)}) \leftarrow \widehat{P}(y_{1:T}|\theta^{(i-1)})$
    \Statex \hspace{\algorithmicindent}\textbf{end if}
\EndFor
\end{algorithmic}
\end{algorithm}

The block particle filter in Algorithm~\ref{BPF} differs from the standard bootstrap particle filter \citep{gordon_novel_1993}, hereafter the standard particle filter, which applies a single resampling step to the full particle system $\{X_{t-L+1:t}^{(j)}\}_{j=1}^{N_x}$ at each time $t$ using the joint importance weights
\begin{align}
    w_t^{(j)} := w_t(X_{t-L:t-1}^{(j)}, \tilde{X}_{t-L+1:t}^{(j)})= \prod_{a=1}^{A} w_{a,t}^{(j)},
    \label{eq:joint_weight}
\end{align}
which combine the likelihood contributions from all $A$ age-group blocks. The standard particle filter gives the estimator
\begin{align}\label{eq:app_unb_mglik}
    \hat{P}(y_{1:t}|\theta)=\prod_{s=1}^{t}\left(\frac{1}{N_x}\sum_{j=1}^{N_x} w_s^{(j)}\right),
\end{align}
which is unbiased for the marginal likelihood \citep{andrieu_particle_2010}, unlike the block-localised estimator $\tilde{P}$ in Eq.~\ref{eq:app_bloc_mglik}. To motivate the block approach, we compare the two filters in terms of particle efficiency and likelihood estimation quality. A useful efficiency diagnostic is the effective sample size \citep[ESS,][]{doucet_tutorial_}, defined at each time $t$ as
\begin{equation}
    \mathrm{ESS}_t =\left(\sum_{j=1}^{N_{x}}w^{(j)}_{t}\right)^2 
\bigg/\sum_{k=1}^{N_x} \left(w_t^{(k)}\right)^2,
    \label{eq:ess}
\end{equation}
% where $\bar{w}_t^{(j)}=w_t^{(j)}/\sum_{k=1}^{N_x}w_t^{(k)}$ are the normalised importance weights.

\begin{figure}[H]
    \centering
    \includegraphics[width=0.69\linewidth]{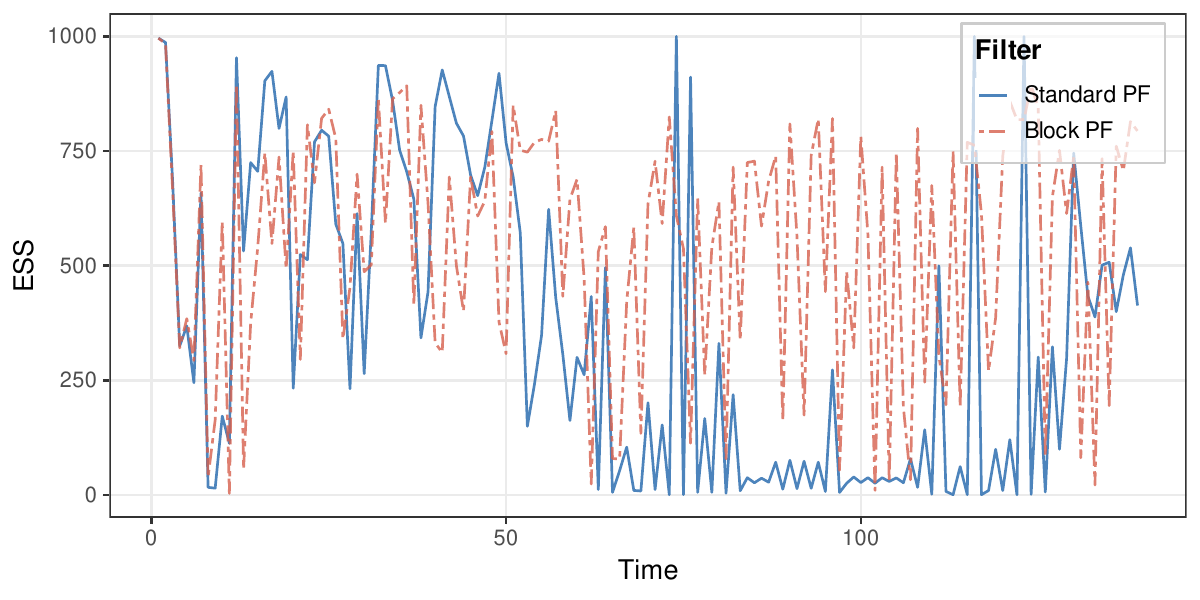}
    \caption{\footnotesize Effective sample size (ESS) over time for the standard particle filter (blue, solid) and the block particle filter (red, dashed). Both filters use $N_x = 1\,000$ particles, with model parameters set to the median values of their respective prior distributions.}
    \label{fig:ess_Nx}
\end{figure}

\begin{figure}[H]
    \centering
    \includegraphics[width=0.9\linewidth]{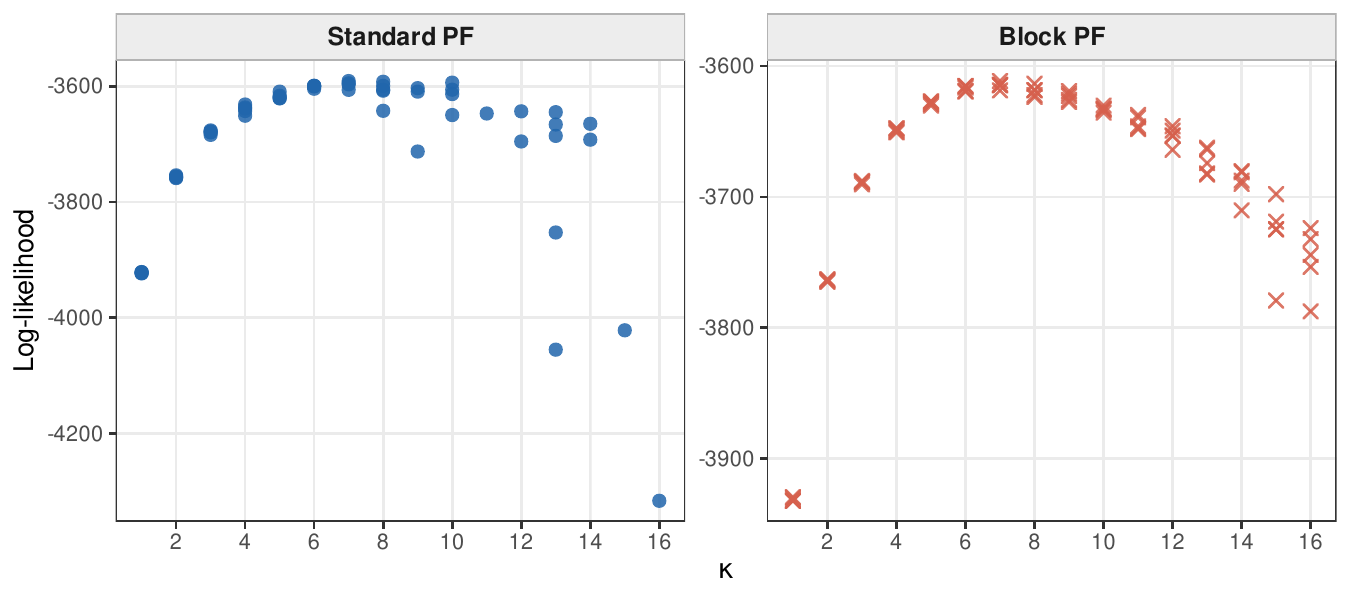}
\caption{\footnotesize Log-likelihood profiles as a function of the dispersion parameter $\kappa$. The standard particle filter (left panel, blue circles) uses $N_x = 20\,000$ particles, whereas the block particle filter (right panel, red crosses) uses $N_x = 1\,000$ particles. The remaining model parameters are fixed at the mean of their respective prior distributions. Five independent replicates are shown for each method.}
\label{fig:profile_loglik}
\end{figure}

Figure~\ref{fig:ess_Nx} compares the effective sample size for the standard and block particle filters. Both filters show similar ESS variability during the first half of the run, but from around $t\approx60$ onward, the standard particle filter suffers severe weight degeneracy, with $\mathrm{ESS}$ repeatedly collapsing to near zero and remaining there for extended periods before spiking back up after resampling. The block particle filter, by contrast, maintains a substantially higher ESS throughout the run and rarely collapses this severely, reflecting more consistent particle diversity. This arises because the joint importance weight~\eqref{eq:joint_weight} is a product over age groups, such that local weight mismatches accumulate multiplicatively and concentrate probability mass on a small number of particles. Figure~\ref{fig:profile_loglik} shows the profile log-likelihood over $\kappa$. The standard particle filter uses $N_x=20\,000$ particles, whereas the block particle filter uses only $N_x=1\,000$. Despite this twenty-fold reduction, the block particle filter produces a stable likelihood profile across replicates, indicating improved particle efficiency while maintaining the coupled transmission dynamics through joint propagation.

This behaviour is also reflected in Table~\ref{tab:ll_pf_comparison}. The standard particle filter exhibits substantial Monte Carlo variability at low particle numbers, with a standard deviation of $1.64\times10^{5}$ at $N_x=1\,000$. This decreases as $N_x$ increases but is still $4.31$ at $N_x=20\,000$, whereas the block particle filter has standard deviations below $1.1$ at all particle numbers considered. At $N_x=20\,000$, the block particle filter's mean log-likelihood is about 19 units below that of the standard particle filter. This is expected, since block-wise resampling and fixed-lag truncation yield an approximation to the likelihood rather than an unbiased estimator in Eq.~\eqref{eq:app_unb_mglik}, so inference targets the block-localised posterior $\tilde P(\theta|y_{1:t})$. The stable likelihood profile in Figure~\ref{fig:profile_loglik} and the recovery of known parameters in the simulation study (Section~\ref{app:supp_sim}) indicate that this approximation supports reliable inference in the proposed age-structured model.

\begin{table}[H]
\centering
\caption{Mean (standard deviation) of the log-likelihood over 10 runs for different particle numbers $N_x$, comparing the standard particle filter (PF) and the block particle filter (BPF). Results are reported for fixed parameters set to the mean values of their respective prior distributions.}
\label{tab:ll_pf_comparison}

\begin{tabular}{lcccc}
\toprule
 & $N_x=1\,000$ & $N_x=5\,000$ & $N_x=10\,000$ & $N_x=20\,000$ \\
\midrule
Standard PF 
& $-124324\ (164300)$ 
& $-24083\ (28920)$ 
& $-8521\ (6971)$ 
& $-3597\ (4.31)$ \\

Block PF 
& $-3613\ (0.27)$ 
& $-3616\ (0.15)$ 
& $-3616\ (1.02)$ 
& $-3616\ (0.81)$ \\
\bottomrule
\end{tabular}
\end{table}

\begin{figure}[H]
  \centering
  \includegraphics[width=\linewidth]{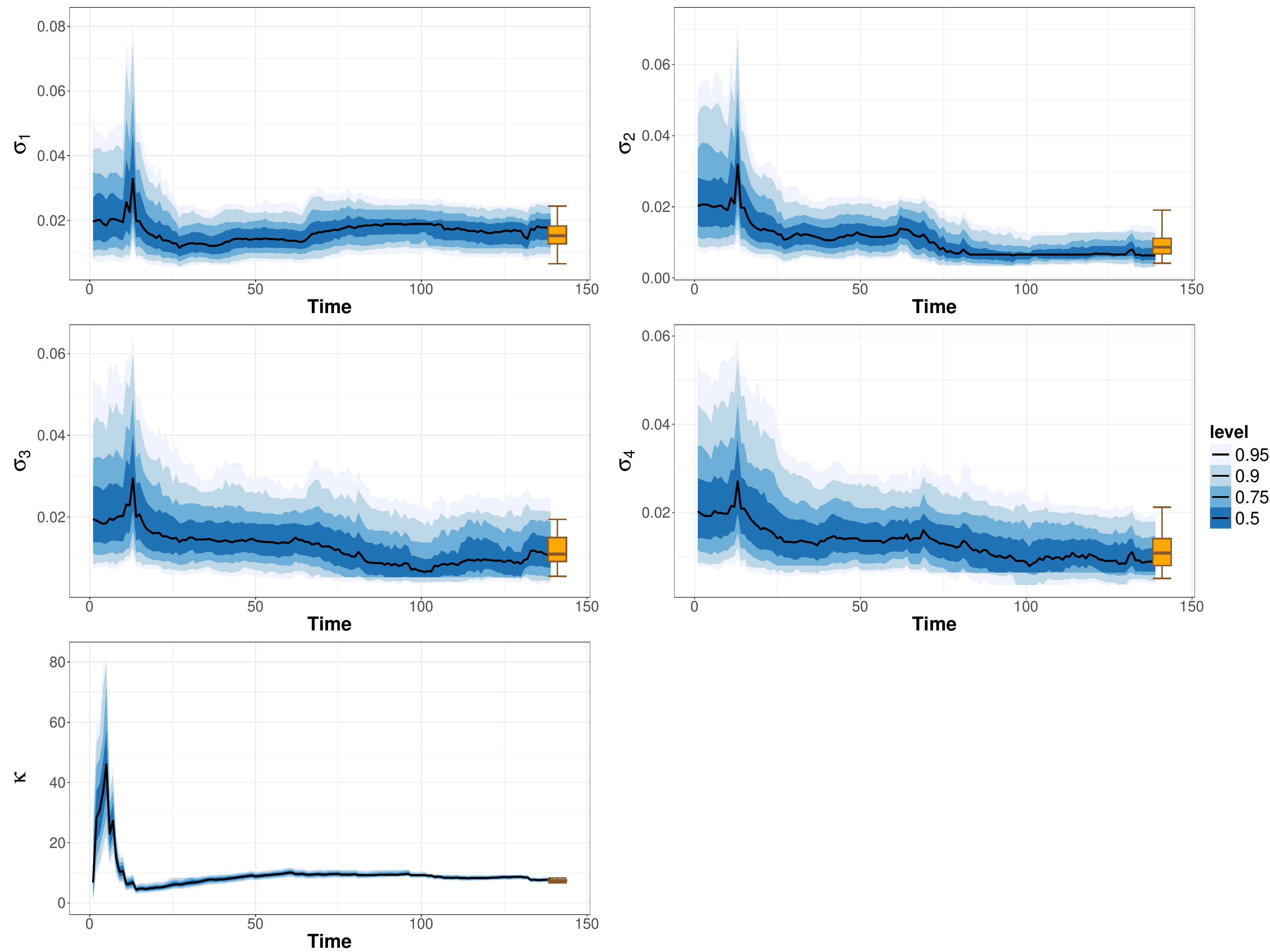}
  \caption{\footnotesize Marginal filtering estimates for the RW2 innovation standard deviations $\sigma_a$ ($a = 1, \dots, 4$) and the dispersion parameter $\kappa$.  Shaded bands correspond to pointwise credible intervals at the $50\%$, $75\%$, $90\%$, and $95\%$ levels; the boxplot at the final time point summarises the marginal posterior distribution obtained from the PMMH sampler.}
  \label{fig:supp_param_covid}
\end{figure}

\begin{figure}[H]
    \centering
    \includegraphics[width=\linewidth]{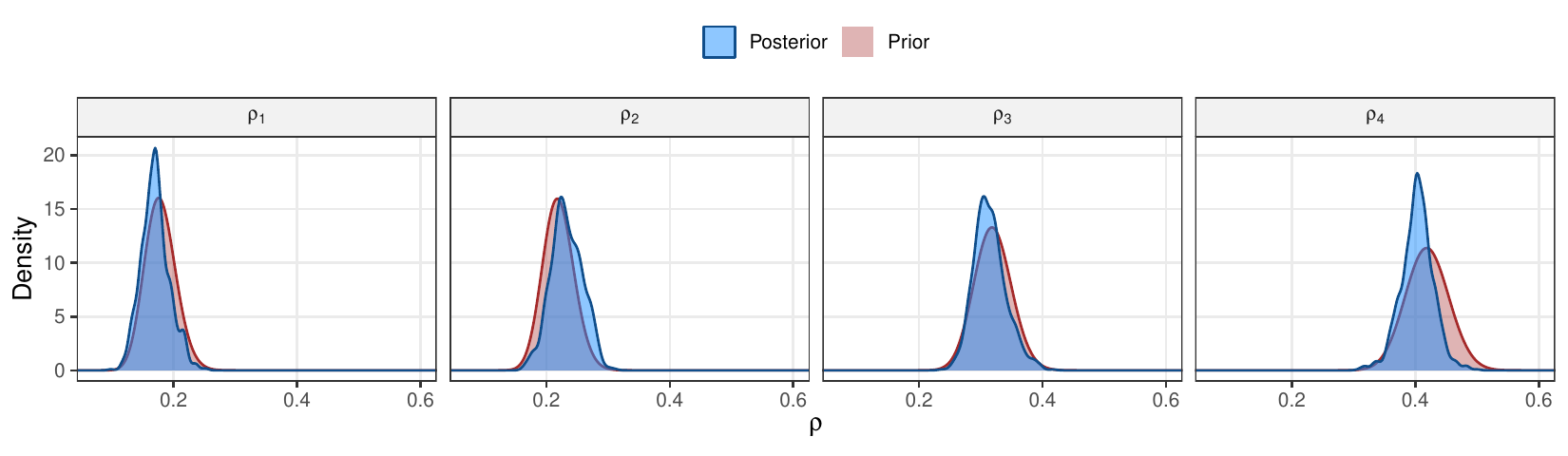}
    \caption{\footnotesize Prior vs. posterior plots for the ascertainment rate $\rho_a, ~~a=1,\cdots, 4.$}
    \label{fig:supp_rho}
\end{figure}

Figure~\ref{fig:supp_param_covid} shows the marginal filtering trajectories for the RW2 innovation standard deviations $\sigma_a$ ($a=1,\ldots,4$) and the dispersion parameter $\kappa$ obtained using SMC$^2$, together with their marginal posterior distributions at the final time point. The corresponding PMMH posterior distributions are in close agreement with those obtained using SMC$^2$ at the final time point, providing reassurance that the SMC$^2$ approximation is reliable and that the block particle filter supports valid posterior inference in this setting. The ascertainment rates $\rho_a$ exhibit limited posterior variation over the study period, and their prior and final-time posterior distributions are shown in Figure~\ref{fig:supp_rho}. This limited updating suggests that the case data alone provide relatively little information about these parameters over the study period.

\begin{figure}[H]
    \centering
    \includegraphics[width=\linewidth]{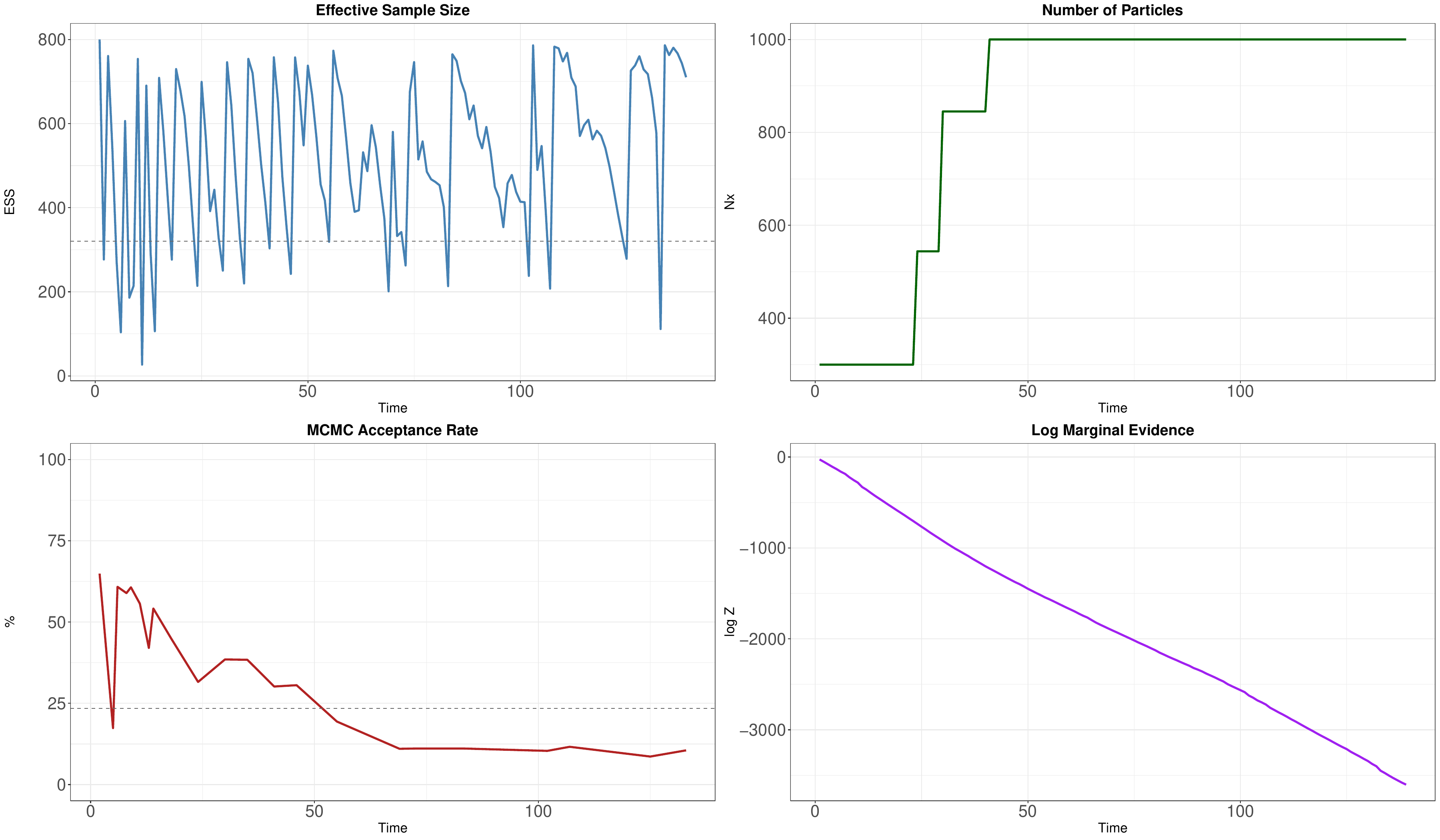}
    \caption{\footnotesize
    SMC$^2$ diagnostics for the block particle filter over the assimilation period. Top left: effective sample size $\mathrm{ESS}$ of the parameter particles, which triggers a resample-move step whenever it falls below the threshold ($0.4N_{\theta}$). Top right: number of particles $N_x$ used by the inner BPF at each time step. Bottom left: acceptance rate of the PMwG. Bottom right: running estimate of the log marginal model evidence.}
    \label{fig:diag_smc2}
\end{figure}

Figure~\ref{fig:diag_smc2} reports SMC$^2$ diagnostics. The effective sample size of the parameter particles repeatedly falls below the threshold $0.4N_\theta$, triggering resample-move steps. The PMwG acceptance rate starts at about 60\% and declines to around 10\% by $t\approx 70$, where it remains for the rest of the run. Acceptance rates of this order are not unusual for pseudo-marginal samplers, whose optimal acceptance rate is lower than the 23.4\% of exact random-walk Metropolis because of noise in the likelihood estimate, but the decline indicates that the moves become less efficient as data accumulate. The number of state particles $N_x$ increases from $300$ to its cap of $1\,000$ by $t\approx 40$ and is constant thereafter, so the variance of the log-likelihood estimator is no longer controlled at later time points, which may contribute to the decline in acceptance.

\section{Sensitivity analyses}
\label{app:sensitivity}

We assess the robustness of posterior inference to three sources of modelling uncertainty: the prior specification for the initial age-specific susceptibility, the assumed epidemiological delay distributions, and the structure of the contact matrix $\mathbf{M}$.

\subsection{Sensitivity to the prior on initial susceptibility}
\label{app:sensitivity_prior}

For the initial value of the age-specific susceptibility parameter $\beta_{a,1}$, the baseline analysis assumes a $\mathrm{Beta}(1,10)$ prior, which we compare with a $\mathrm{LogNormal}(-2.45,0.3)$ alternative. The latter was calibrated using the mean initial $R_0\approx 1.25$ estimate obtained from EpiNow2 divided by the largest eigenvalue of the contact matrix, thereby providing an epidemiologically informed initial value for the susceptibility parameter. The spread parameter was selected to represent moderate uncertainty around this estimate. 

\begin{figure}[H]
    \centering
    \includegraphics[width=\linewidth]{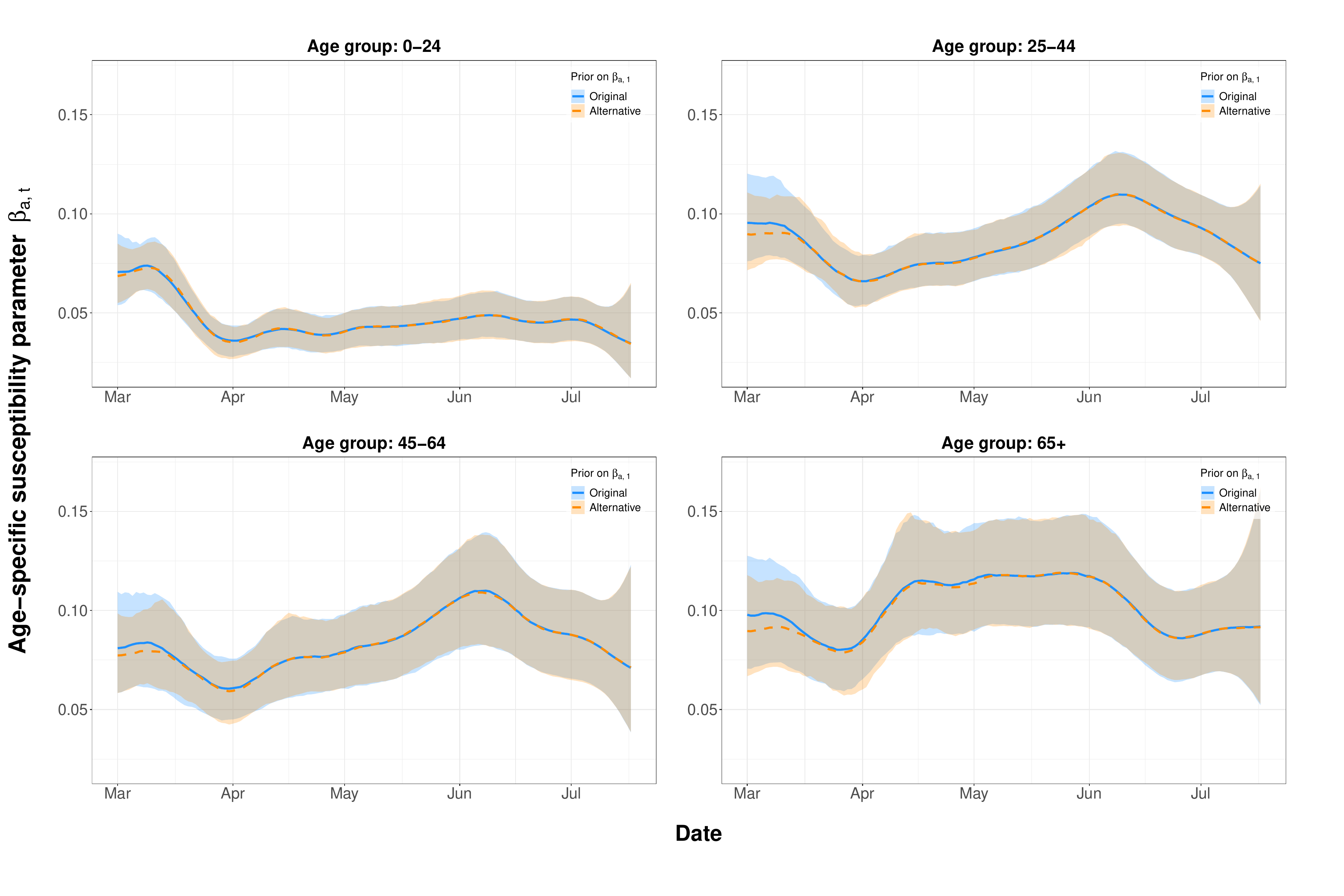}
\caption{\footnotesize
Sensitivity of posterior estimates of the age-specific susceptibility parameter $\beta_{a,t}$, $a=1,\ldots,4$, to the prior specification for its initial value $\beta_{a,1}$. Results are shown under $\mathrm{Beta}(1,10)$ (blue) and $\mathrm{LogNormal}(-2.45,0.3)$ (orange) priors. Shaded bands denote 95\% credible intervals.}
\label{fig:Rt_age_prior_sens}
\end{figure}

Figure~\ref{fig:Rt_age_prior_sens} compares the resulting posterior estimates of the age-specific susceptibility $\beta_{a,t}$, which scales the exposure of individuals in group $a$ to infection net of the contact structure captured by $\mathbf{M}$. The inferred trajectories are very similar under the two prior choices, with only slight differences at the start of the time series, suggesting limited sensitivity to the prior specification over most of the study period.

\subsection{Sensitivity to epidemiological delay distributions}
\label{app:sensitivity_delays}

\begin{figure}[H]
  \centering
  \includegraphics[width=0.75\linewidth]{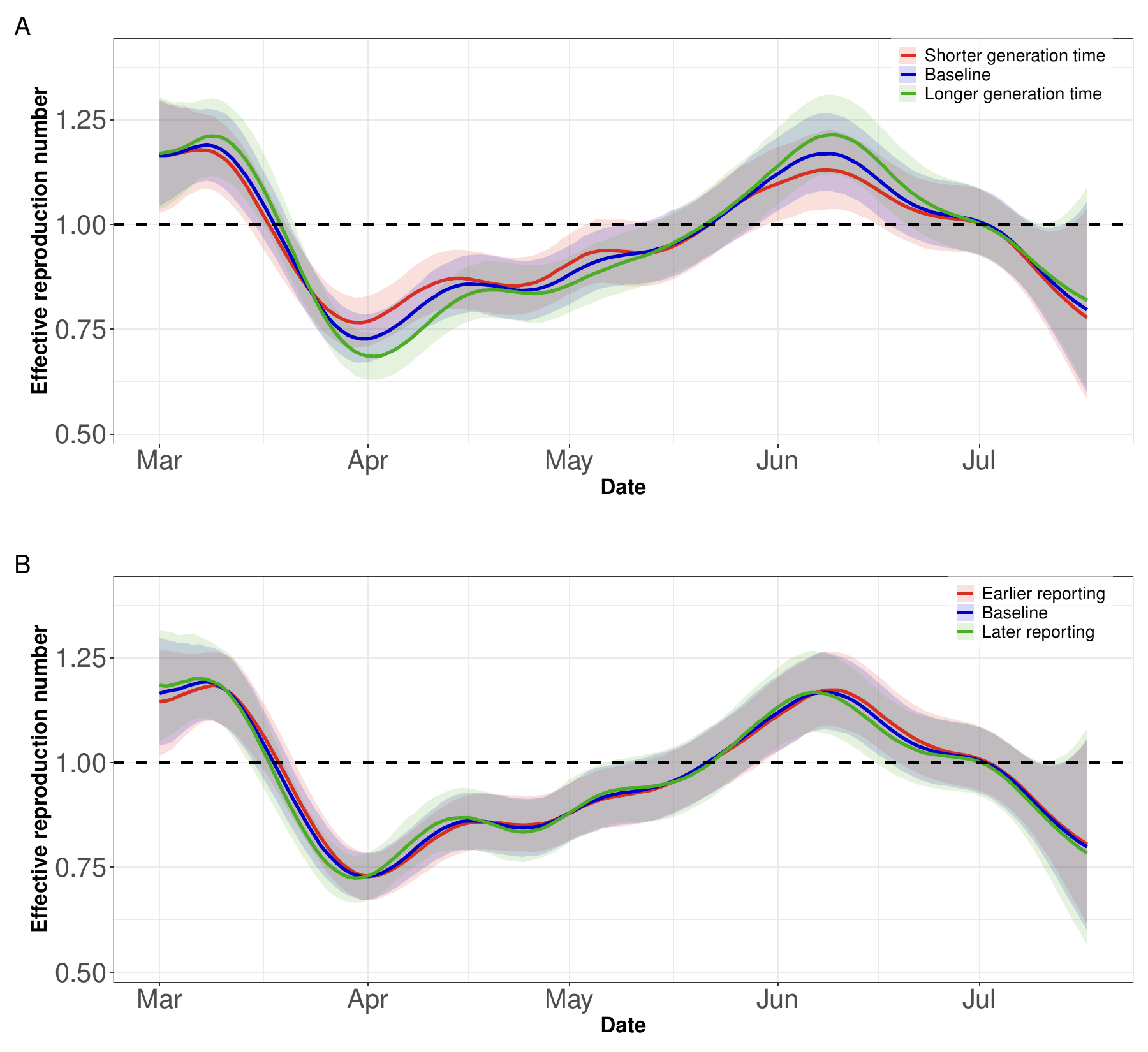}
\caption{\footnotesize
Sensitivity of the estimated effective reproduction number, $R_t$, to assumptions on epidemiological delay distributions. Panel A shows sensitivity to the generation time distribution under alternative specifications obtained by proportionally scaling the mean and standard deviation (shorter, baseline, and longer scenarios). Panel B shows the corresponding sensitivity to the infection-to-report delay distribution, where only the onset-to-report component is varied while the incubation period is held fixed. In both panels, solid lines represent posterior medians and shaded regions indicate 95\% credible intervals.}
  \label{fig:Rt_sens}
\end{figure}

For each alternative specification below, the full particle filtering and next-generation matrix procedure is rerun using the corresponding delay distributions. We first consider sensitivity to the generation time distribution, specified at baseline as a discretised Gamma distribution with mean 3.3 days and standard deviation 2.1 days (truncated at 14 days). Alternative specifications are obtained by scaling both the mean and standard deviation by factors of 0.85 (shorter generation time) and 1.2 (longer generation time), preserving the coefficient of variation. As shown in Figure~\ref{fig:Rt_sens}A, the generation time has a noticeable effect on the posterior $R_t$ estimates. Longer generation times amplify deviations of $R_t$ from one: they give lower values at the April trough and higher values at the early June peak, whereas the three specifications nearly coincide when $R_t$ is close to one (early March and July). The timing and shape of the epidemic trajectory are unchanged.

We then assess sensitivity to the infection-to-report delay. The onset-to-report delay is modelled as a discretised Gamma distribution truncated at 10 days, with baseline mean 2.0 days and standard deviation 1.0 days. We consider two alternative scenarios representing earlier reporting (mean 1.5, standard deviation 0.8) and later reporting (mean 2.8, standard deviation 1.4). The incubation period is fixed throughout as a discretised Gamma distribution with mean 3.0 days and standard deviation 2.3 days (truncated at 14 days). As in the main analysis, the infection-to-report delay distribution is obtained by convolving the incubation period with the onset-to-report distribution. Posterior $R_t$ estimates are shown in Figure~\ref{fig:Rt_sens}B and are nearly indistinguishable across the three scenarios, indicating robustness to these alternative delay assumptions. 

\subsection{Sensitivity to the contact matrix}
\label{app:sensitivity_contact}

\begin{figure}[H]
    \centering
    \includegraphics[width=0.9\linewidth]{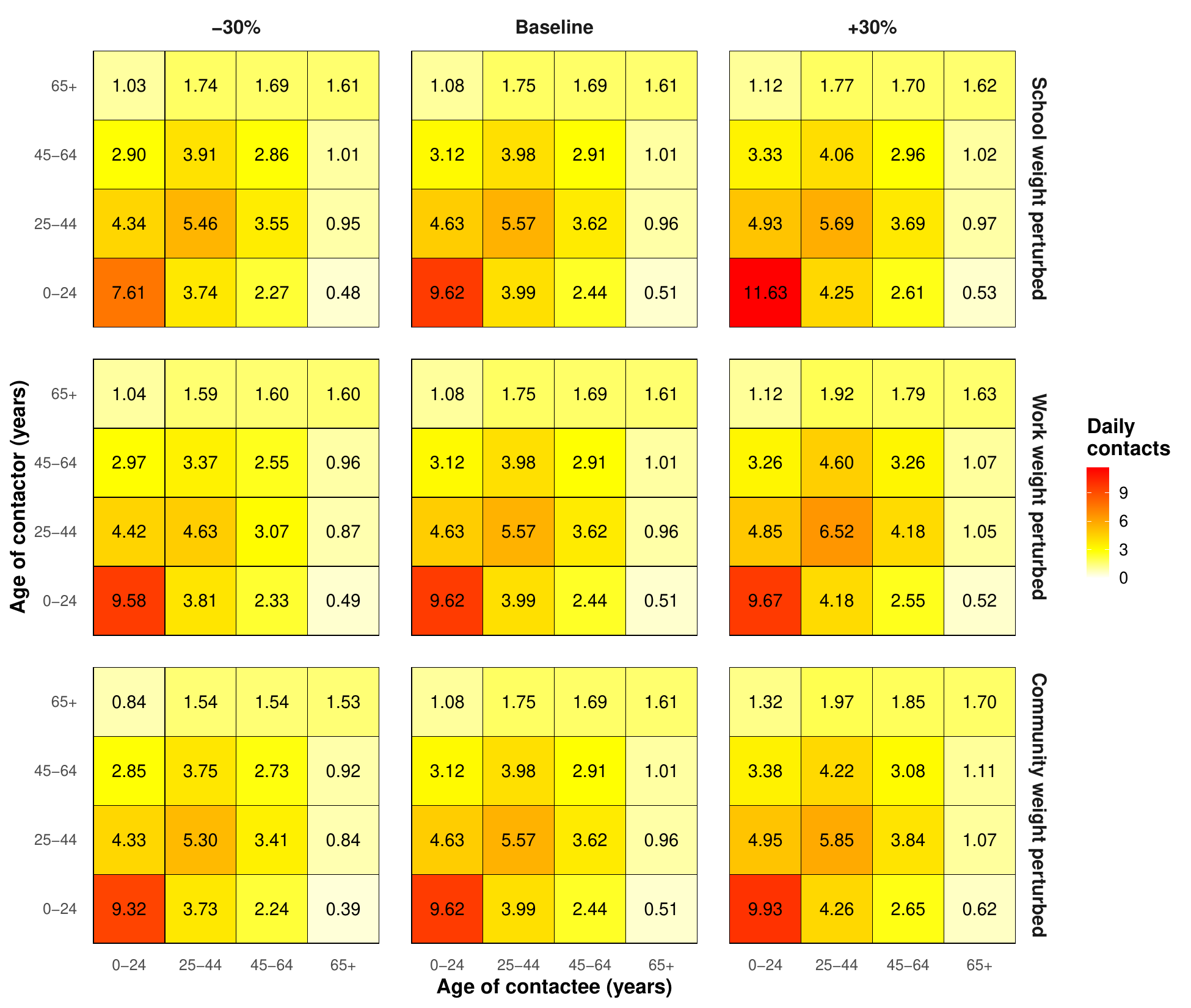}
    \caption{\footnotesize
    Contact matrices used in the sensitivity analysis. Rows correspond to the setting whose weight is perturbed, and columns to a $-30\%$, baseline, and $+30\%$ perturbation. Entries give the average daily number of contacts from the age group of the contactor (rows) to that of the contactee (columns), on a common colour scale.}
    \label{fig:M_sens}
\end{figure}

Finally, we assess the sensitivity of the inferred transmission patterns to the contact structure $\mathbf{M}$. Recall that $\mathbf{M}$ is constructed as a weighted sum of four setting-specific contact matrices,
\begin{equation}
\mathbf{M} = w_H \mathbf{M}^{(H)} + w_S \mathbf{M}^{(S)} + w_W \mathbf{M}^{(W)} + w_C \mathbf{M}^{(C)},
\end{equation}
with baseline weights corresponding to household ($w_H=4.11$), school ($w_S=11.41$), work ($w_W=8.07$), and community ($w_C=2.79$) contacts, taken from the contact matrices in \citet{mistry_inferring_2021} for Ireland. Because the renewal process depends on $\beta_{a,t}$ and $\mathbf{M}$ only through their product, a uniform rescaling of all weights is absorbed entirely by $\beta_{a,t}$ and is uninformative about the age structure of mixing. We therefore perturb the school, work and community weights one at a time,
\begin{equation}
w_k^{(\delta)} = w_k \cdot (1 + \delta), \qquad k \in \{S, W, C\},
\end{equation}
with $\delta \in \{-\epsilon,+\epsilon\}$ and $\epsilon=0.30$, while the other three weights, including the household weight, remain at their baseline values. Household contacts are not perturbed, as they are relatively well characterised by sociodemographic data, whereas contacts in schools, workplaces and community settings are subject to considerable variation \citep{vizi_age_2026}. These perturbations change the relative age structure of $\mathbf{M}$ (Figure~\ref{fig:M_sens}). For each perturbed matrix $\mathbf{M}^{(k,\delta)}$, we re-run the particle filter to re-estimate the latent states, including $\beta_{a,t}$, under the perturbed contact structure, with the static parameters fixed at their baseline posterior. From the resulting particle trajectories, we recompute at each time $t$ the next-generation matrix
\begin{equation}
\mathbf{K}^{(k,\delta)}(t) = \mathrm{diag}(\boldsymbol\beta_t) \left(\mathbf{M}^{(k,\delta)}\right)^{\top},
\end{equation}
the effective reproduction number $R_t = \rho\big(\mathbf{K}^{(k,\delta)}(t)\big)$, and the age-specific reproduction contributions $R_{a,t} = \sum_b K^{(k,\delta)}_{ba}(t)$.

\begin{figure}[H]
    \centering
    \includegraphics[width=\linewidth]{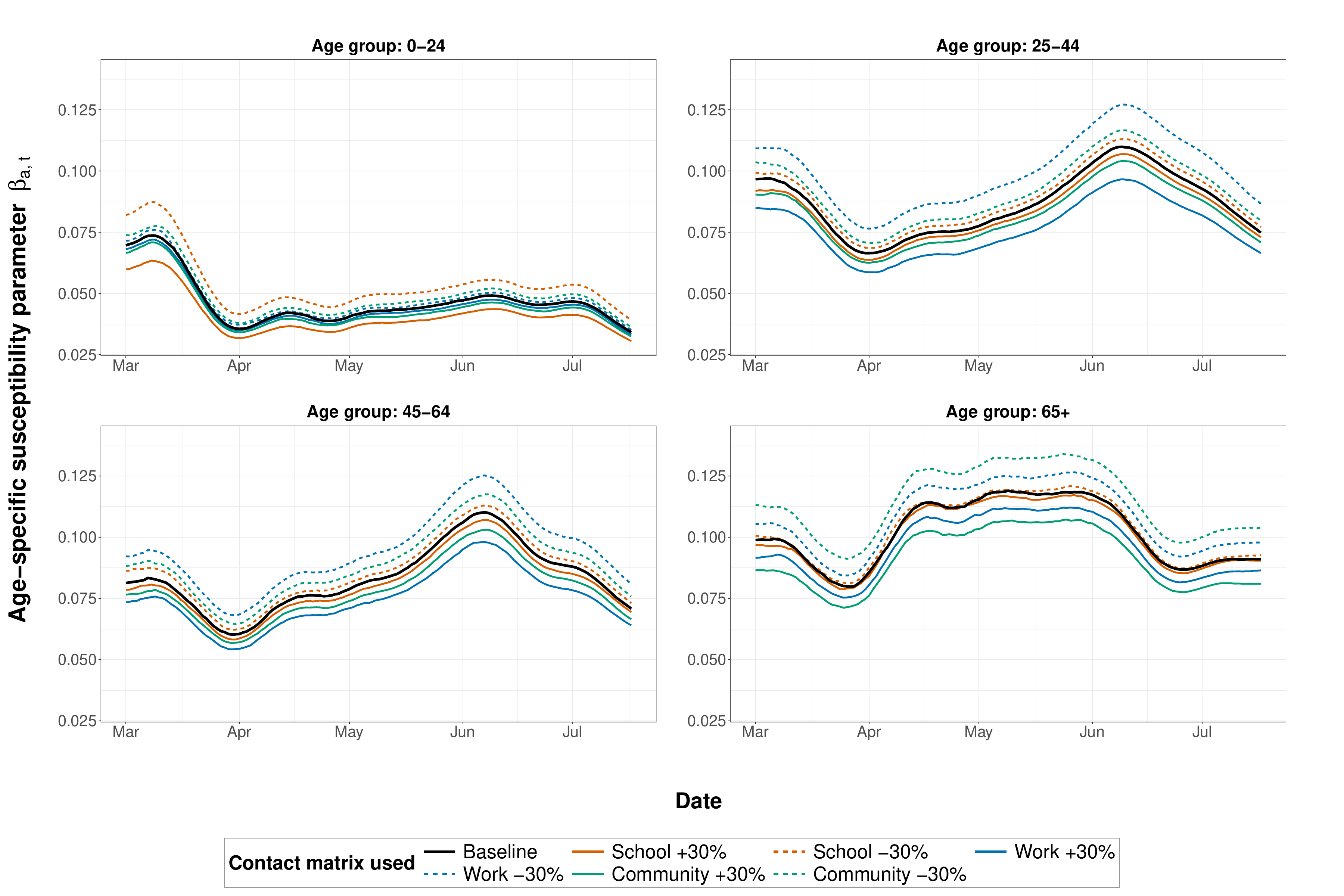}
    \caption{\footnotesize
    Sensitivity of the posterior median of the age-specific susceptibility parameter $\beta_{a,t}$, $a=1,\ldots,4$, to one-at-a-time perturbations of the contact matrix setting weights. The black line shows the baseline; colours indicate the perturbed setting (school, work, community), with solid lines for $+30\%$ and dashed lines for $-30\%$.}
    \label{fig:beta_contact_sens}
\end{figure}

Figure~\ref{fig:beta_contact_sens} shows the resulting posterior median estimates of $\beta_{a,t}$. Each perturbation shifts $\beta_{a,t}$ in the direction opposite to the change in contacts, and mainly in the age groups affected: school for the 0--24 group, and work and community for the three older groups (community most strongly for the 65+ group). The shifts reach about 20\% at most. This reflects the structural non-identifiability of $\beta_{a,t}$ and $\mathbf{M}$ in isolation, as $\beta_{a,t}$ absorbs changes in the contacts of its age group. The age ordering of $\beta_{a,t}$ is preserved in all scenarios, with the highest values in the 65+ group and the lowest in the 0--24 group.

\begin{figure}[H]
    \centering
    \includegraphics[width=\linewidth]{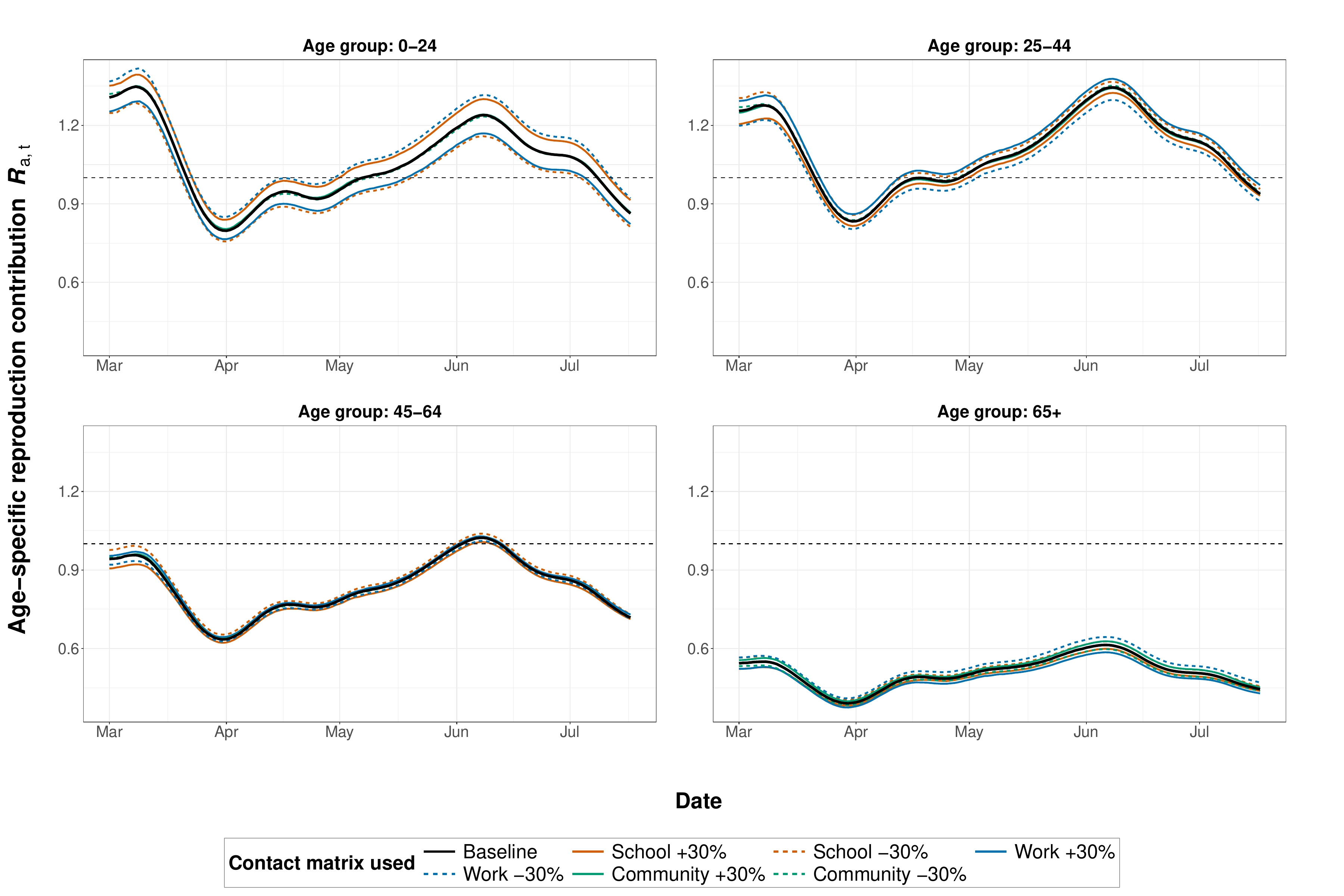}
    \caption{\footnotesize
    Sensitivity of the posterior median of the age-specific reproduction contribution $R_{a,t}$ to one-at-a-time perturbations of the contact matrix setting weights. Each panel corresponds to one age group. Line colours and types are as in Figure~\ref{fig:beta_contact_sens}; the dashed horizontal line marks $R_{a,t}=1$.}
    \label{fig:Rat_contact_sens}
\end{figure}

\begin{figure}[H]
    \centering
    \includegraphics[width=0.75\linewidth]{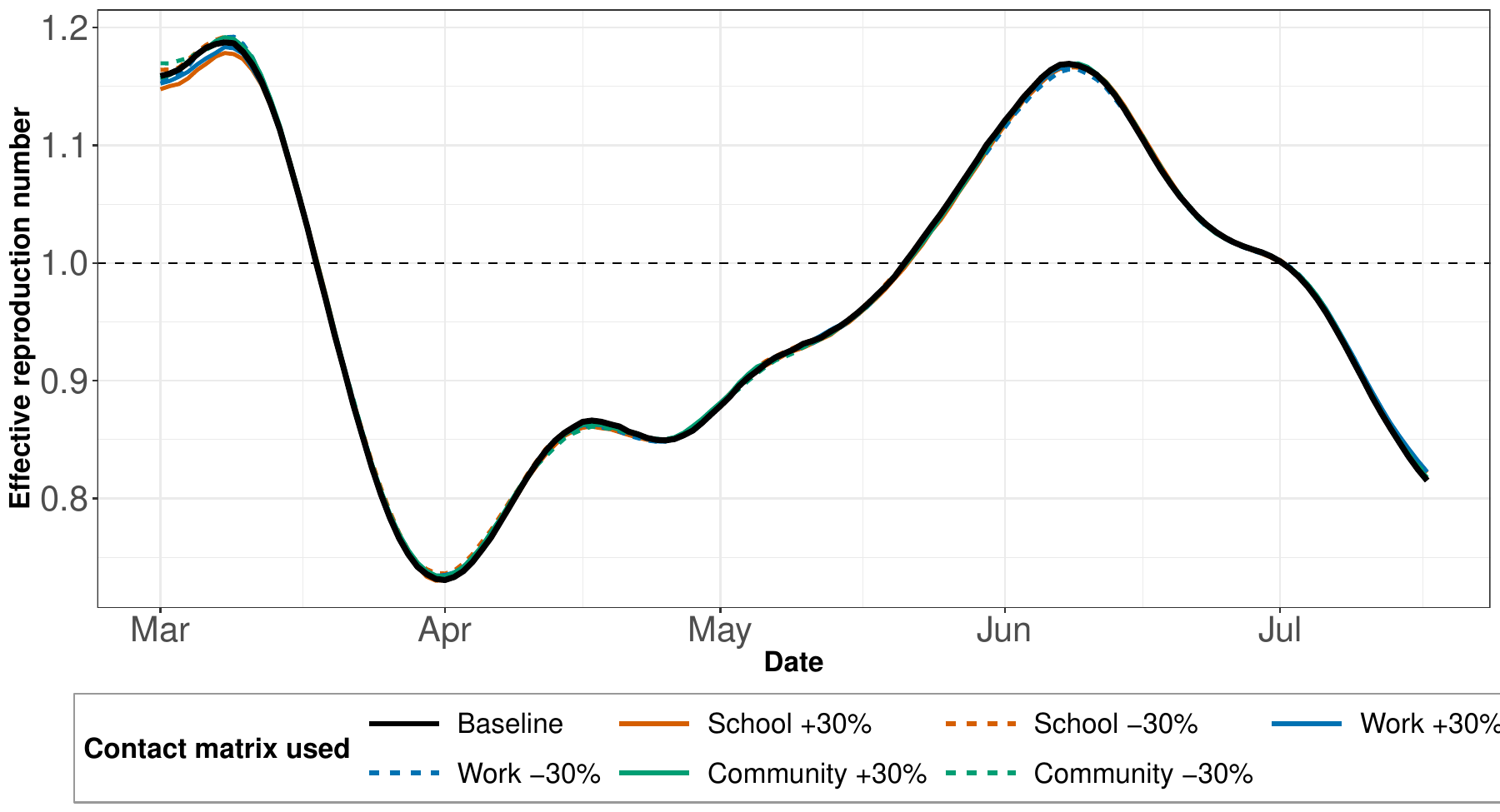}
\caption{\footnotesize
Sensitivity of the posterior median of the effective reproduction number $R_t$ to one-at-a-time perturbations of the contact matrix setting weights. Line colours and types are as in Figure~\ref{fig:beta_contact_sens}; the dashed horizontal line marks $R_t=1$.}
\label{fig:Rt_contact_sens}
\end{figure}

Because this compensation operates on the product entering the next-generation matrix, $R_{a,t}$ and $R_t$ are much less affected. Figure~\ref{fig:Rat_contact_sens} shows that $R_{a,t}$ keeps the same trajectory in all scenarios, with peaks in early March and early June and a trough in early April. The largest changes occur in the 0--24 group, where $R_{a,t}$ varies by up to about 6\% (upwards under school $+30\%$ and work $-30\%$, downwards under school $-30\%$ and work $+30\%$), shifting the dates on which it crosses one by up to about 10 days. The 65+ group is also comparatively sensitive, varying by up to about 9\% under the work $\pm30\%$ scenarios, whereas the 25--44 group changes by at most about 2--4\% and the 45--64 group is almost unchanged. The 0--24 and 25--44 groups remain the largest contributors to transmission, and the 65+ group remains the furthest from one throughout, peaking at about 0.65 under the work $-30\%$ scenario. Figure~\ref{fig:Rt_contact_sens} shows that the median $R_t$ is nearly indistinguishable across scenarios, with only small differences in the first days of March. These perturbations change the relative weight of settings while keeping their internal age patterns and the time-constant structure of $\mathbf{M}$ fixed, so they assess sensitivity to the composition of contacts rather than to arbitrary misspecification.

\newpage
\section{Additional results for the homogeneous model}\label{app:simple_model}

The homogeneous model is a simplified version of the heterogeneous formulation presented in Section~\ref{sec:model}, in which age-group stratification is removed and the population is treated as a single compartment. The time-varying effective reproduction number $R_t$ is assigned a second-order random walk prior on the log scale:
\begin{subequations}\label{eq:rw2_Rt}
\begin{align}
R_1 &\sim \mathcal{N}(1.5,\, 0.2^2), \\
R_2 | R_1 &\sim \exp\!\bigl(\log(R_1) + \varepsilon_2\bigr), \\
R_t | R_{t-1}, R_{t-2}
  &\sim \exp\!\bigl(2\log(R_{t-1}) - \log(R_{t-2}) + \varepsilon_t\bigr),
  \quad t \geq 3,
\end{align}
\end{subequations}
where $\varepsilon_t \sim \mathcal{N}(0,\sigma^2)$ are independent innovations and $\sigma>0$ controls the degree of temporal smoothing. The latent infections evolve according to the renewal process
\begin{equation}\label{eq:renewal_homogeneous}
I_t | R_t,\, I_{1:t-1}
  \sim \mathrm{Poisson}\!\left(
        R_t \sum_{\tau=1}^{m_g} I_{t-\tau}\, g_\tau
      \right).
\end{equation}
The observation model is identical to that of the heterogeneous specification, retaining both the negative-binomial likelihood and the day-of-the-week reporting effect. The seven reporting factors $\omega_d$ are constructed from six unconstrained latent variables $\tilde{\omega}_d \sim \mathrm{LogNormal}(0,0.2)$ and mapped to a simplex-constrained scale via
\begin{equation}\label{eq:dow_transform}
\omega_d =
\frac{7,\tilde{\omega}_d}{\sum_{d'=1}^{6}\tilde{\omega}_{d'}+1}, \quad d=1,\ldots,6,
\qquad
\omega_7 =
\frac{7}{\sum_{d'=1}^{6}\tilde{\omega}_{d'}+1},
\end{equation}
ensuring $\omega_d>0$ and $\sum_{d=1}^7 \omega_d = 7$. This normalisation centres the factors around one, enabling interpretation as multiplicative deviations from the weekly mean while maintaining identifiability.

\begin{figure}[H]
  \centering
  \includegraphics[width=\linewidth]{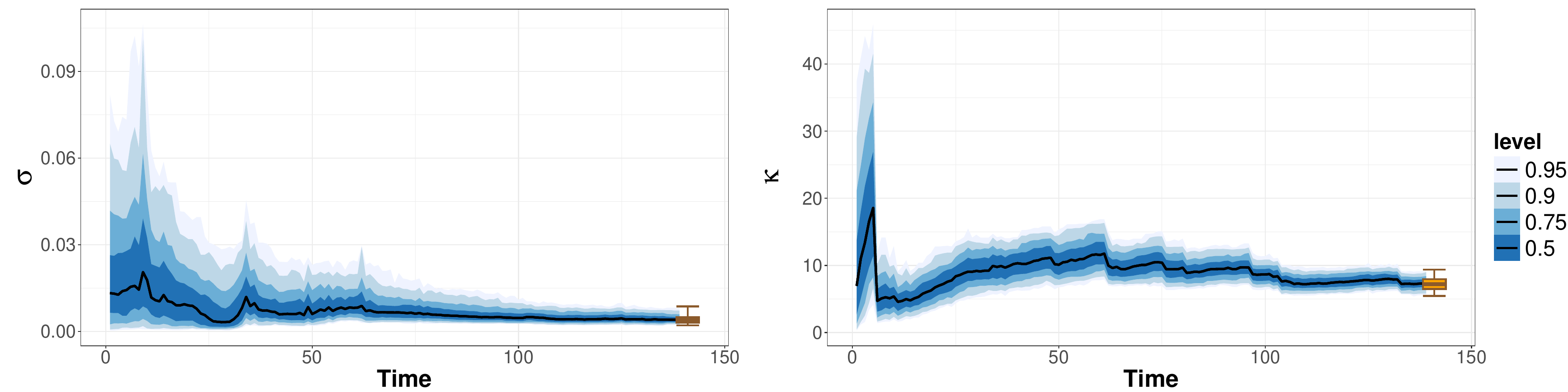}
\caption{\footnotesize
Marginal filtering estimates of the RW2 innovation standard deviations $\sigma$ and the dispersion parameter $\kappa$. The boxplot at the final time point summarises the marginal posterior distribution obtained from the PMMH.}
  \label{fig:params_simple}
\end{figure}

\begin{figure}[H]
  \centering
  \includegraphics[width=\linewidth]{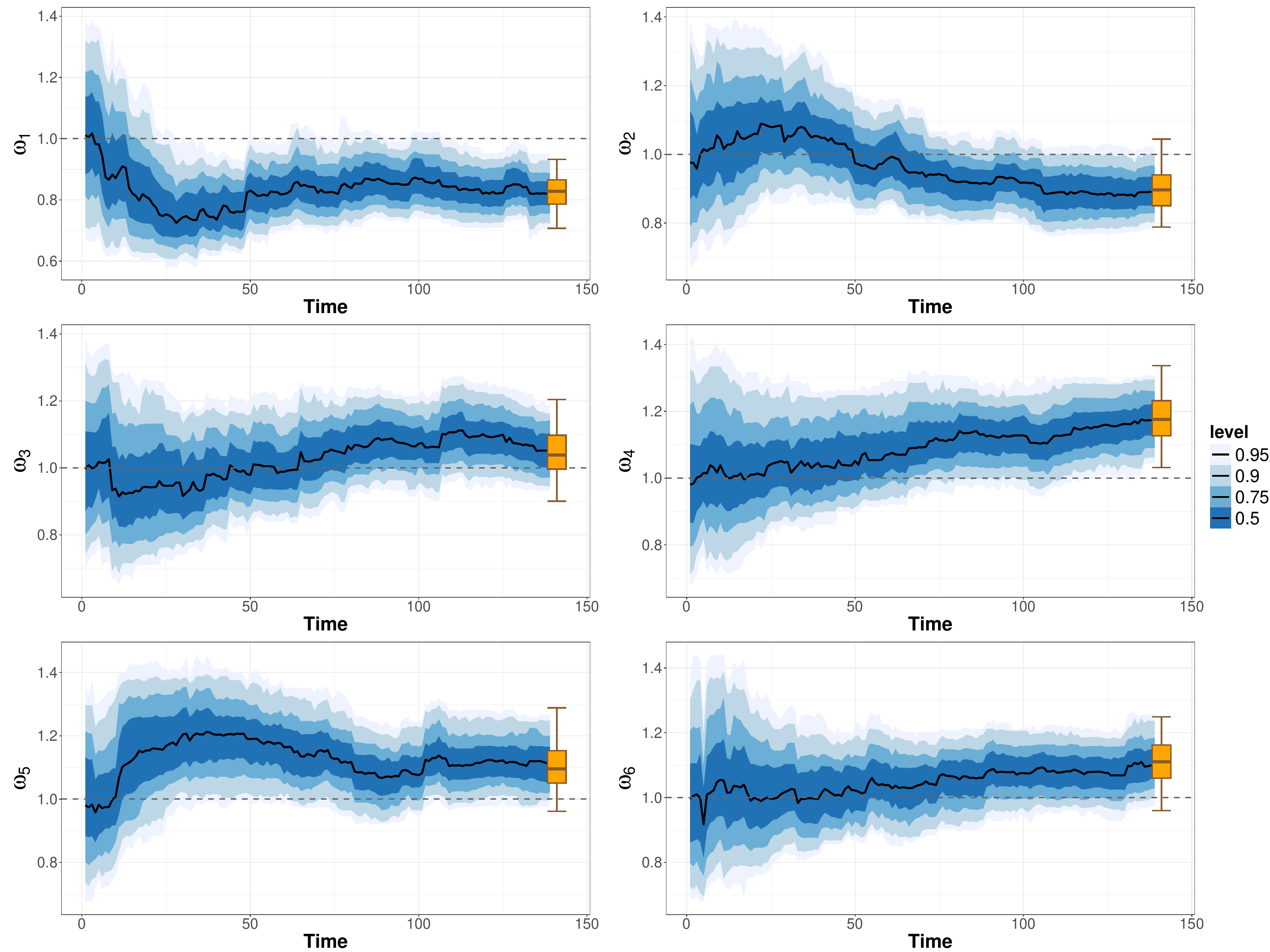}
\caption{\footnotesize
Marginal filtering estimates of the day-of-the-week effect with $\omega_7=7-\sum_{d=1}^{6}\omega_d$. The boxplot at the final time point summarises the marginal posterior distribution obtained from the PMMH.}
  \label{fig:dow_simple}
\end{figure}

\begin{figure}[H]
    \centering
    \includegraphics[width=0.7\linewidth]{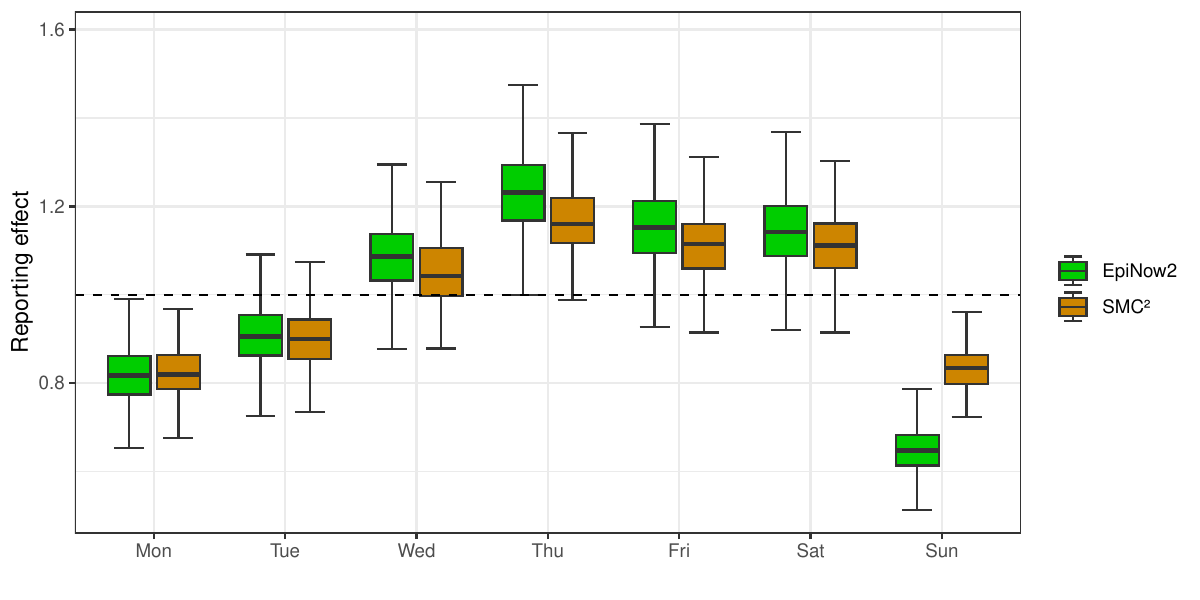}
\caption{\footnotesize Posterior estimates of the day-of-the-week reporting effects from the homogeneous model, inferred using SMC$^2$ at the final time point $T$ (orange) and using EpiNow2 (green). Factors below unity indicate days with systematically lower reported counts relative to the weekly mean, whereas factors above unity indicate higher reported counts.}
    \label{fig:week_effect_simple}
\end{figure}

Figures~\ref{fig:params_simple} and~\ref{fig:dow_simple} show the marginal filtering distributions of $\sigma$, $\kappa$, and the day-of-the-week factors $\omega_d$ obtained using SMC$^2$, while Figure~\ref{fig:week_effect_simple} compares the resulting day-of-the-week factors at the final observation time with the corresponding EpiNow2 estimates. The filtering distributions of $\sigma$ and the dispersion parameter $\kappa$ closely resemble those obtained under the heterogeneous model, with posterior uncertainty decreasing as additional observations become available. The estimated day-of-the-week effects recover a clear weekly reporting pattern, with lower reported incidence on Sundays and Mondays and higher counts from Wednesday to Saturday, peaking on Thursday, consistent with well-known reporting artefacts such as reduced testing and processing at the start of the week. The two methods agree closely except on Sunday, where SMC$^2$ yields a modestly higher reporting effect than EpiNow2. Across all parameters, the SMC$^2$ and PMMH estimates are in close agreement, providing further evidence that the SMC$^2$ approximation is accurate for this model.

\newpage
\section{Simulation experiment}
\label{app:supp_sim}

We evaluate the proposed age-structured renewal model and sequential Bayesian inference framework using a controlled simulation study based on synthetic epidemic data. The simulation spans $T=100$ days and considers four age groups (0--24, 25--44, 45--64, and 65+), with the same contact structure as in Section~\ref{subsec:age}. Smooth age-specific susceptibility trajectories $\beta_{a,t}$ are generated as deterministic functions combining periodic variation with gradual exponential decay, inducing heterogeneous transmission dynamics across age groups. Conditional on these trajectories, infections are simulated using the renewal model in Section~\ref{sec:renewal_model}, and reported cases are generated via a Negative Binomial observation model with dispersion parameter $\kappa=15$. The infection-to-report delay distribution follows Table~\ref{tab:model_params}, and day-of-week effects are omitted for simplicity.

\begin{figure}[H]
  \centering
  \includegraphics[width=0.86\linewidth]{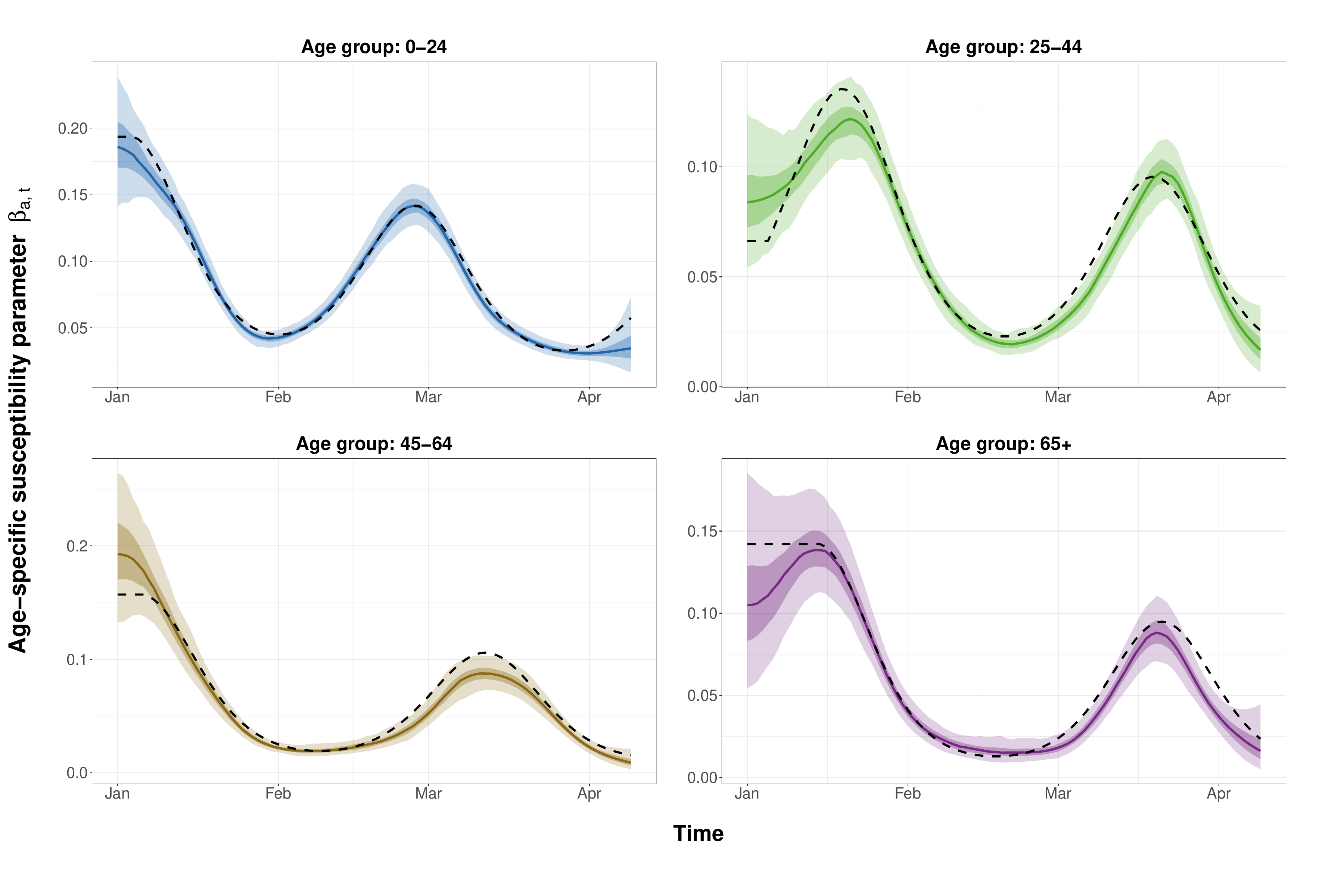}
\caption{\footnotesize
Filtered estimates of the age-specific susceptibility trajectories $\beta_{a,t}$ $a=1,\ldots,4$, from the simulation study. Solid lines denote posterior medians and shaded regions indicate 50\% and 95\% posterior credible intervals. Dashed black lines show the ground truth used to generate the data.
}
  \label{fig:supp_beta}
\end{figure}

\begin{figure}[H]
  \centering
  \includegraphics[width=0.86\linewidth]{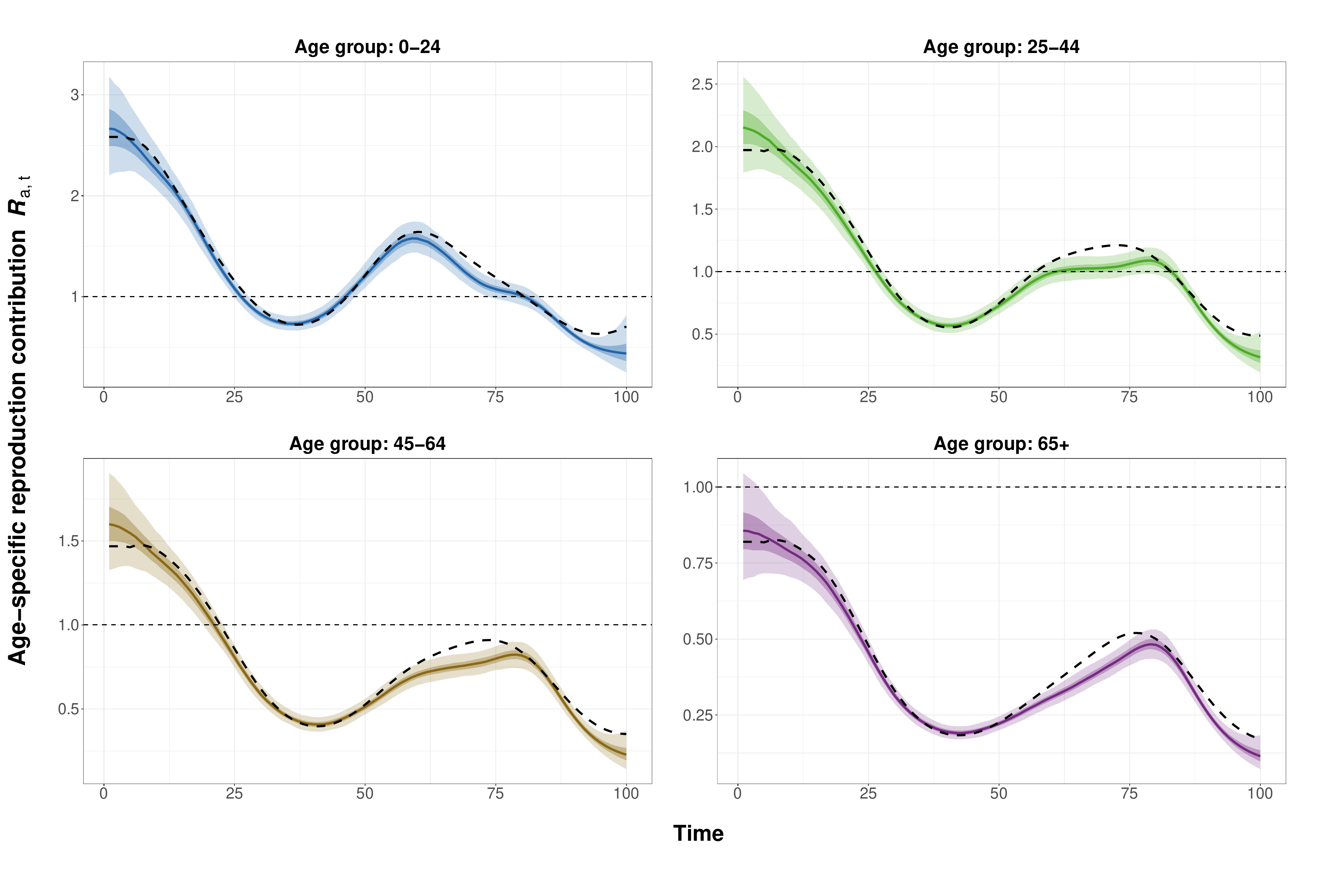}
\caption{\footnotesize
Marginal filtered estimates of the age-specific reproduction number contribution $R_{a,t}$, $a=1,\ldots,4$, from the simulation study. Solid lines denote posterior medians and shaded regions indicate 50\% and 95\% posterior credible intervals. Dashed black lines show the true simulated trajectories.
}
  \label{fig:supp_Rat}
\end{figure}

\begin{figure}[H]
  \centering
  \includegraphics[width=0.86\linewidth]{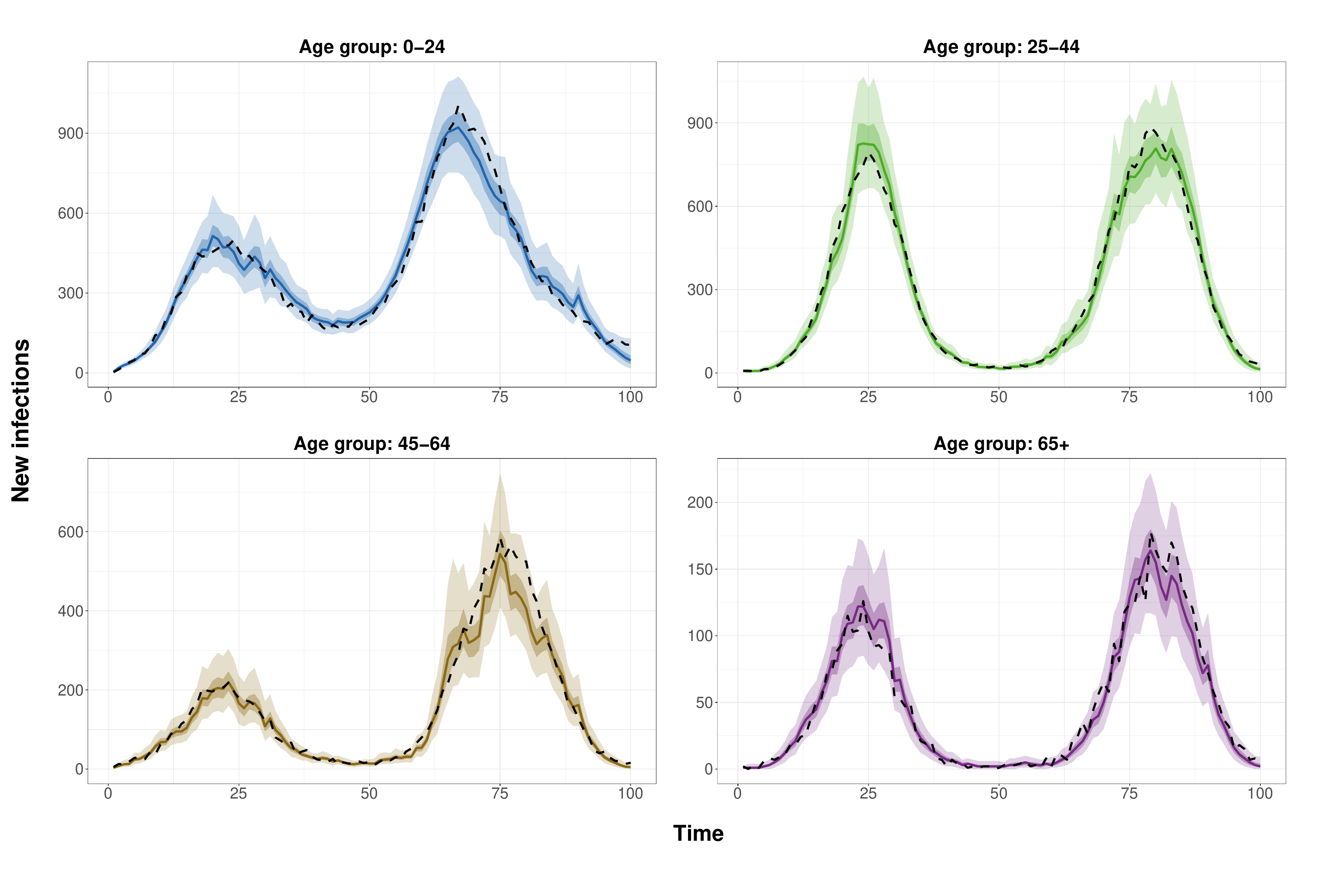}
  \caption{\footnotesize Marginal filtered estimates of latent daily infections $I_{a,t}$ for $a = 1, \dots, 4$. Solid lines denote posterior medians and shaded regions indicate 50\% and 95\% posterior credible intervals. Dashed black lines indicate the true infection trajectories.}
  \label{fig:supp_I}
\end{figure}

Figures~\ref{fig:supp_beta}--\ref{fig:supp_I} report filtered estimates of the age-specific susceptibility $\beta_{a,t}$, reproduction contribution $R_{a,t}$, and latent infections $I_{a,t}$ obtained via the SMC$^2$ algorithm, integrating over parameter uncertainty. In all age groups, posterior medians closely track the true latent trajectories, with the true values consistently lying within the 95\% credible intervals. These results indicate accurate recovery of both the latent infection process and the source-specific transmission contributions implied by the model.

\begin{figure}[H]
  \centering
  \includegraphics[width=0.95\linewidth]{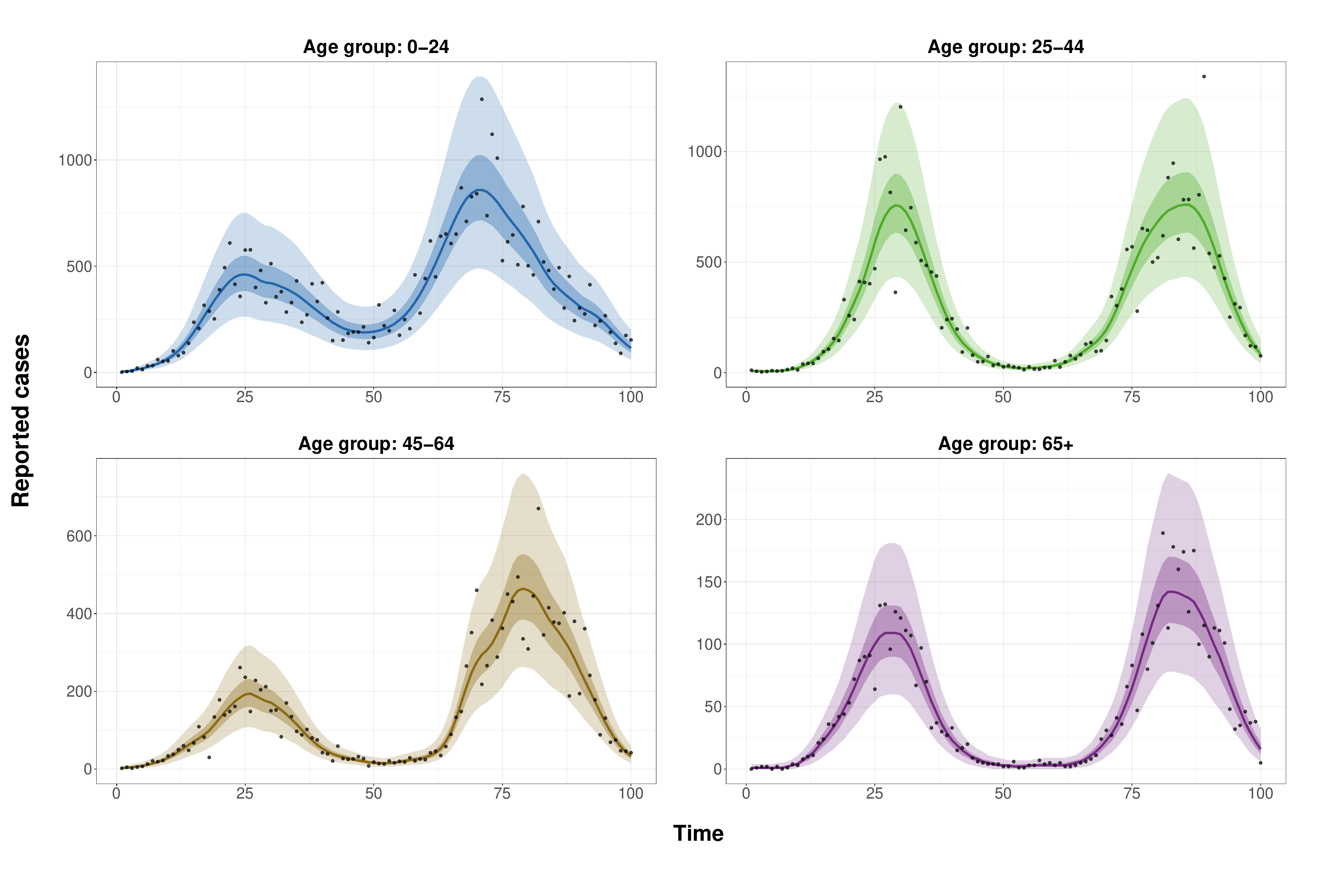}
  \caption{\footnotesize Posterior predictive distributions for daily reported cases $C_{a,t}$, $a = 1, \dots, 4$. Solid lines denote posterior medians and shaded regions indicate 50\% and 95\% posterior credible intervals. Black points correspond to observed synthetic case counts.}
  \label{fig:supp_C}
\end{figure}

To assess the adequacy of the observation model, we use posterior samples of the time-varying parameters together with filtered latent states to generate posterior predictive distributions of daily reported cases through the full observation process. This procedure propagates uncertainty in both latent states and parameters at each time step. Figure~\ref{fig:supp_C} shows that the observed synthetic case counts are consistently well captured by the predictive distributions across all age groups, with the realised trajectories lying well within the 95\% predictive intervals. This indicates that the model adequately represents both the age-specific reporting structure and the assumed level of overdispersion.

\begin{figure}[H]
    \centering
    \includegraphics[width=0.9\linewidth]{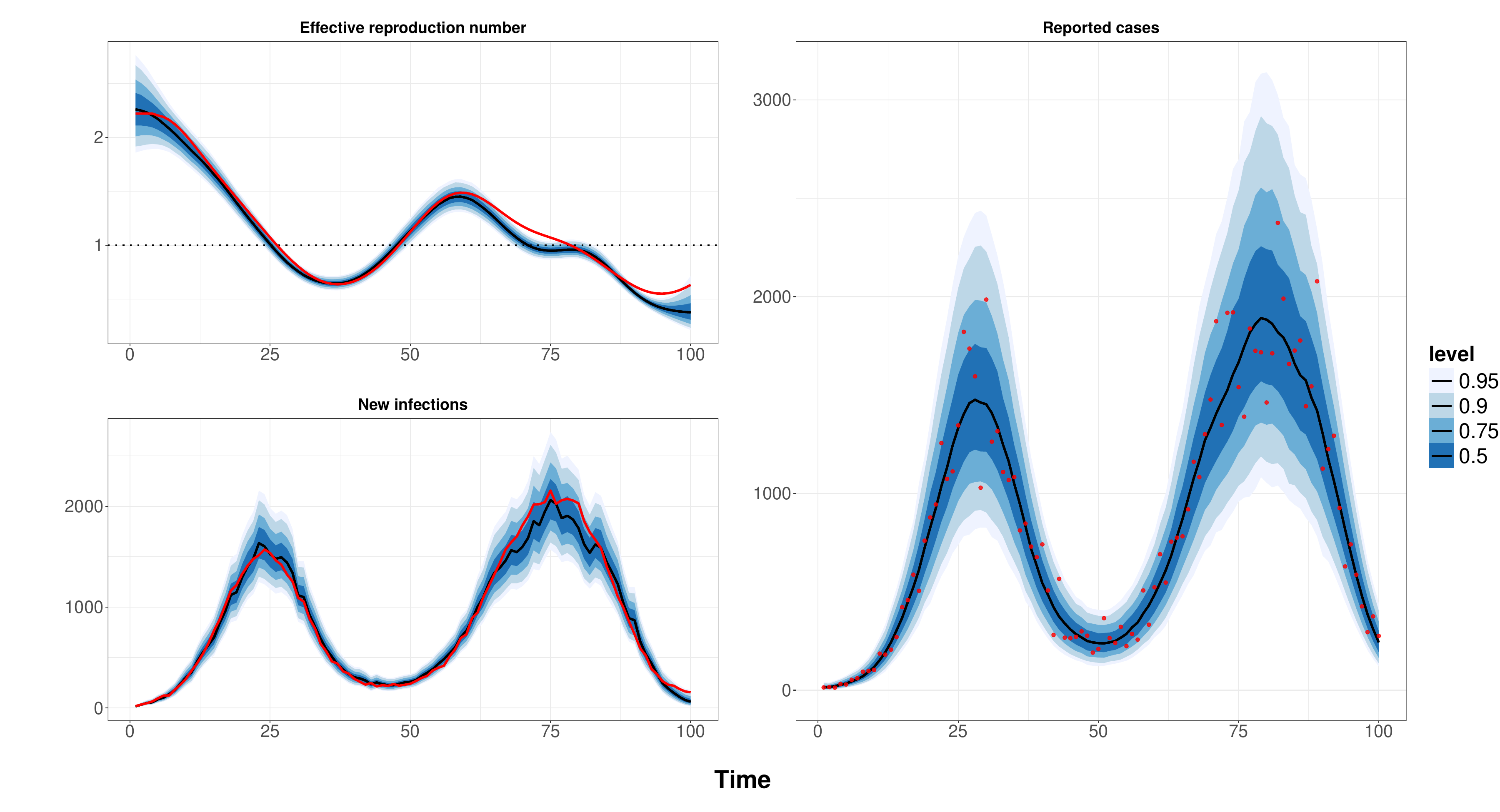}
    \caption{\footnotesize Marginal filtering distributions for the effective reproduction number $R_t$, total latent infections $I_t$, and reported cases $C_t$ aggregated across age groups. Solid black lines indicate posterior medians. Red lines correspond to the true simulated trajectories for $R_t$ and $I_t$, while red points denote observed synthetic case counts for $C_t$.}
    \label{fig:supp_total}
\end{figure}

\begin{figure}[H]
  \centering
  \includegraphics[width=0.87\linewidth]{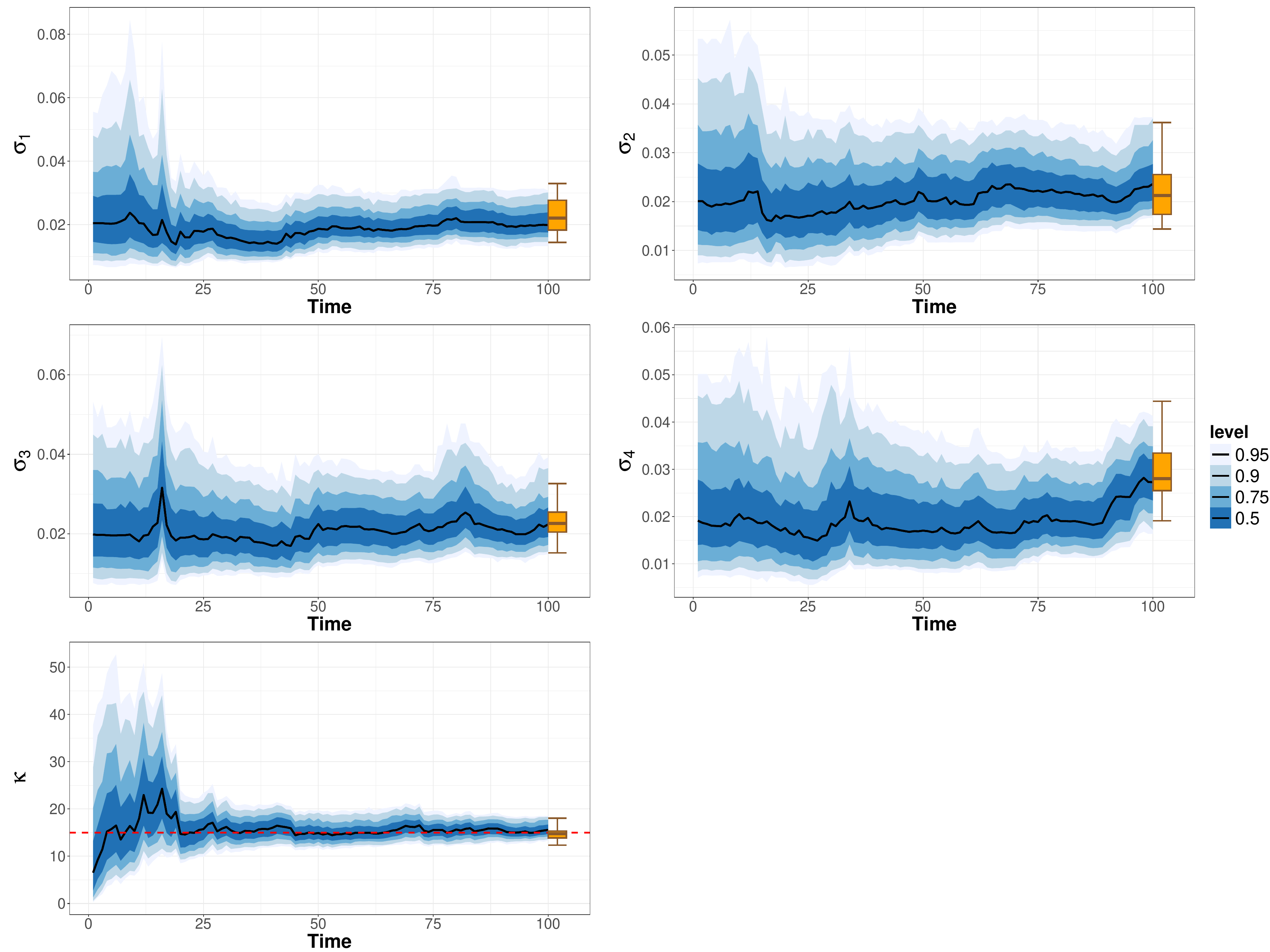}
  \caption{\footnotesize Marginal filtering estimates for the RW2 innovation standard deviations $\sigma_a$ ($a = 1, \dots, 4$) and the dispersion parameter $\kappa$ in the simulated data. The dashed red line indicates the true value of $\kappa$. The boxplot at the final time point summarises the marginal posterior distribution obtained from the PMMH.}
  \label{fig:supp_param_sim}
\end{figure}

Figure~\ref{fig:supp_total} summarises the corresponding marginal filtering distributions for the effective reproduction number $R_t$, the total number of latent infections $I_t$, and the aggregated reported cases $C_t$ across all age groups. In all cases, the posterior medians closely track the true simulated trajectories, with appropriate uncertainty quantification over time. Figure~\ref{fig:supp_param_sim} presents the marginal filtering estimates of the random-walk innovation standard deviations $\sigma_a$ ($a = 1, \dots, 4$) and the dispersion parameter $\kappa$. The filtering distributions of $\kappa$ concentrate around the true parameter values used in the data-generating process, demonstrating successful joint learning of both latent states and observation-level parameters. The boxplot at the final time point summarises the marginal posterior distribution obtained from the PMMH.

\end{document}